\documentclass[apj]{emulateapj}
\usepackage{apjfonts}
\usepackage{times}
\usepackage{CJK}

\usepackage{amsmath}
\usepackage{cleveref}
\usepackage{hyperref}
\usepackage{graphicx}
\usepackage{natbib}
\usepackage{lineno,hyperref}
\usepackage{amssymb}
\usepackage{lineno}

\usepackage{color}

\slugcomment{Accepted for publication in ApJ}
\shorttitle{}
\shortauthors{Le et al.}

\newcommand{\Hb}{H{$\beta$}}
\newcommand{\Ha}{H{$\alpha$}}
\newcommand{\FeII}{\ion{Fe}{2}}

\newcommand{\OIII}{[\ion{O}{3}]}
\newcommand{\OII}{[\ion{O}{2}]}
\newcommand{\NII}{[\ion{N}{2}]}
\newcommand{\SII}{[\ion{S}{2}]}

\newcommand{\mbh}{M$_{\rm BH}$}

\newcommand{\kms}{km~s$^{\rm -1}$}
\newcommand{\ergs}{erg s$^{-1}$}
\def\gsim{\mathrel{\rlap{\lower4pt\hbox{\hskip1pt$\sim$}}
    \raise1pt\hbox{$>$}}}         

\def\lsim{\mathrel{\rlap{\lower4pt\hbox{\hskip1pt$\sim$}}
    \raise1pt\hbox{$<$}}}         

\begin{document}
\begin{CJK*}{UTF8}{gbsn}

\title{\OIII\ 5007\AA\ Emission Line Width as a Surrogate for $\sigma_{*}$ in Type 1 AGNs?}

\author{Huynh Anh N. Le (黎阮黄英) $^{1,2,*}$}
\author{Yongquan Xue (薛永泉) $^{1,2,*}$}
\author{Xiaozhi Lin (林晓鸷) $^{1,2}$}
\author{Yijun Wang (王倚君) $^{3,4}$}

\affil{$^{1}$ CAS Key Laboratory for Research in Galaxies and Cosmology, Department of Astronomy, University of Science and Technology of China, Hefei 230026, China; lha@ustc.edu.cn; xuey@ustc.edu.cn; xzlin@ustc.edu.cn \\
$^{2}$ School of Astronomy and Space Science, University of Science and Technology of China, Hefei 230026, China} 
\affil{$^{3}$ Department of Astronomy, Nanjing University, Nanjing 210093, China; wangyijun@nju.edu.cn  \\
$^{4}$ Key Laboratory of Modern Astronomy and Astrophysics (Nanjing University), Ministry of Education, Nanjing 210093, China}
\altaffiltext{*}{Author to whom any correspondence should be addressed.}

\begin{abstract}

We present a study of the relation between the \OIII\ 5007\AA\ emission line width ($\rm \sigma_{[OIII]}$) and stellar velocity dispersion ($\sigma_{*}$), utilizing a sample of 740 type 1 active galactic nuclei (AGNs) with high-quality spectra at redshift z $<$ 1.0. We find the broad correlation between the core component of \OIII\ emission line width ($\rm \sigma_{[OIII,core]}$) and $\sigma_{*}$ with a scatter of 0.11~dex for the low redshift (z $<$ 0.1) sample; for redshift (0.3 $<$ z $<$ 1.0) AGNs, the scatter is larger, being 0.16~dex. We also find that the Eddington ratio ($\rm L_{bol}/L_{Edd}$) may play an important role in the discrepancies between $\rm \sigma_{[OIII,core]}$ and $\sigma_{*}$. As the $\rm L_{bol}/L_{Edd}$ increases, $\rm \sigma_{[OIII,core]}$ tends to be larger than $\sigma_{*}$. By classifying our local sample with different minor-to-major axis ratios, we find that $\sigma_{*}$ is larger than $\rm \sigma_{[OIII,core]}$ for those edge-on spiral galaxies. In addition, we also find that the effects of outflow strength properties such as maximum outflow velocity ($\rm V_{max}$) and the broader component of \OIII\ emission line width and line shift ($\rm \sigma_{[OIII,out]}$ and $\rm V_{[OIII,out]}$) may play a major role in the discrepancies between $\rm \sigma_{[OIII,core]}$ and $\sigma_{*}$. The discrepancies between $\rm \sigma_{[OIII,core]}$ and $\sigma_{*}$ are larger when $\rm V_{max}$, $\rm V_{[OIII,out]}$, and $\rm \sigma_{[OIII,out]}$ increase. Our results show that the outflow strengths may have significant effects on the differences between narrow-line region gas and stellar kinematics in AGNs. We suggest that caution should be taken when using $\rm \sigma_{[OIII,core]}$ as a surrogate for $\sigma_{*}$. In addition, the substitute of $\rm \sigma_{[OIII,core]}$ for $\sigma_{*}$ could be used only for low luminosity AGNs.  
\end{abstract}


\section{INTRODUCTION} \label{section:intro}

Black-hole mass (\mbh) is a fundamental driving factor in active galactic nuclei (AGNs). The study of the correlations between \mbh\ and properties of their host galaxies is important in understanding the physical nature and the growth history of AGNs. Over the last two decades, there have been many studies for investigating the co-evolution between supermassive black holes and their host galaxies \citep[e.g.,][]{Ferrarese00, Gebhardt00, MJ02, Tremaine+02, Woo06, Woo+10, Xue+10, Kormendy&Ho13, Le+14, Sun+15, Shankar+16, Xue+17, Shankar+19, Le+19, Woo+19b, Lin+22, Ayubinia+22}. The tight correlations between \mbh\ and their host galaxies suggest that black holes and their hosts are co-evolving and regulating each other. These scaling relations could be explained by either mutual merger processes or AGN feedback (\citealp{Silk98}; \citealp{Peng+06}). 

In spite of the tight correlations between black holes and their host galaxies, it is difficult for us to measure the host galaxy properties due to AGNs often outshining their host galaxies. In particular, the measurement of stellar velocity dispersions ($\sigma_{*}$) is challenging because of the contamination of AGN continuum and emission lines in the observed spectra (particularly for high redshift AGNs). To overcome this issue, some studies have suggested using the line width of \OIII\ 5007\AA\ ($\rm \sigma_{[OIII]}$) as a surrogate for the stellar velocity dispersion ($\sigma_{*}$) by assuming that the ionized gas kinematics of the narrow-line region (NLR) is followed the gravitational potential of the bulge of the host galaxy \citep[e.g.,][]{Wilson85, Terlevich90, Whittle92, Nelson+96, Nelson20, Boroson03, Shields+03, Greene05, Rice+06, Netzer+07, Salviander07, Salviander13}. By analyzing in detail the \OIII\ kinematics, \citet{Wilson85} is the first study which found that $\rm \sigma_{[OIII]}$ is correlated well with $\sigma_{*}$. Later on, using a large sample of 75 Seyfert galaxies, \citet{Nelson+96} found that there is a strong correlation between $\sigma_{*}$ and \OIII\ profile. By using 21 radio-quiet quasars, \citet{Bonning05} found that $\rm \sigma_{[OIII]}$ is on average consistent with $\sigma_{*}$, converted from the Faber-Jackson relation. Using a large sample from the Sloan Digital Sky Survey (SDSS, \citealp{York+20}), \citet{Salviander15} found a similar result as that of \citet{Bonning05}, supporting for the use of $\rm \sigma_{[OIII]}$ as a surrogate for $\sigma_{*}$ in statistical studies.    

The use of $\rm \sigma_{[OIII]}$ to replace for $\sigma_{*}$ is appealing because it is easy for us to detect and measure $\rm \sigma_{[OIII]}$ for low and high redshift AGNs. Thus, the \OIII\ emission line plays an important role in the study of cosmic evolution between \mbh\ and their host galaxies. However, some studies have also found that the \OIII\ emission line often shows an asymmetric profile (e.g., the blue wing component) which could be interpreted as signatures of the non-gravitational potential such as outflows \citep[e.g.,][]{Heckman84, Bae+14, Woo+16, Le+17}. Because of this reason, some studies have suggested using only the core component of \OIII\ emission line ($\rm \sigma_{[OIII,core]}$) as a surrogate for $\sigma_{*}$ \citep[e.g.,][]{Greene05, Komossa07, Bennert+18}. \citet{Greene05} found that after correcting for the blue wing component in the \OIII\ profile, $\sigma_{*}$ is traced well by $\rm \sigma_{[OIII]}$, albeit with large scatter. In addition, some other studies also proposed to use other NLR emission lines which have lower ionization potential compared to that of \OIII, suffering less from outflow effects, such as \OII\ 3727\AA, \NII\ 6583\AA\ and \SII\ 6716\AA,3731\AA\ \citep[e.g.,][]{Phillips86, Greene05, Komossa07, Ho09, Bennert+18}. However, the use of those emission lines as a surrogate for $\sigma_{*}$ also shows large scatter as the case of using $\rm \sigma_{[OIII,core]}$. Although the correlation between $\rm \sigma_{[OIII,core]}$ and $\sigma_{*}$ has been investigated by many studies, the driving factor for the large scatter between both components is still unclear. 

Most of the studies of comparison between $\rm \sigma_{[OIII]}$ and $\sigma_{*}$ are carried out by indirect comparison between the \mbh$-$$\rm \sigma_{[OIII]}$ and \mbh$-$$\rm \sigma_{*}$ relations \citep[e.g.,][]{Nelson20, Boroson03, Komossa07, Grupe04, Wang01}. Until now, there are a few studies in which the authors directly and simultaneously compare $\rm \sigma_{[OIII]}$ and $\sigma_{*}$. Using 1749 type 2 Seyfert galaxies, \citet{Greene05} studied the direct comparison between $\rm \sigma_{[OIII]}$ and $\sigma_{*}$. They found that $\rm \sigma_{[OIII,core]}$ is consistent with $\sigma_{*}$ after removing for the asymmetric blue wing in the \OIII\ profile, and the ratio $\rm \sigma_{[OIII]}$$/$$\sigma_{*}$ is larger as the Eddington ratio increases. \citet{Rice+06} analyzed 24 low redshift targets of mostly type 2 AGNs using spatially resolved spectra of Hubble Space Telescope and found that the scatter between $\rm \sigma_{[OIII]}$ and $\sigma_{*}$ is complicated and cannot be corrected by using a simple relation. \citet{Ho09} did a similar study as that of \citet{Greene05} but for the \NII\ 6583\AA\ emission line by using 345 galaxies from the Palomar spectroscopy survey, and found consistent results as those of \citet{Greene05}. For a larger sample, \citet{Woo+16} used 39,000 type 2 AGNs at z $<$ 0.3 to compare directly $\rm \sigma_{[OIII]}$ and $\sigma_{*}$. They found that $\rm \sigma_{[OIII]}$ fitted with double Gaussian is larger than $\sigma_{*}$ by a factor of 1.3-1.4, suggesting that the non-gravitational component (e.g., outflows) is relatively comparable to the gravitational potential. Recently, by using the SDSS sample of 611 hidden type 1 AGNs, \citet{Eun+17} investigated that $\sigma_{*}$ is consistent with velocity dispersions of \NII\ and $\rm \sigma_{[OIII,core]}$. \citet{Bennert+18} used 59 long-slit, high signal-to-noise (S/N) and spatially resolved Keck spectra to have a direct comparison of the relationship between $\rm \sigma_{[OIII,core]}$ and $\sigma_{*}$. They found large scatter in the correlation between $\rm \sigma_{[OIII,core]}$ and $\sigma_{*}$ in their sample and suggested that the use of $\rm \sigma_{[OIII,core]}$ as a surrogate for $\sigma_{*}$ should take caution when applied for individual targets. 

Besides the studies of comparison between $\rm \sigma_{[OIII]}$ and $\sigma_{*}$, studying the velocity offset of \OIII\ emission line ($\rm V_{[OIII]}$) is an interesting topic for exploring AGN-driven outflows. Many works have investigated $\rm V_{[OIII]}$ with respect to the systemic velocity of the host galaxy or lower ionization lines such as \OII\ or \SII\ \citep[e.g.,][]{Crenshaw00, Zamanov02, Eracleous04, Boroson05, Hu08, Komossa08, Zhang11, Bae+14}. Using high-quality spatially resolved spectra, \citet{Crenshaw00} studied the acceleration radial velocity based on the peak of \OIII\ emission line and interpreted their results using the AGN-driven biconical radial outflow model. Using single-aperture spectra, the velocity offset of the peak of \OIII\ has been investigated statistically \citep{Zamanov02, Eracleous04, Boroson05, Hu08}. \citet{Boroson05} found that more than half of their sources show blueshifted of the \OIII\ peaks with respect to the systemic velocity based on other lower ionization lines. They suggested that the blueshifted $\rm V_{[OIII]}$ may be governed by AGN-properties such as \mbh\ and Eddington ratio. Interestingly, the blueshifts of higher ionization lines such as \OIII\ are larger compared to that of lower ionization lines, e.g., \OII, \SII, \NII\ \citep{Eracleous04, Boroson05, Hu08}. By using a small sample of narrow-line Seyfert 1 galaxies with strong blueshifted \OIII, \citet{Komossa08} found that the velocity shift of the core component of \OIII\ emission line ($\rm V_{[OIII,core]}$) shows a moderate correlation with Eddington ratio and optical \FeII\ strength. \citet{Zhang11} found similar results but a weak correlation between $\rm V_{[OIII,core]}$ and Eddington ratio by using homogenous samples of radio-quiet Seyfert 1 galaxies. Using a large sample, $\sim$23,000 type 2 AGNs, \citet{Bae+14} found that half of their sources show $\rm V_{[OIII]}$ > 20 \kms, and the fractions of such $\rm V_{[OIII]}$ in type 1 and type 2 AGNs are similar after considering orientation effects. In short, the velocity of \OIII\ emission line does show some offset with respect to the systemic velocity. Particularly, the core component of \OIII\ also shows some offset and has weak or moderate correlations with AGN properties (e.g., Eddington ratio). It is therefore important that we should consider analyzing $\rm V_{[OIII]}$ when studying the correlation between $\rm \sigma_{[OIII,core]}$ and $\sigma_{*}$. $\rm V_{[OIII]}$ with respect to the systemic velocity may have impacts on the discrepancies between $\rm \sigma_{[OIII,core]}$ and $\sigma_{*}$.

In general, the correlations between $\rm \sigma_{[OIII,core]}$ and $\sigma_{*}$ mean that they are tracing the same velocity field of the gravitational potential. However, there is large scatter in these correlations and the driving factors for these discrepancies are still unclear. Therefore, a detailed study using a high S/N spectral sample is necessary for robust measurements of both $\rm \sigma_{[OIII]}$ and $\sigma_{*}$, and crucial for understanding the physically driven properties of gas and stellar kinematics in type 1 AGNs. In particular, for high redshift AGNs, we often utilize $\rm \sigma_{[OIII]}$ as a surrogate for $\sigma_{*}$ since it is challenging to measure $\sigma_{*}$ for high redshift objects. It is necessary to have a proper study for the relations between $\rm \sigma_{[OIII]}$ and $\sigma_{*}$ for high redshift AGNs. 

In this paper, we present a detailed comparison between $\rm \sigma_{[OIII]}$ and $\sigma_{*}$ for not only local AGNs but also for high redshift AGNs. For local AGNs (z $<$ 0.1), the sample from \citet{Bennert+18} with long-slit and high-quality spatially resolved $\sigma_{*}$ is a suitable sample for studying the relation between $\rm \sigma_{[OIII]}$ and $\sigma_{*}$ that has robust measurements of both quantities. In this study, by using the 59 local AGNs from \citet{Bennert+18}, we investigate in detail the discrepancy between $\rm \sigma_{[OIII]}$ and $\sigma_{*}$ as a function of AGN properties. Then, we apply the same approach for studying the local sample which has integrated spectra, observed with a larger aperture size (3$\arcsec$) by SDSS, including 611 hidden type 1 AGNs from \citet{Eun+17}. For high redshift AGNs (0.3 $<$ z $<$ 1.0), we select 18 high S/N Keck spectra \citep[]{Woo06, Woo08} and 52 co-added high S/N spectra from the SDSS Reverberation Mapping (SDSS-RM) project \citep{Shen+15}. Finally, our sample includes 740 targets at redshift z $<$ 1.0, covering the broad dynamic ranges of $\mathrm{\lambda L_{5100} \sim10^{41.5} - 10^{46.0}}$ \ergs and 6.5 $<$ $\log($\mbh) $<$ 9.5. The high-quality spectra of this sample enable robust measurements of $\rm \sigma_{[OIII]}$ and $\sigma_{*}$, hence providing us a good opportunity for studying in detail the direct relationship between both components and investigating for the physically driving factors of the discrepancies between them. 

In Section \ref{section:obs}, we describe the sample selection. Sections \ref{section:meas} and \ref{section:result} present the measurements and results, respectively. The discussions and summary are presented in Sections \ref{section:discuss} and \ref{section:sum}, respectively. The following cosmological parameters are used throughout the paper: $H_0 = 70$~km~s$^{-1}$~Mpc$^{-1}$, $\Omega_{\rm m} = 0.30$, and $\Omega_{\Lambda} = 0.70$.

\section{Sample Selection}\label{section:obs} 

In this work, we combined the aforementioned low and high redshift samples (z $<$ 1.0) which allow for robust measurements of $\sigma_{*}$ and the \OIII\ emission line widths ($\rm \sigma_{[OIII,core]}$ and $\rm \sigma_{[OIII,broad]}$) to have a detailed study on the correlations between gas and stellar kinematics not only for local AGNs but also for high redshift ones. 

Firstly, for the local AGN sample (z $<$ 0.1), we selected 59 targets of \citet{Bennert+18} which have observed using Low Resolution Imaging Spectrometer (LRIS, \citealp{Oke+95}) at the Keck telescope. Those targets have high S/N spectra which provide robust measurements of spatially resolved $\sigma_{*}$ within the bulge radius \citep{Harris+12}. In addition, for studying the relation between $\sigma_{*}$ and $\rm \sigma_{[OIII,core]}$ measured from the integrated SDSS spectra (aperture size 3$\arcsec$), we utilized 611 hidden type 1 AGNs (hereafter, local SDSS sample) from \citet{Eun+17} which have reliable measurements of $\sigma_{*}$ and $\rm \sigma_{[OIII,core]}$. 

Secondly, for high redshift AGNs (0.3 $<$ z $<$ 1.0), we chose 18 sources from \citet[]{Woo06,Woo08}. The sample was also observed using LRIS-Keck, and has high S/N spectra which allowed for subtracting the contamination of AGN continuum component in the stellar feature, leading to reliable estimation of $\sigma_{*}$. To enlarge the dynamic range of AGN luminosity, we selected 88 co-added sources with co-added high S/N spectra from the SDSS-RM project \citep{Shen+15} which have robust measurements of $\sigma_{*}$. Among them, we removed 36 sources which have large noisy distortion surrounding the \OIII\ and \Hb\ emission line profiles. Then, the remaining SDSS-RM sample has 52 sources which have reliable determinations of both $\sigma_{*}$ and $\rm \sigma_{[OIII]}$. In total, we obtained 70 AGNs (hereafter, high redshift sample) which have a unique broad luminosity dynamic range and are suitable for studying the relation between $\sigma_{*}$ and $\rm \sigma_{[OIII,core]}$ at redshift 0.3 $<$ z $<$ 1.0. With the advantages of high-quality spectral sample observed with LRIS-Keck and co-added high S/N SDSS-RM spectra, we carry out  a detailed study of the direct relation between $\sigma_{*}$ and $\rm \sigma_{[OIII,core]}$ for high redshift AGNs. 

\begin{figure*}
	\includegraphics[width=0.38\textwidth]{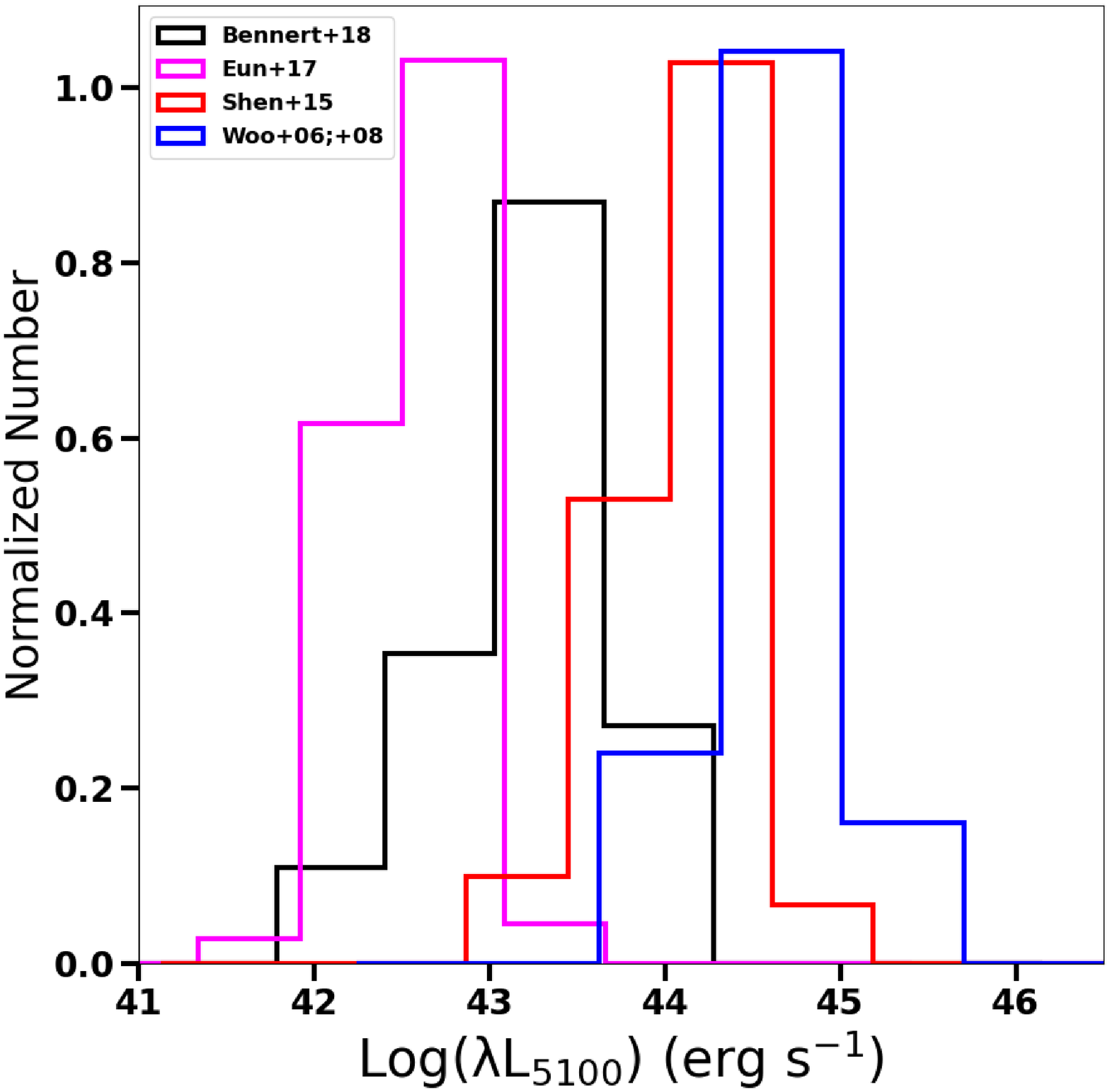}
	\includegraphics[width=0.38\textwidth]{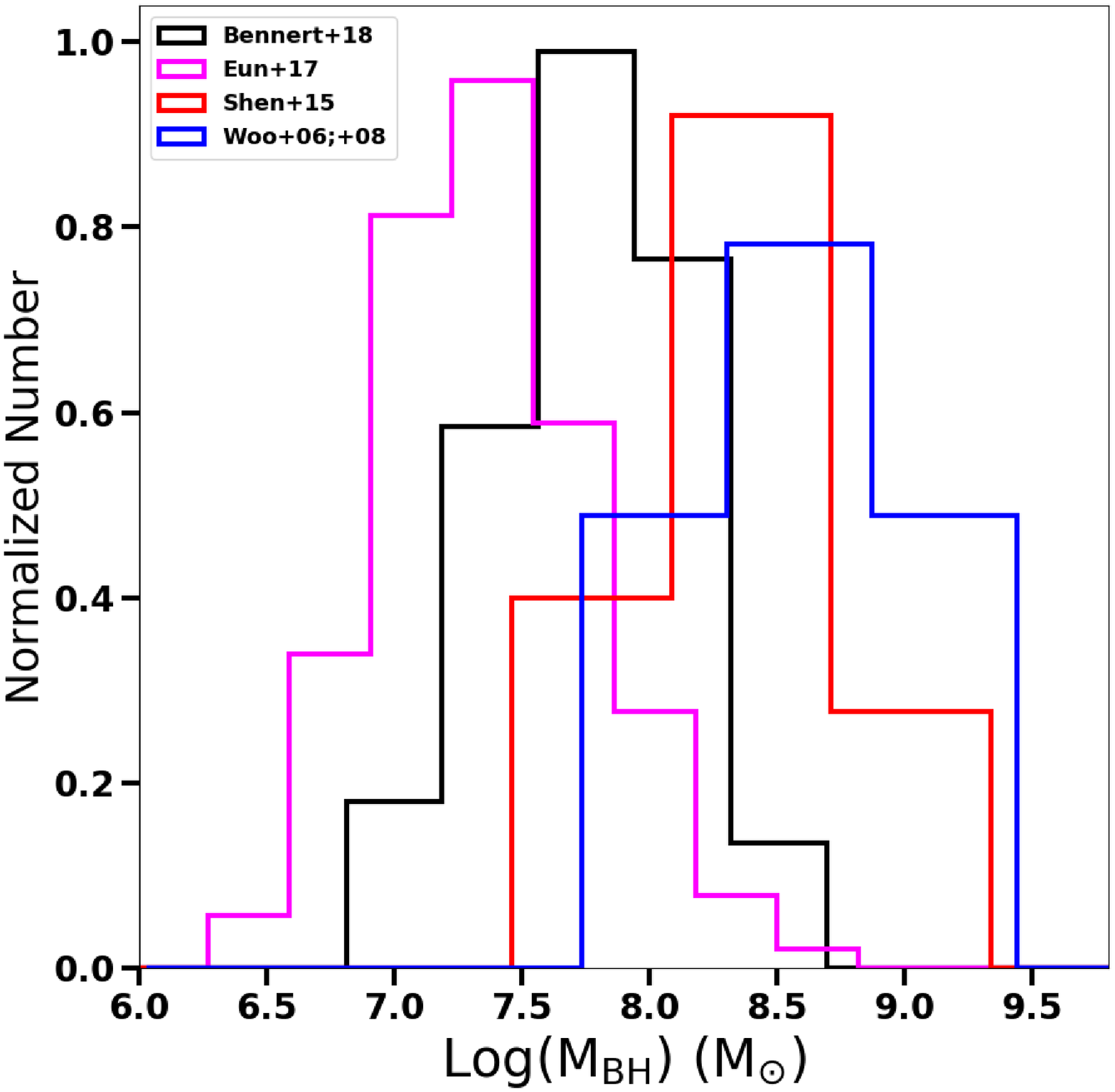}	
	\centering
	\caption{Distributions of continuum luminosity at 5100 \AA\ (left panel) and black-hole mass (right panel). The magenta, black, blue, and red histograms show the distributions of the samples of \citet{Eun+17}, \citet{Bennert+18}, \citet{Woo06,Woo08}, and \citet{Shen+15}, respectively.
	\label{fig:hist}}
\end{figure*}

Finally, our sample included 740 local and high redshift AGNs, covering the broad dynamic ranges of $\mathrm{\lambda L_{5100} \sim10^{41.5} - 10^{46.0}}$ \ergs\ and 6.5 $<$ $\log($\mbh) $<$ 9.5 (see Section 3.1). Figure \ref{fig:hist} shows the distributions of $\mathrm{M_{\rm BH}}$ and continuum luminosity $\mathrm{L_{5100}}$ of our sample. 

\section{Measurements}\label{section:meas}

\subsection{Spectral Analysis and AGN property measurements}\label{section:property}

In our study, we have four different subsamples of \citet[]{Woo06,Woo08}, \citet{Shen+15}, \citet{Eun+17} and \citet{Bennert+18}. In this section, we describe the spectral analysis processes of those subsamples to obtain pure emission spectra for measuring AGN properties. 

Similar to 18 targets of \citet[]{Woo06,Woo08}, 59 AGNs from \citet{Bennert+18} were also observed with LRIS-Keck. So, the data reduction and spectral analysis of both samples were similar to each other. In our previous work in \citet{Woo+18}, we have updated the multi-component spectral analysis of 18 sources in \citet[]{Woo06,Woo08} following the proredure of \citet{Park+15}. Similarly, \citet{Bennert+18} also applied the method of \citet{Park+15} in their spectral analysis. Here, we briefly describe the recipe of our optical spectral analysis. A model with a combination of the pseudo-continuum including a single power law and an \FeII\ component based on the I Zw 1 \FeII\ template \citep{BG92}, as well as a host-galaxy component adopted from the stellar template of the Indo-US spectral library in \citet{Valdes04} was applied to fit the observed optical spectra within the wavelength ranges of 4430\AA\ $-$ 4770\AA\ and 5080\AA\ $-$ 5450\AA. The best-fit model was determined using the $\chi^{2}$ minimization from MPFIT. After subtracting the pseudo-continuum from the observed spectra, a sixth order Gauss-Hermite series was applied to fit the broad component of the \Hb\ emission line. The narrow component of \Hb\ was modeled separately by using the \OIII\ 5007\AA\ best-fit model. From the best-fit model, we measured line width ($\rm FWHM_{H\beta}$), line dispersion ($\rm \sigma_{H\beta}$), the luminosity of the \Hb\ line ($\rm L_{H\beta}$), and the monochromatic luminosity at 5100\AA\ ($\rm L_{5100}$). The updated measured properties of 18 targets in \citet[]{Woo06,Woo08} can be found in Table 1 of \citet{Woo+18}, and the resulting measurements of 59 sources of \citet{Bennert+18} are shown in Table 1 of \citet{Bennert+15}. 

\begin{figure*}
	\includegraphics[width=0.65\textwidth]{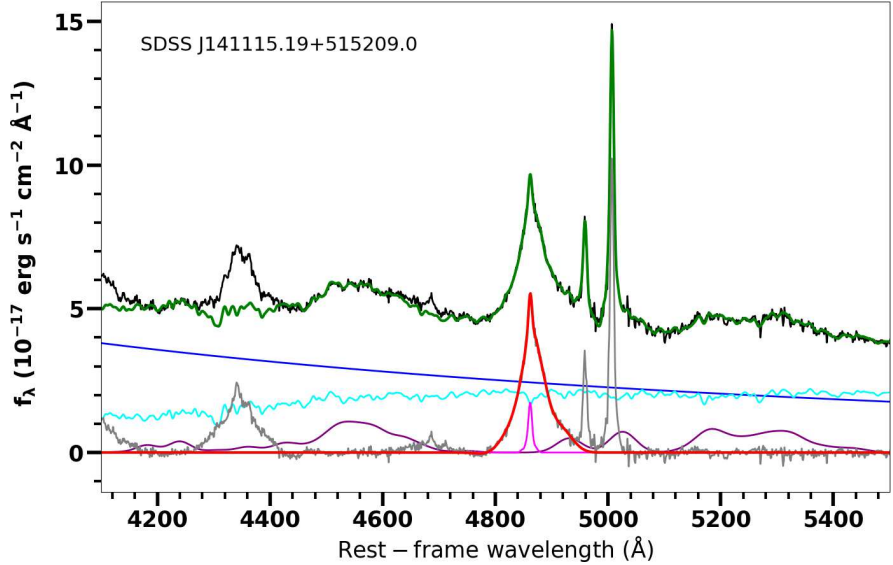}
	\centering
	\caption{Example of multi-component fitting result for the SDSS spectrum, SDSS J141115.19$-$515209.0 from the sample of \citet{Shen+15}. The rest-frame SDSS spectrum is in thick black. The total model (green) includes power-law continuum (blue), \ion{Fe}{2} model (purple), and stellar model (cyan). The continuum subtracted emission line is displayed in gray and the \Hb\ line broad and narrow models are presented in red and pink, respectively.
\label{fig:spectra_fit}}
\end{figure*}

For 52 sources from \citet{Shen+15}, the authors have provided the co-added integrated spectra in their public catalog. To obtain the pure emission line spectra from the provided co-added spectra, we performed multi-component spectral analysis for 52 SDSS-RM spectra similarly as that for targets observed with LRIS-Keck. Figure~\ref{fig:spectra_fit} shows an example of multi-component fitting results for a target from the SDSS-RM sample.


The spectral analysis of the local SDSS sample is presented in detail in Section 2 of \citet{Eun+17}. This sample is basically selected from the type 2 AGN catalog of \citet{Bae+14}. After subtracting the SDSS spectra from the best-fit stellar population model using the penalized pixel-fitting code (pPXF, \citealp{Cappellari04}), the authors identified the hidden type 1 AGNs among the type 2 AGNs based on a careful analysis of the broad component of \Ha\ emission line. 

\begin{figure}
	\includegraphics[width=0.49\textwidth]{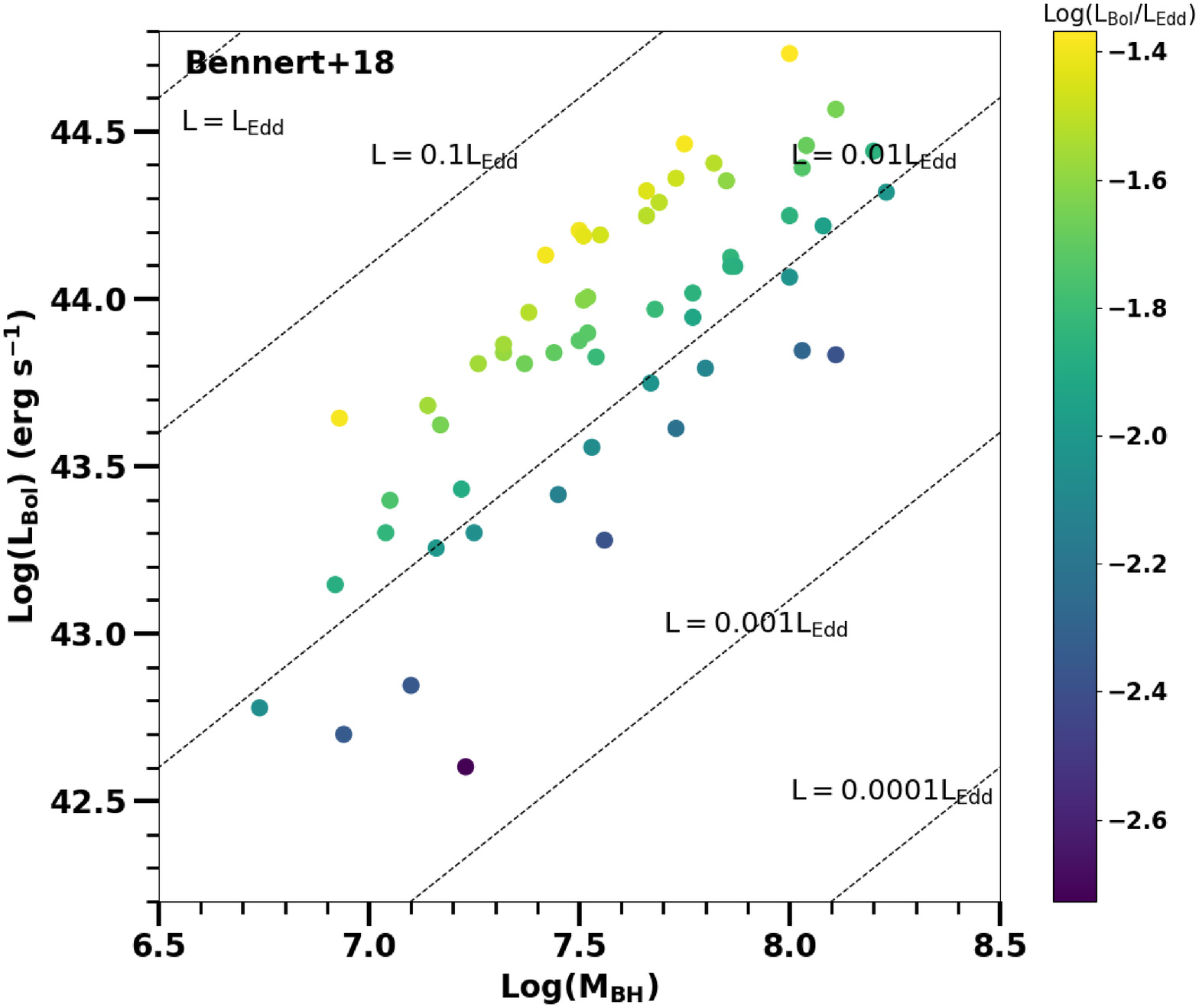}
	\includegraphics[width=0.49\textwidth]{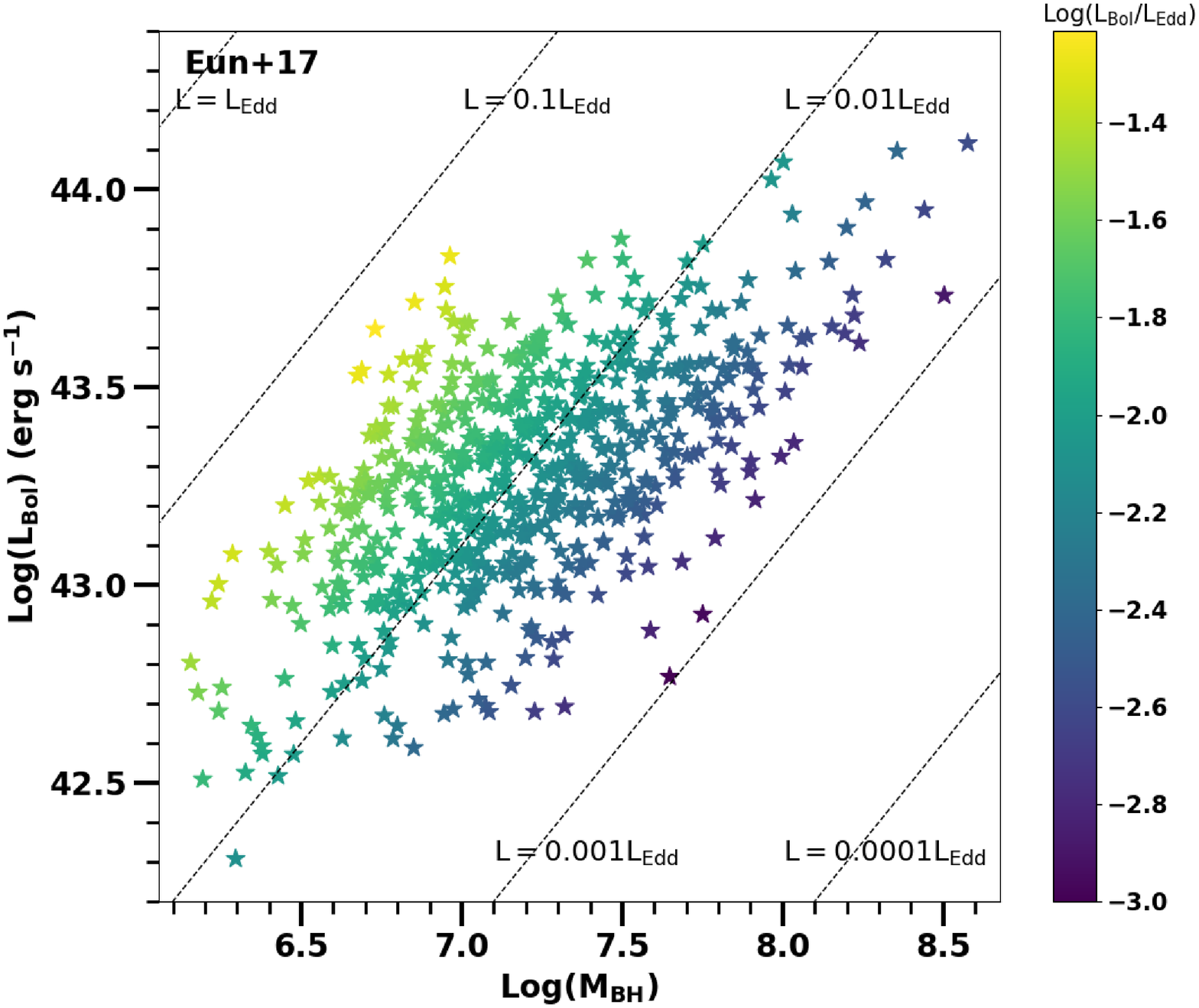}
	\includegraphics[width=0.49\textwidth]{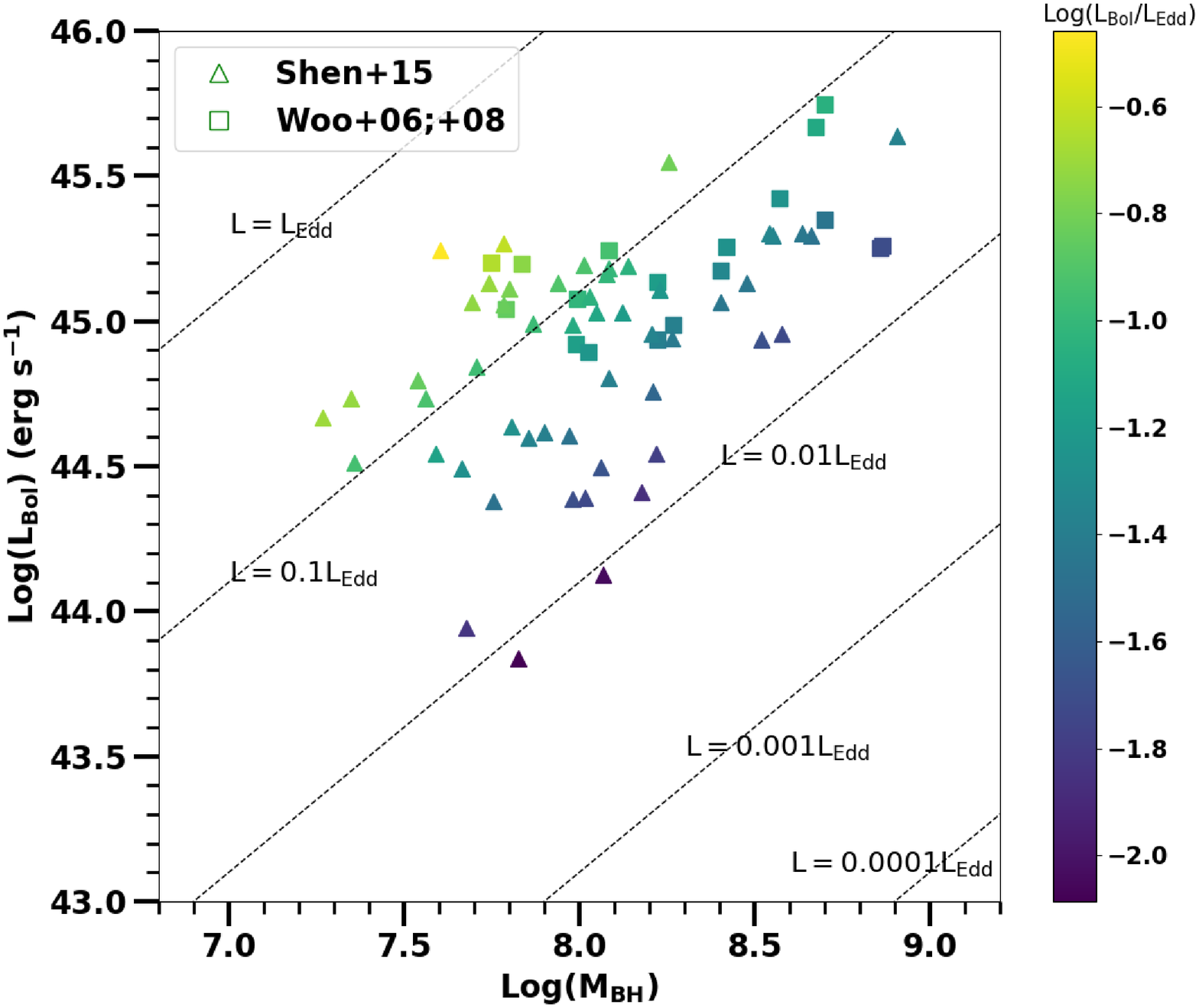}
	\centering
	\caption{Distributions of \mbh, $\rm L_{bol}$, and Eddington ratio for those subsamples of \citet{Bennert+18}, \citet{Eun+17}, \citet[]{Woo06,Woo08} and \citet{Shen+15}, repsectively. The dash lines indicate different Eddington ratios. 
	\label{fig:mbh_bol_edd}}
\end{figure}

To have a consistent measurement of \mbh\ for all sources in our sample, we applied the single-epoch method, using the virial theorem and \Hb\ size-luminosity relation based on the result of \citet{Bentz+13} (see equation 1 in \citealp{Le+20}) for calculating \mbh\ for the high redshift sample (\citealp[]{Woo06,Woo08} and \citealp{Shen+15}). As in our previous work, we chose $\rm \sigma_{H\beta}$ and the monochromatic luminosity at 5100\AA\ ($\rm L_{5100}$) for determining $\rm M_{BH}$ \citep{Le+20}. We used the virial factor (f = 4.47) from \citet{Woo15} for the masses based on $\rm \sigma_{H\beta}$. From \mbh\ measurements, we determined Eddington ratio for our sample. The bolometric luminosity is calculated from the monochromatic luminosity at 5100\AA\ ($\rm L_{bol}$ = 10$\times$$\rm L_{5100}$). For \mbh\ measurements of the local SDSS sample, \citet{Eun+17} used the single-epoch method to determine \mbh\ for their sample using \Hb\ and \Ha\ emission lines. We simply adopted their measurements in our analysis. Figure \ref{fig:mbh_bol_edd} displays the distributions of \mbh, $\rm L_{bol}$, and Eddington ratio for all subsamples of \citet{Bennert+18}, \citet{Eun+17}, \citet[]{Woo06,Woo08} and \citet{Shen+15}, respectively.

\subsection{Stellar Velocity Dispersion and Velocity Shift}\label{section:stellar}

With the high S/N spectra of our selected sample, $\sigma_{*}$ has been robustly measured by previous works of \citet[]{Woo06,Woo08}, \citet{Shen+15}, \citet{Eun+17}, and \citet{Bennert+18}. Basically, the authors used similar approaches for estimating $\sigma_{*}$ by comparing in the pixel space between the observed spectra and the Gaussian-broadened stellar template spectra (G and K giants). Therefore, in this work, we simply adopted those measurements for our analysis. Here, we briefly explain their measurements. 

\citet[]{Woo06,Woo08} measured $\sigma_{*}$ for their sample by using stellar absorption lines around Mg b (5175\AA) and Fe (5270\AA). After subtracting the AGN Fe emission, the authors compared in the pixel space the pure observed spectra with five Gaussian-broadened stellar template spectra (G8, G9, K0, K2, and K5 giants). Then the final $\sigma_{*}$ estimations were based on the best-fit template. 

For SDSS-RM targets, \citet{Shen+15} used the {\it{vdispfit}} package and the direct-pixel-fitting code Penalized Pixel-Fitting (pPXF) to measure $\sigma_{*}$. The measurement was obtained within the wavelength range of 4125-5350 \AA\ containing the G band (4304\AA), Mg Ib (5167\AA, 5173\AA, 5184\AA), and Fe (5270\AA). Given the advantage of high S/N co-added spectra, the authors obtained robust measurements for 88 targets among the total of 212 AGNs used in their sample. We neglected some sources which have noisy distortion surrounding the \Hb\ and \OIII\ emission lines for a reliable comparison with $\sigma_{*}$. Finally, we selected 52 targets from \citet{Shen+15} in our study. 

Similar to \citet[]{Woo06,Woo08}, \citet{Bennert+18} compared their observed spectra with the Gaussian-broadened stellar templates of G, K, A0, and F2 stars from the Indo-US survey within three spectral regions of Ca H and Ca K (3969\AA\ and 3934\AA), Mg Ib (5167\AA, 5173\AA, 5184\AA) and Ca II (8498\AA, 8542\AA, 8662\AA). From the high-quality long-slit spectra of LRIS-Keck, the sample of \citet{Bennert+18} allowed for the robust measurements of spatially resolved $\sigma_{*}$ within the bulge radius (obtained from the surface photometry fitting of SDSS image, \citealp{Harris+12}). 

Finally, for the local SDSS hidden type 1 AGN sample, \citet{Eun+17} used $\sigma_{*}$ available from the MPA-JHU \citep{Abazajian09} catalog. We adopted those $\sigma_{*}$ values in our analysis.  

All of the $\sigma_{*}$ used in this work were corrected for the instrumental resolution of LRIS-Keck ($\sim$58 \kms) and SDSS ($\sim$65 \kms), respectively.

The high S/N spectra of our selected samples also allow us to measure the stellar absorption line velocity ($\rm V_{*}$) accurately and provide a reliable systemic velocity for analyzing the 
kinematics of \OIII\ emission line. Therefore, we adopted $\rm V_{*}$ as systemic velocity in this work. We used the provided $\rm V_{*}$ from \citet[]{Woo06,Woo08}, \citet{Shen+16}, and \citet{Eun+17}. For 59 targets of \citet{Bennert+18}, since the authors did not provide $\rm V_{*}$, we adopted the measured $\rm V_{*}$ of this sample (high S/N spectra, 38 sources) from \citet{Sexton+20} using the SDSS spectra. \citet{Sexton+20} developed the Bayesian decomposition analysis for SDSS spectra (BADASS) using Python. BADASS is a powerful fitting package for obtaining all spectral components of SDSS spectra simultaneously and accurately by utilizing a Markov-Chain Monte Carlo (MCMC) fitting technique.

\subsection{Gas Emission Line Kinematics}\label{section:oiii}

In this section, we present the fitting model to measure $\rm \sigma_{[OIII]}$ and $\rm V_{[OIII]}$. In addition, we also measure outflow strength properties, e.g., maximum outflow velocity ($\rm V_{max}$), to study the effects of outflow strength on the relations between $\rm \sigma_{[OIII]}$ and $\sigma_{*}$ . 

\subsubsection{Line Width and Line Shift Measurements}

Following the results of \citet{Bennert+18}, a double Gaussian is the best approach to fit the \OIII\ emission line. In this fitting, a Gaussian is fitted to the narrow (core) component of the \OIII\ line ($\rm \sigma_{[OIII,core]}$). While, another Gaussian is modeled for the broader (outflow) component of the \OIII\ profile ($\rm \sigma_{[OIII,out]}$). Based on the best-fit model of the \OIII\ emission line, we calculated the second moment ($\rm \sigma_{[OIII]}$) as follows:

\begin{equation}\label{eq:secmom}
\sigma_{\rm [OIII]} ^{2}=\frac{\int \lambda^{2} f_{\lambda }d\lambda }{\int f_{\lambda } d\lambda }- \lambda _{0}^{2},
\end{equation}
Here, $f_{\lambda}$ is the flux at each wavelength. $\rm \lambda _{0}$ is the first moment (centroid wavelength) of the emission line which is calculated as follows:

\begin{equation}\label{eq:firstmom}
\lambda _{0} = \frac{\int \lambda f_{\lambda }d\lambda }{\int f_{\lambda } d\lambda },
\end{equation}

\begin{figure*}
	\includegraphics[width=0.45\textwidth]{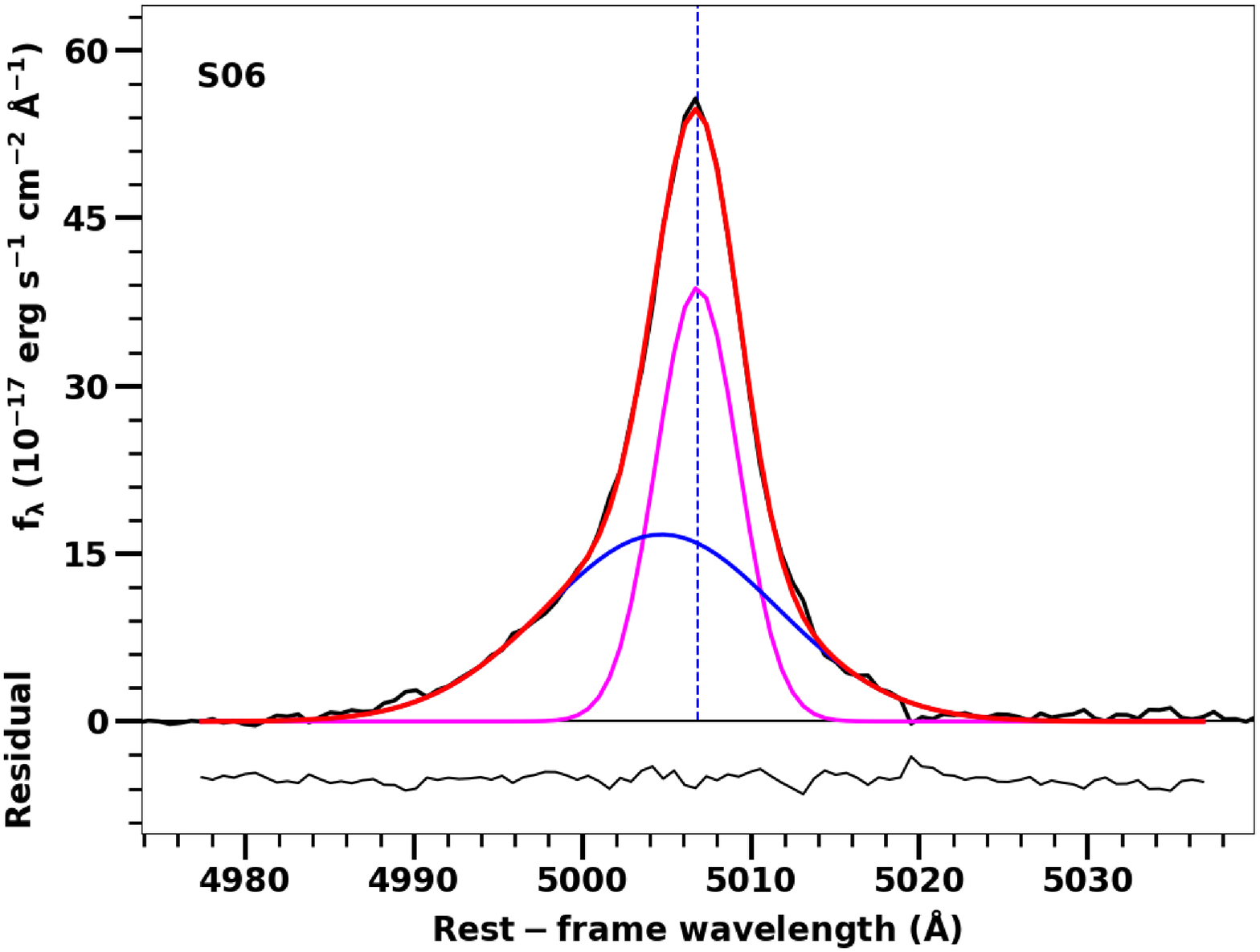}
	\includegraphics[width=0.45\textwidth]{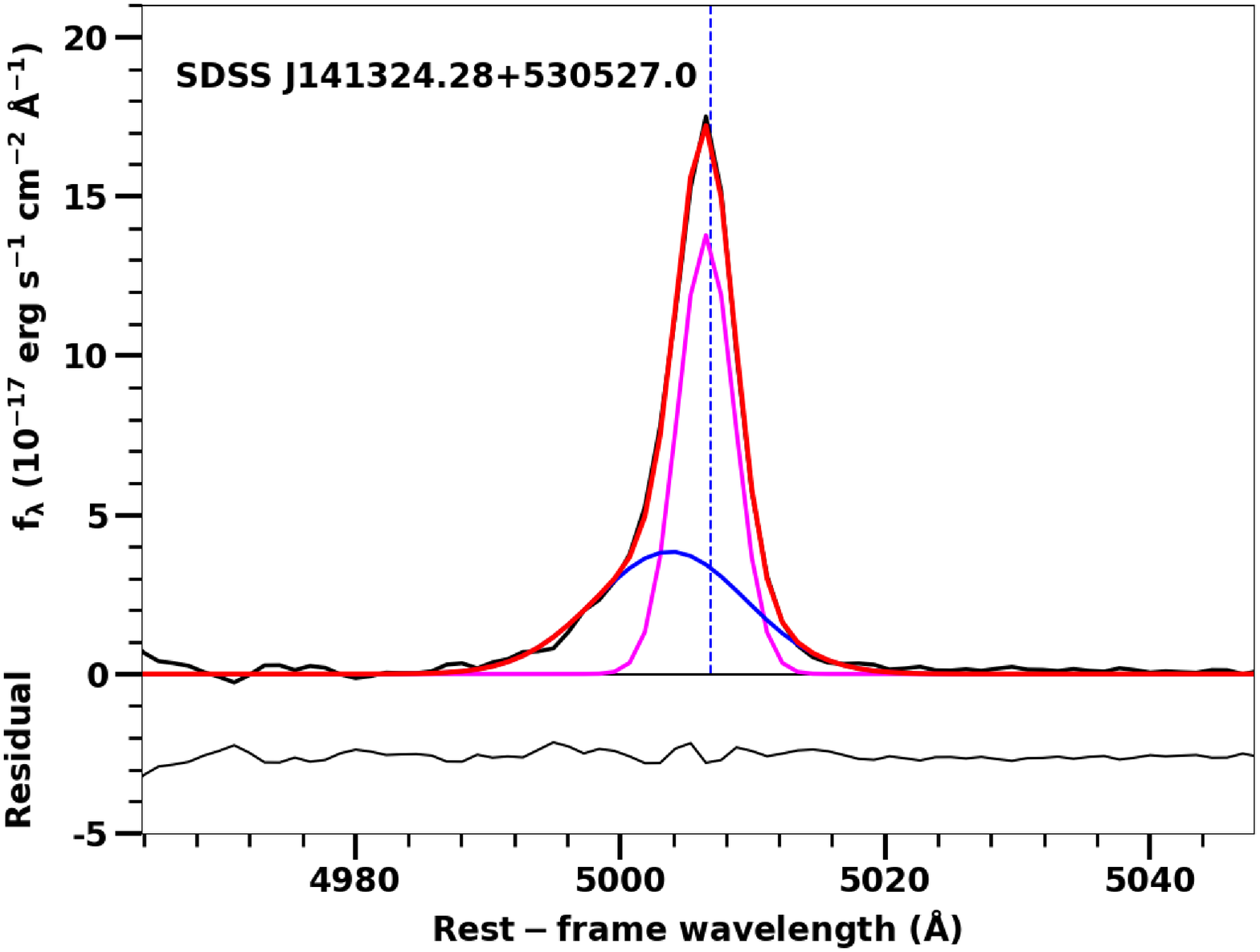}
	\centering
	\caption{Examples of \OIII\ 5007\AA\ fitting models for S06 and SDSS J141324.28$+$530527.0 from the samples of \citet[]{Woo06,Woo08} and \citet{Shen+15}, respectively. The total model is shown in red. The core component of \OIII\ is fitted by a Gaussian model and displayed in pink. The broader component (blue) is also fitted by a Gaussian function. The dash-line shows the rest-frame position of \OIII\ at 5007\AA. 
	\label{fig:oiii_fit}}
\end{figure*}

In our analysis, after subtracting the pseudo-continuum from the observed spectra, we also applied a double Gaussian to model the \OIII\ emission line for 18 targets of \citet[]{Woo06,Woo08} and 52 targets of \citet{Shen+15}. However, in some targets, if the peak of the broader component model of the \OIII\ emission line is lower than the noise level in the continuum (i.e., the amplitude peak-to-noise ratio < 3), we discarded the broad component as noise and used a single Gaussian function to fit the \OIII\ profile. Figure \ref{fig:oiii_fit} presents examples of \OIII\ fitting models of S06 and SDSS J141324.28$+$530527.0 from the samples of \citet[]{Woo06,Woo08} and \citet{Shen+15}, respectively. The measured \OIII\ dispersions $\rm \sigma_{[OIII,core]}$ and $\rm \sigma_{[OIII,out]}$ were also corrected for the instrumental resolution of LRIS-Keck ($\sim$58 \kms) and SDSS ($\sim$65 \kms), respectively. 

As mentioned in Section \ref{section:stellar}, we adopted $\rm V_{*}$ as the systemic velocity. We measured the velocity shifts based on the peaks of the core ($\rm V_{[OIII,core]}$) and the broad ($\rm V_{[OIII,out]}$) components of \OIII\ emission line.

Similar to our analysis, \citet{Eun+17} used a double Gaussian to model the \OIII\ emission line for the local SDSS sample. Section 3.1 of \citet{Eun+17} presents their emission line fitting model in detail. We adopted the measurements in their provided catalog in our study. 

For the 59 targets of \citet{Bennert+18}, since the authors did not release the spectra, we simply adopted their measurements of $\rm \sigma_{[OIII,core]}$ (see Table 1 in \citealp{Bennert+18}). In addition, we also used the measured \OII\ dispersion from \citet{Bennert+18} for further analysis in Section \ref{section:result}. For the velocity shift measurements, we adopted the measured $\rm V_{[OIII,core]}$ and $\rm V_{[OIII,out]}$ with respect to the stellar systemic velocity of this sample from \citet{Sexton+20} using the SDSS spectra. Among the 59 targets of \citet{Bennert+18}, \citet{Sexton+20} measured $\rm V_{[OIII,core]}$ and $\rm V_{[OIII,out]}$ for 38 targets which show strong outflow signatures in their \OIII\ emission line profiles.

\subsubsection{Maximum Outflow Velocity}

$\rm \sigma_{[OIII,out]}$ is often considered as a tracer of outflow signatures in AGNs (e.g., \citealp{Greene05}, \citealp{Bae+14}, \citealp{Woo+16}, \citealp{Le+17}). In our analysis, we determined the outflow strengths of our sample such as $\rm \sigma_{[OIII,out]}$ and $\rm V_{max}$. 

Following the relation from \citet{Harrison+14}, we determined $\rm V_{max}$ from the \OIII\ fitting models of our sample as follows:

\begin{equation}
	\rm V_{max} = \rm \Delta V + \rm \frac{W_{80}}{2} ,
	\label{eq:outflow}
\end{equation}
$\rm \Delta V$ is the measured velocity offset of the broader component with respect to that of the core component of the \OIII\ emission line. A negative value of $\rm \Delta V$ means that the broad component is blueshifted, while a positive value shows the redshifted offset. $\rm W_{80}$ presents the width of 80$\%$ of the Gaussian flux of the broad component of \OIII\ emission line, $\rm W_{80}$ $=$ 1.09 $\times$ $\rm FWHM_{[OIII,out]}$.  

Since we do not have the spectra for 59 targets of \citet{Bennert+18}, we adopted the measured outflow velocities of this sample from \citet{Sexton+20} using the SDSS spectra. 

For the samples of \citet[]{Woo06,Woo08}, \citet{Shen+15}, and \citet{Eun+17}, we measured $\rm V_{max}$ for 262 sources which show the broad components in the \OIII\ profiles. Totally, we measured $\rm V_{max}$ for 300 sources in our sample. Table \ref{table:sample} shows the targets and measured properties for the sample of \citet[]{Woo06,Woo08} and \citet{Shen+15}. 

Following the assumption of uncertainties from \citet{Bennert+18}, we adopted 0.4 dex for the uncertainty of black-hole mass measurements. For other measured properties, we applied 0.04 dex as the uncertainties.  

\section{Results}\label{section:result}

In this section, we study the direct relation between the gas emission line widths ($\rm \sigma_{[OIII,core]}$ and $\rm \sigma_{[OII]}$), and $\rm \sigma_{*}$.  
For a detailed study of the difference between those components, we compared the ratios of $\rm \sigma_{[OIII,core]}$$/$$\sigma_{*}$ and $\rm \sigma_{[OII]}$$/$$\sigma_{*}$ as a function of AGN properties such as black-hole mass, bolometric luminosity, Eddington ratio, and $\rm \sigma_{*}$. In addition, we also examined the effects of outflow strength (e.g., $\rm V_{max}$) and velocity shifts ($\rm V_{[OIII,core]}$ and $\rm V_{[OIII,out]}$) on the discrepancies between $\rm \sigma_{[OIII,core]}$ and $\rm \sigma_{*}$. 

\subsection{Gas Emission Line Width vs. Stellar Velocity Dispersion}\label{section:result_sigma} 

\begin{figure*}
	\includegraphics[width=0.485\textwidth]{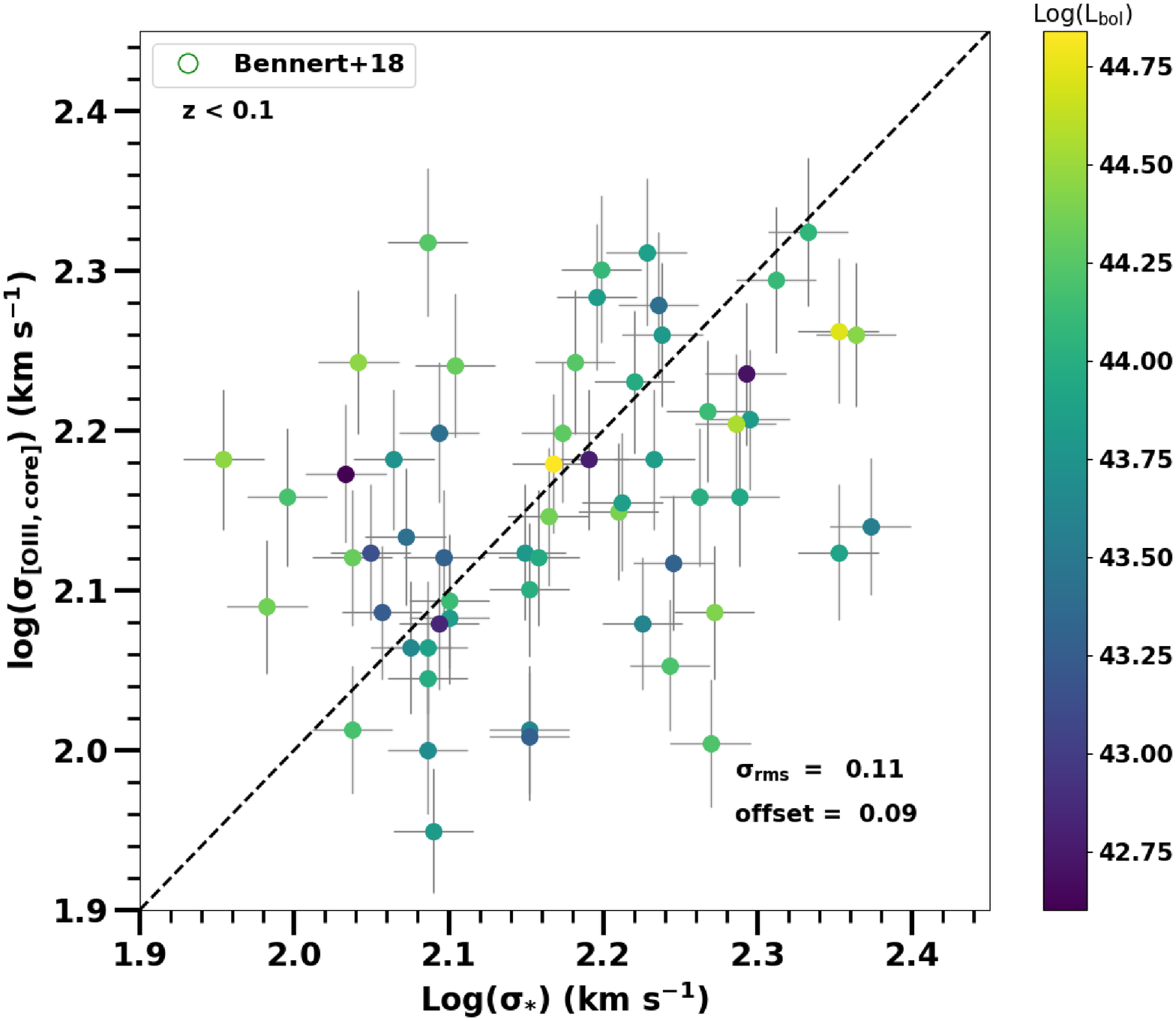}
	\includegraphics[width=0.485\textwidth]{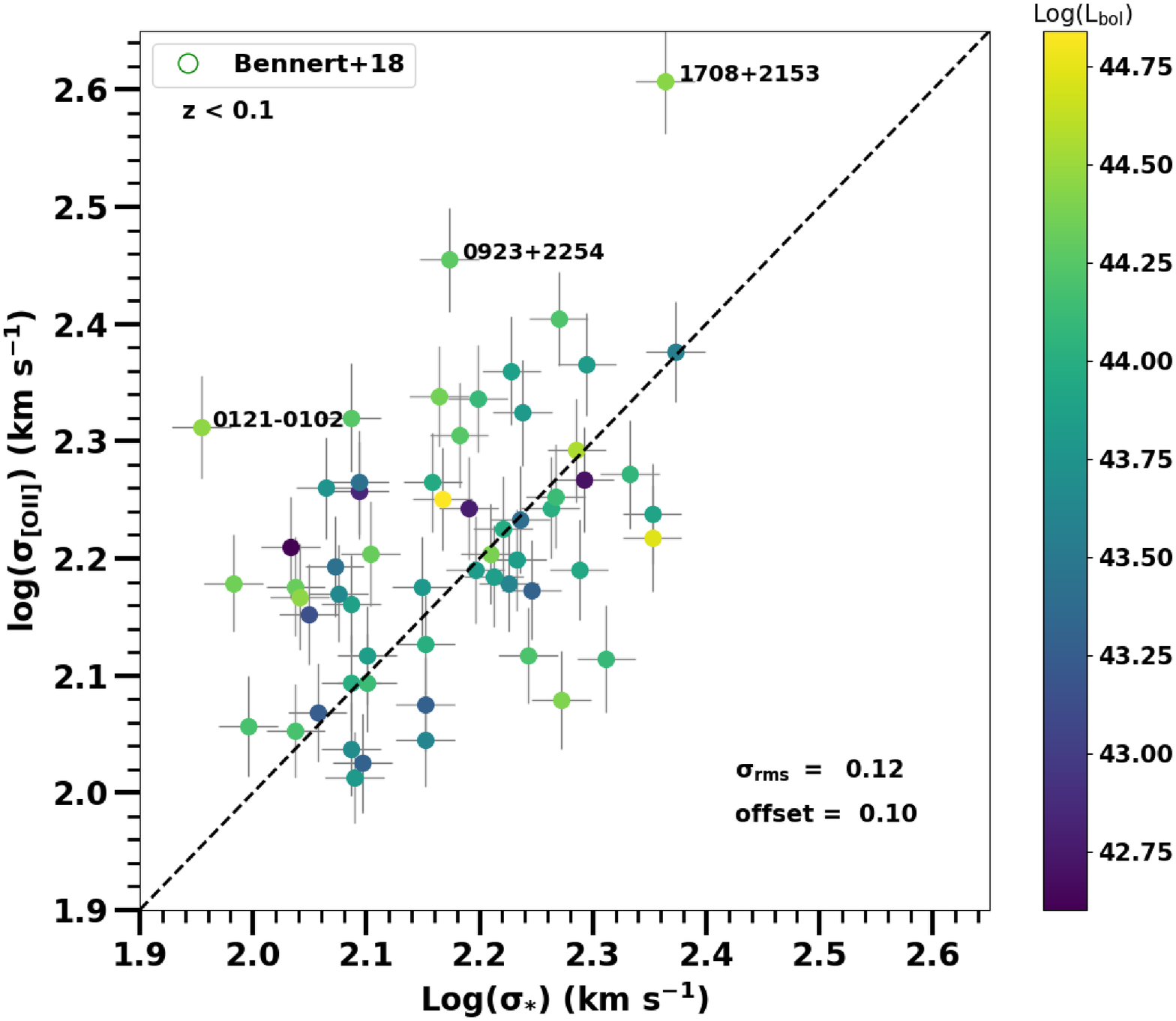}
	\includegraphics[width=0.485\textwidth]{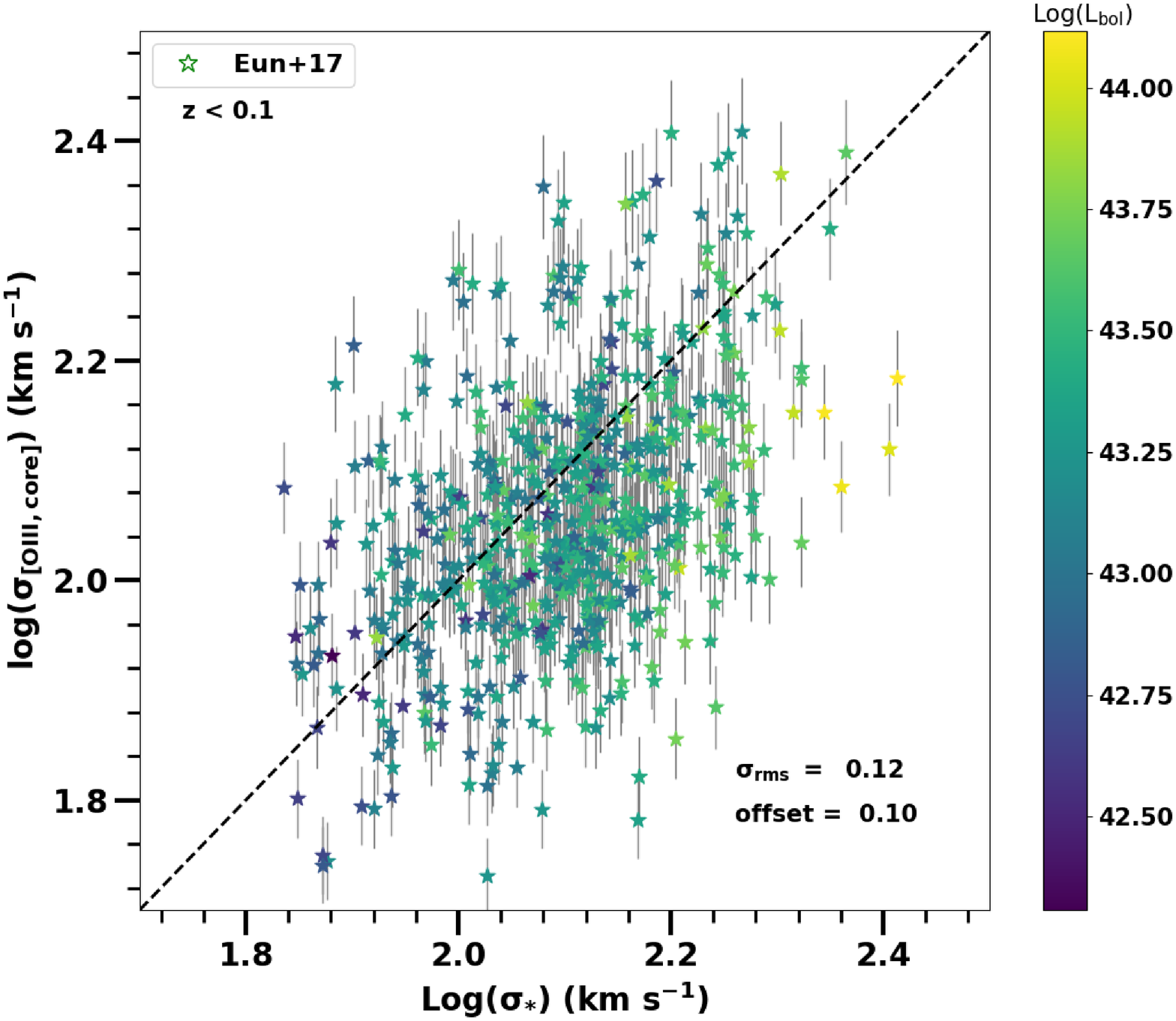}
	\includegraphics[width=0.485\textwidth]{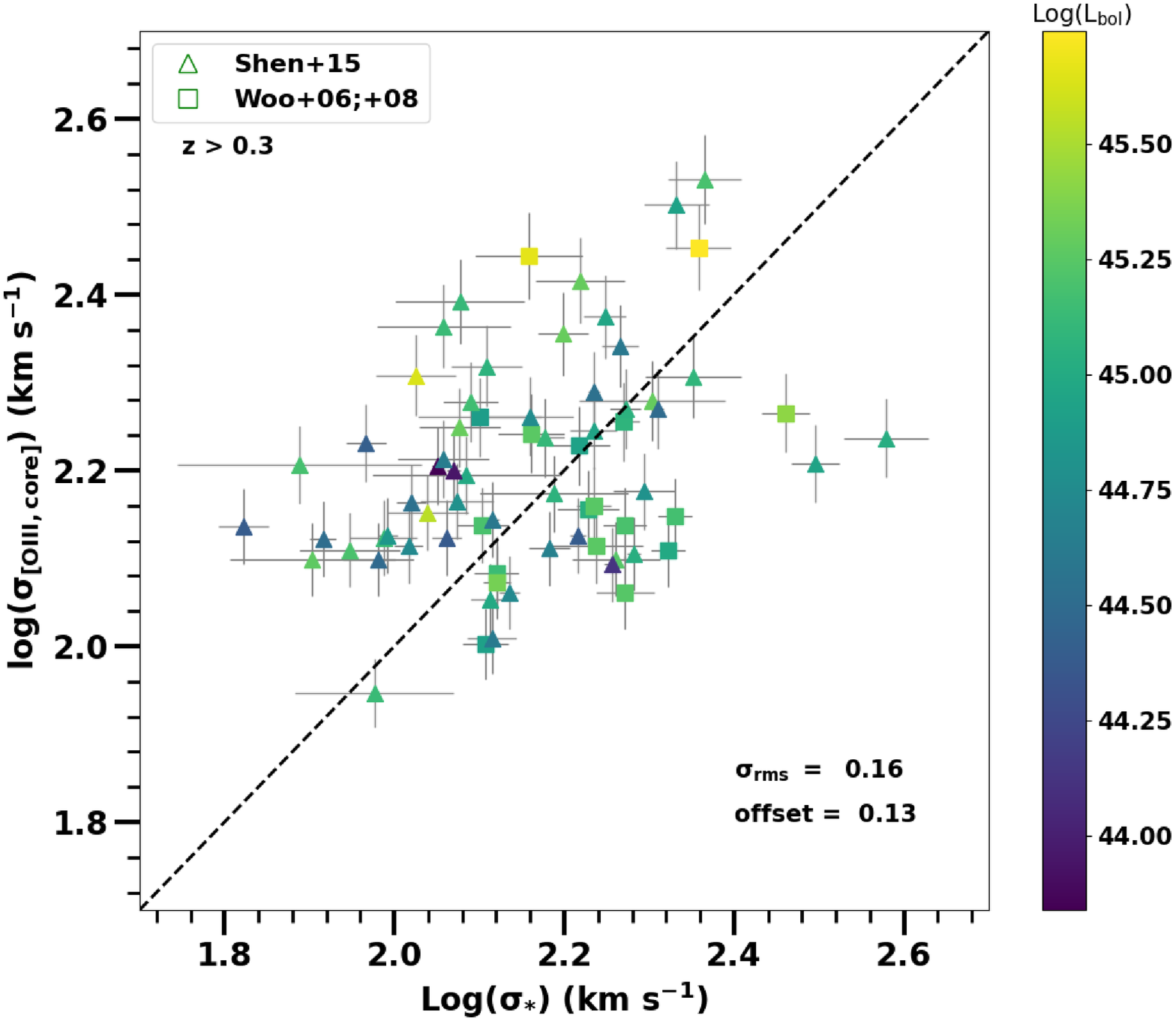}
	\centering
	\caption{Top panels: Comparison between $\rm \sigma_{*}$ and $\rm \sigma_{[OIII,core]}$ (left panel) and $\rm \sigma_{[OII]}$ (right panel) for the sample of \citet{Bennert+18}, respectively. Bottom panels: Comparison between $\rm \sigma_{*}$ and $\rm \sigma_{[OIII,core]}$ for the local SDSS sample \citep{Eun+17} and high redshift targets of \citet[]{Woo06,Woo08} and \citet{Shen+15}. $\rm L_{bol}$ color-scale is displayed. The dash-line shows the one-to-one relation. 	 
\label{fig:compare_sigma}}
\end{figure*}

Figure \ref{fig:compare_sigma} shows the comparison between $\rm \sigma_{*}$ and the gas emission line-widths of $\rm \sigma_{[OIII,core]}$ and $\rm \sigma_{[OII]}$. The local sample from \citet{Bennert+18} is a proper sample to study the direct relations between gas emission line widths and $\rm \sigma_{*}$ because those targets have the long-slit high-quality spatially resolved spectra which allow for robust measurements of $\rm \sigma_{[OIII,core]}$, $\rm \sigma_{[OII]}$ and $\rm \sigma_{*}$. In Figure \ref{fig:compare_sigma}, we found $\rm \sigma_{[OIII,core]}$ and $\sigma_{*}$ are correlated with a scatter of 0.11~dex. Similarly, $\rm \sigma_{[OII]}$ and $\sigma_{*}$ show a correlation with a scatter of 0.12~dex. There are three outliers whose double Gaussian model fitting cannot resolve the \OII\ doublet emission lines (3726\AA, 3729\AA), leading to overestimation of $\rm \sigma_{[OII]}$ (see Section 4.3 of \citealp{Bennert+18}). In the comparison, we displayed our sample color-coded according to $\rm L_{bol}$. We found that the relation between $\rm \sigma_{[OIII,core]}$, $\rm \sigma_{[OII]}$ and $\sigma_{*}$ is better correlated for those low luminosity objects than for higher luminosity sources. When the luminosities of targets increase, the scatters of the correlations between $\rm \sigma_{[OIII,core]}$, $\rm \sigma_{[OII]}$ and $\sigma_{*}$ also grow.


As the aperture size of SDSS is 3$\arcsec$, the integrated spectra of SDSS cover a radius size of $\sim$2.5 kpc of the center of the galaxies with redshift z $\sim$ 0.1. For the higher redshift sample, the SDSS radius size includes $\sim$3.3-7.8 kpc within redshift of z $\sim$ 0.2-1.0. \citet[]{Woo06,Woo08} extracted their LRIS-KECK spectra within 4-5 pixels (1$\arcsec$, $\sim$3-5 kpc at redshift z $\sim$ 0.3-0.5). Since the aperture size of SDSS is large, it is hard to judge whether the measurements of $\rm \sigma_{*}$ and $\rm \sigma_{[OIII,core]}$ are within the bulge radius or also have contribution from the disk component of the galaxy. Without spatially resolved spectra, it is difficult to check the true measurements of $\rm \sigma_{*}$ and $\rm \sigma_{[OIII,core]}$. We will discuss further this issue in Section \ref{section:discuss} that our measurements are reliable within uncertainties. In this section, we simply check the comparison between $\rm \sigma_{*}$ and $\rm \sigma_{[OIII,core]}$ for the local SDSS sample and high redshift sources (Figure \ref{fig:compare_sigma}). Due to the limitation of S/N, we did not obtain $\rm \sigma_{[OII]}$ for those sources. For the local SDSS sample, we found that $\rm \sigma_{[OIII,core]}$ and $\sigma_{*}$ show a correlation with a scatter of 0.12~dex. This result is consistent with that of the high S/N long-slit spatially resolved spectral sample of \citet{Bennert+18}. For the high redshift targets, the correlation displays a scatter of 0.16~dex, indicating a large discrepancy in the relation between $\rm \sigma_{*}$ and $\rm \sigma_{[OIII,core]}$ for high redshift sources. Similar to the local SDSS sample, the low luminosity high redshift sources have smaller scatters between $\rm \sigma_{*}$ and $\rm \sigma_{[OIII,core]}$ compared to that of higher luminosity sources.  

\subsection{Gas and Stellar Kinematics vs. AGN Properties}\label{section:gas_stellar} 

For studying the relation between gas emission line width and $\sigma_{*}$ in more detail, we explored the differences between $\rm \sigma_{[OIII,core]}$ and $\sigma_{*}$ as a function of AGN properties. 

\begin{figure*}
	\includegraphics[width=0.455\textwidth, height=0.42\textwidth]{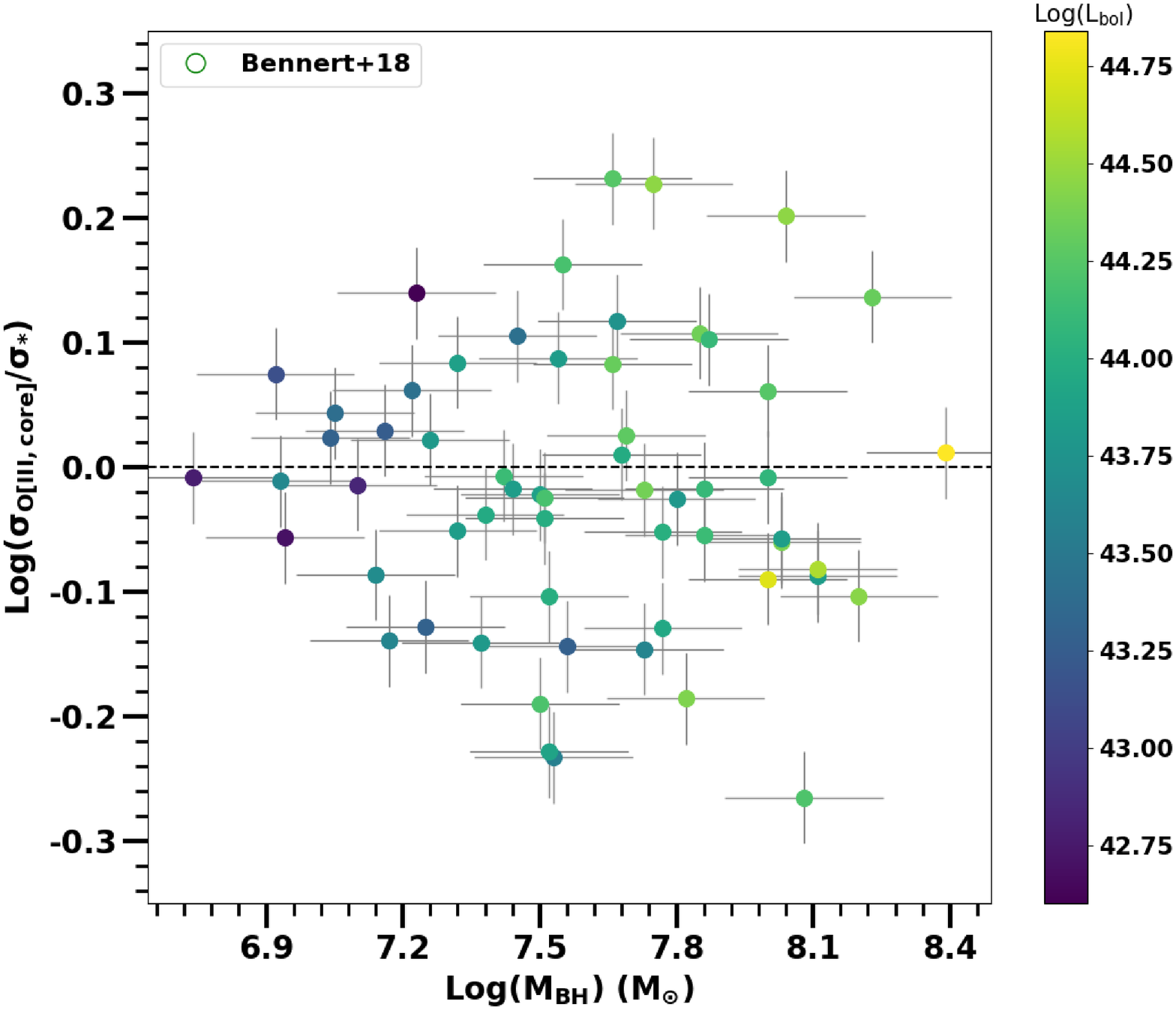}
	\includegraphics[width=0.45\textwidth, height=0.42\textwidth]{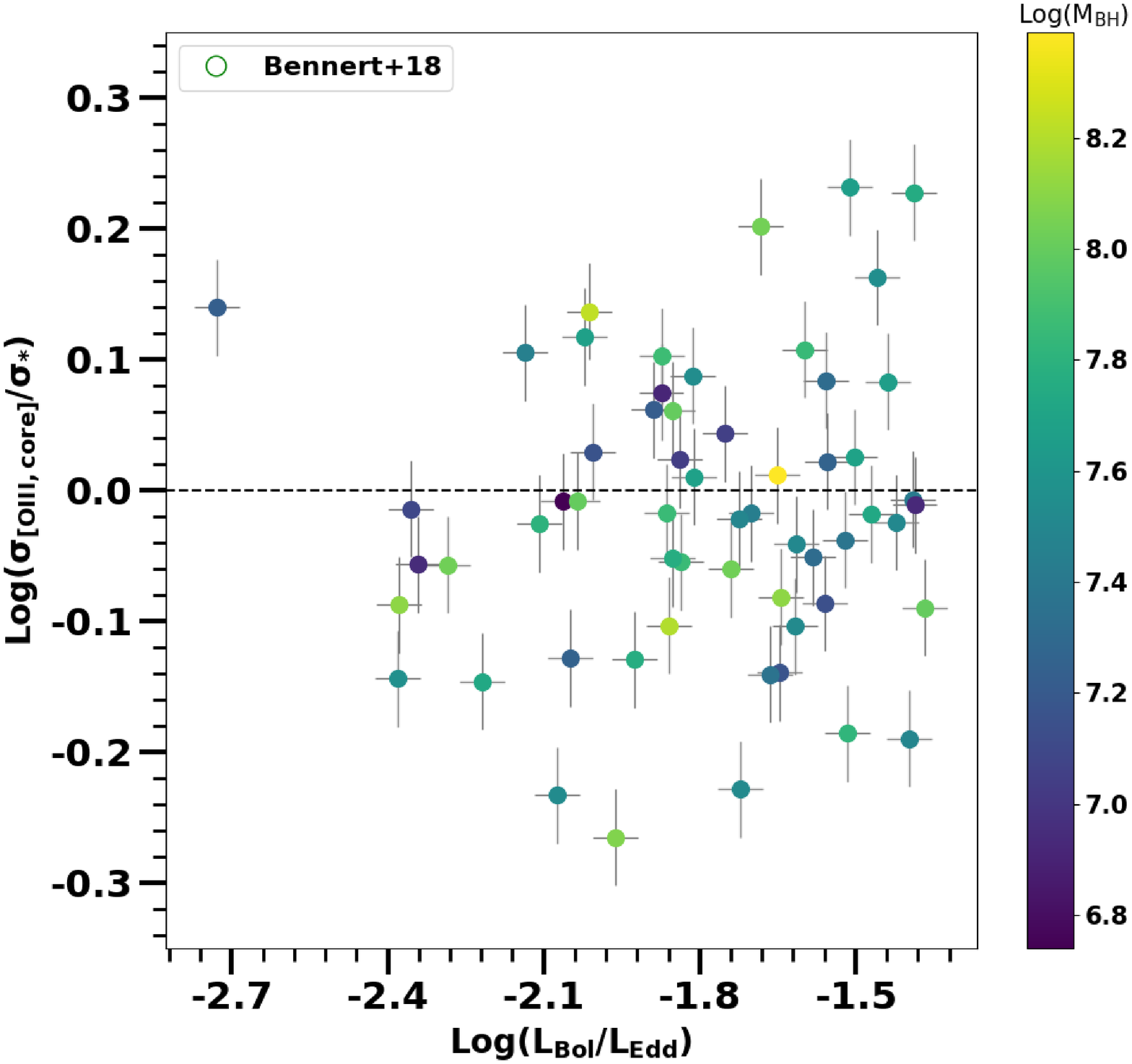}	
	\includegraphics[width=0.45\textwidth, height=0.42\textwidth]{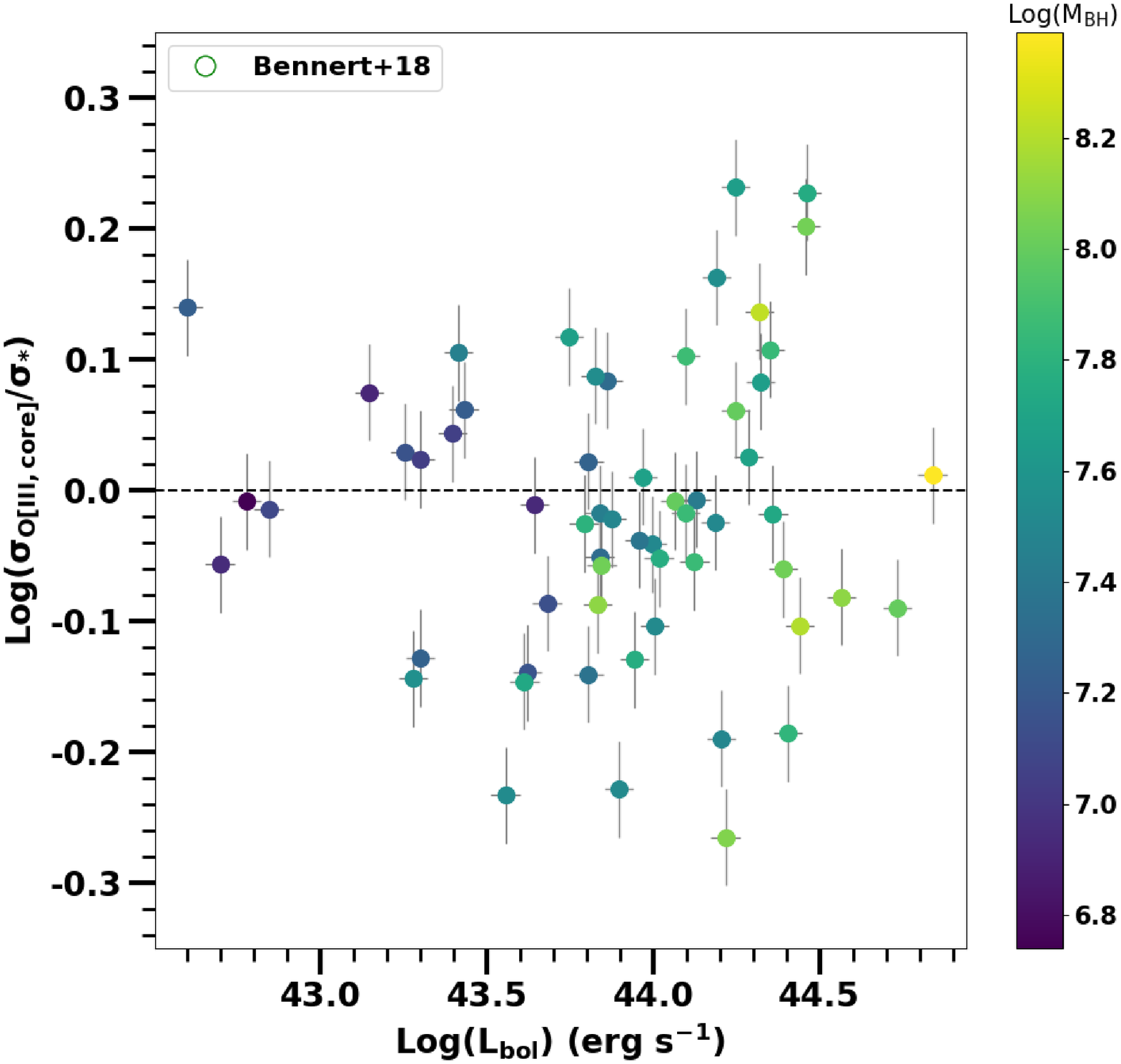}	
	\includegraphics[width=0.45\textwidth, height=0.42\textwidth]{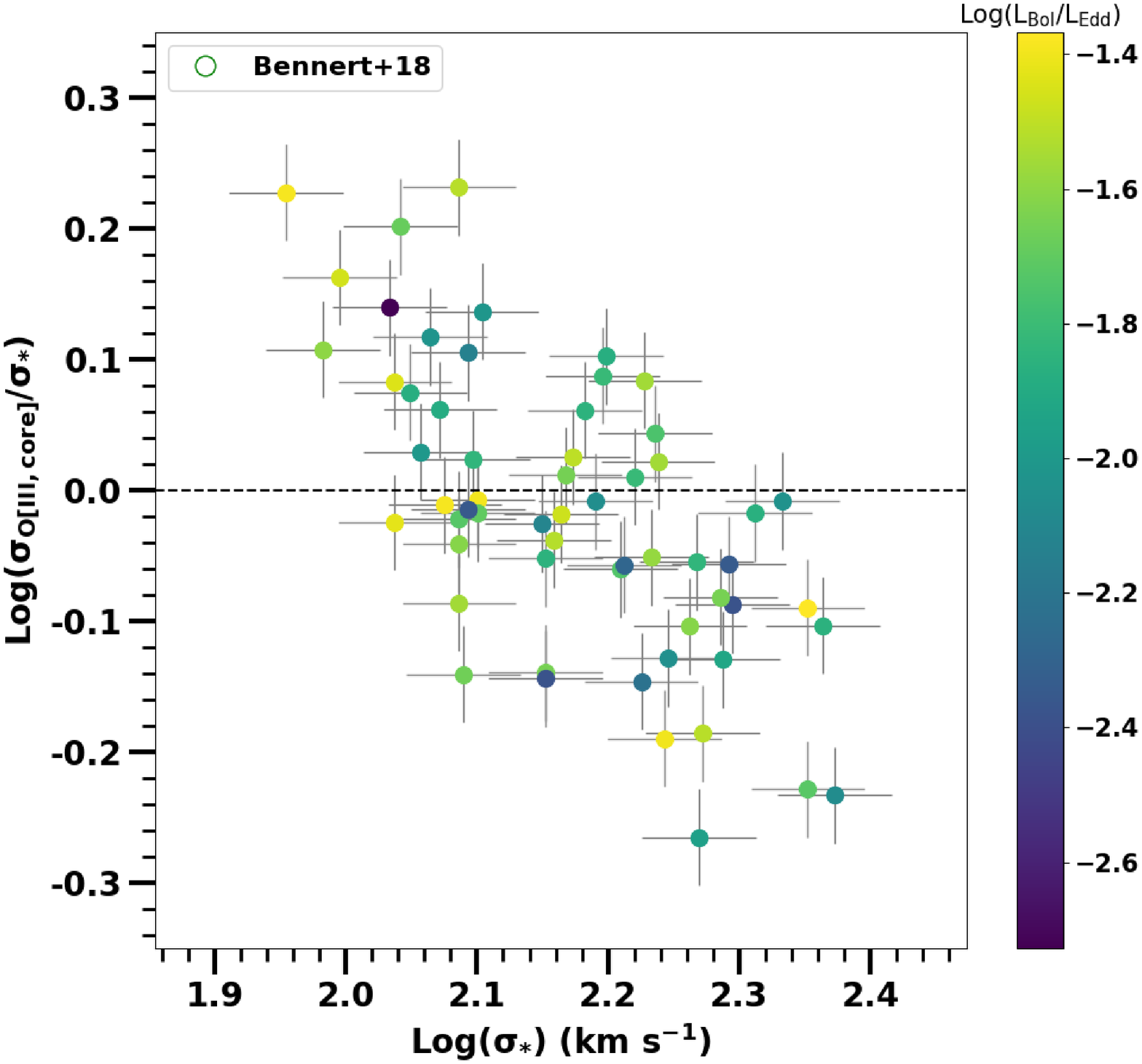}	
	\centering
	\caption{Difference between $\rm \sigma_{[OIII,core]}$ and $\rm \sigma_{*}$ as a function of AGN properties, i.e., $\rm M_{BH}$, $\rm L_{bol}/L_{Edd}$, $\rm L_{bol}$, and $\rm \sigma_{*}$ for the sample of \citet{Bennert+18}. The color-scale is indicated in each color map panel. The dash-line shows the one-to-one relation. 	
\label{fig:comp_bennert_oiii}}
\end{figure*}

\begin{figure*}
	\includegraphics[width=0.47\textwidth, height=0.42\textwidth]{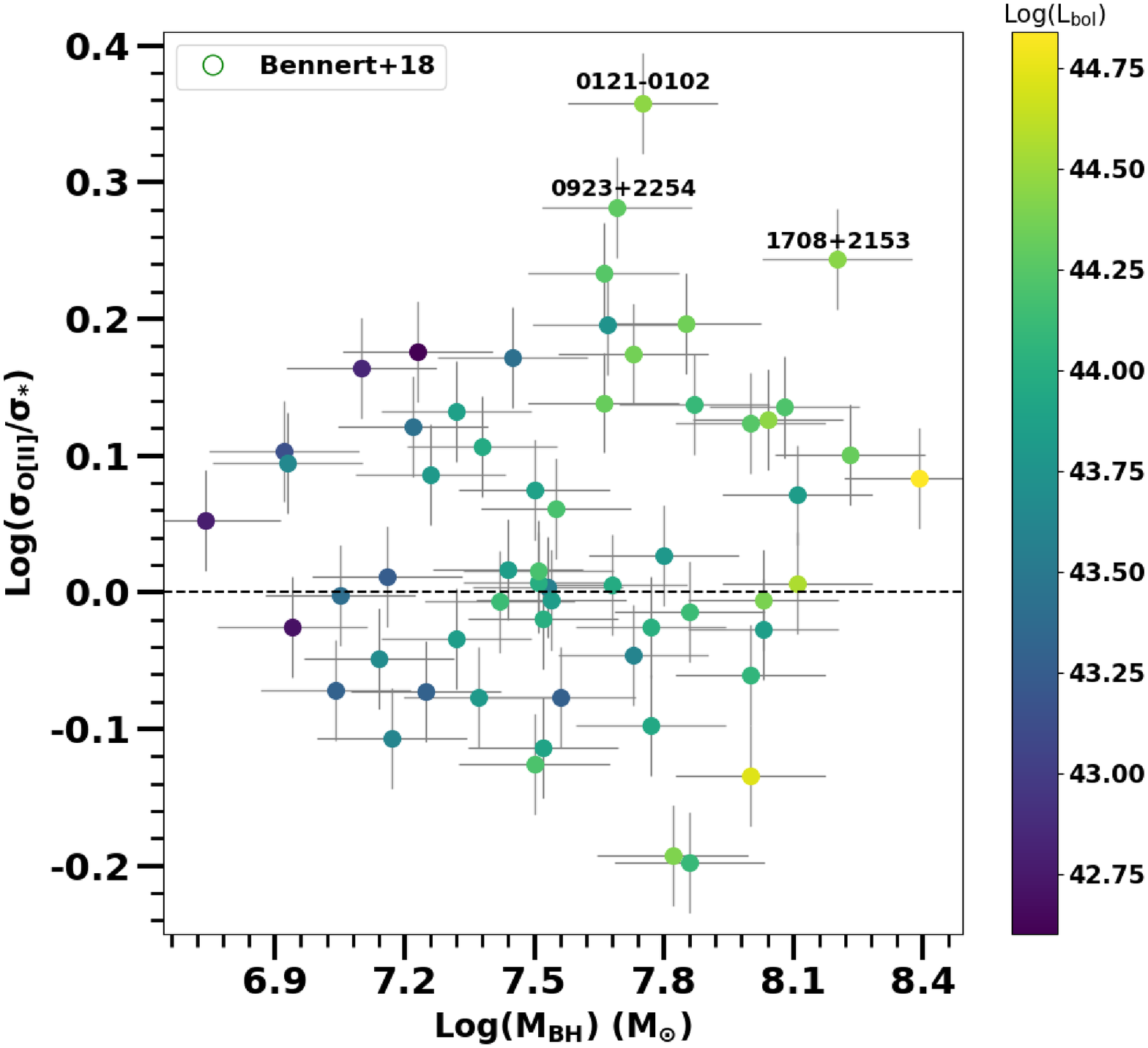}
	\includegraphics[width=0.45\textwidth, height=0.42\textwidth]{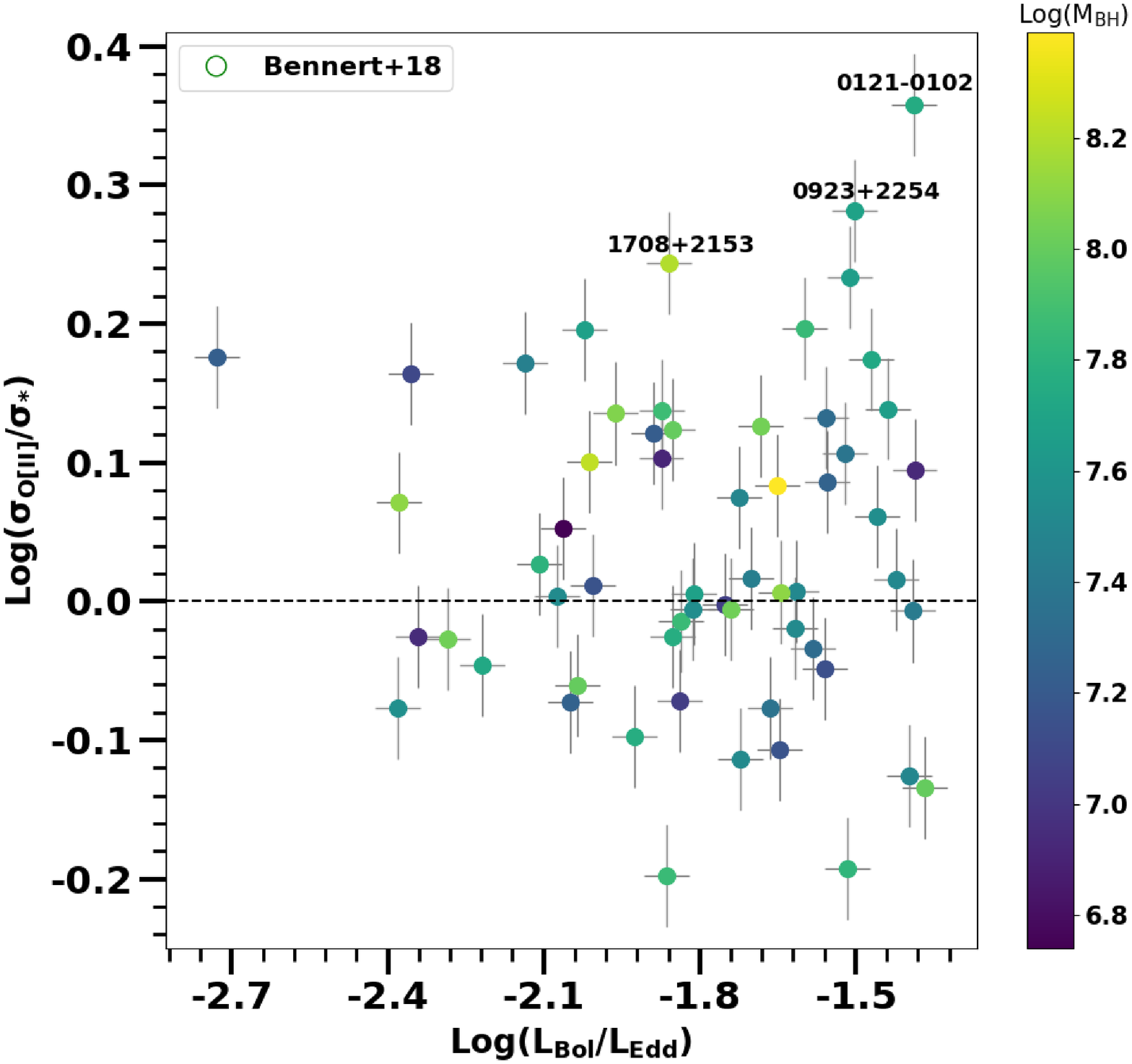}	
	\includegraphics[width=0.45\textwidth, height=0.42\textwidth]{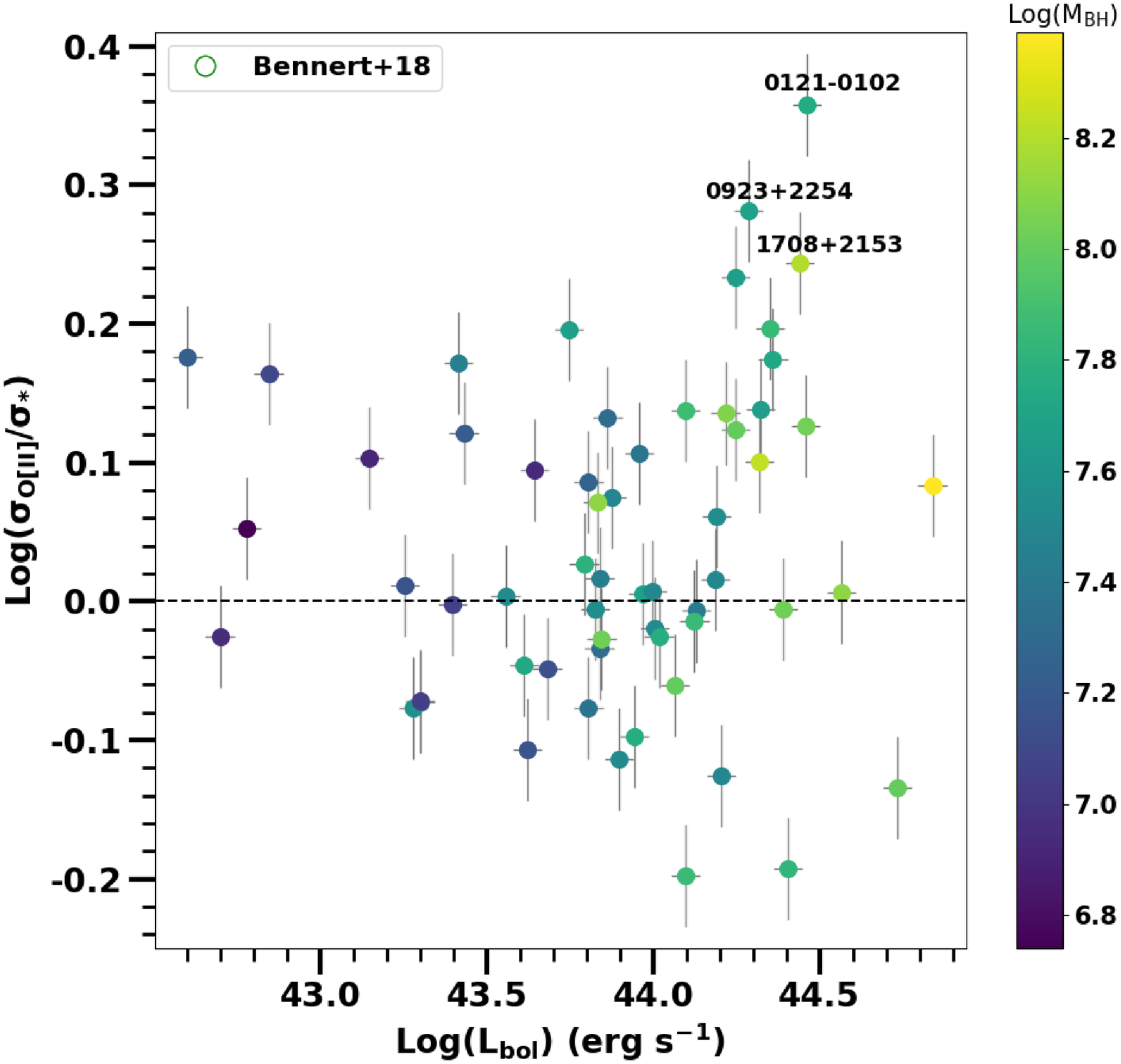}	
	\includegraphics[width=0.45\textwidth, height=0.42\textwidth]{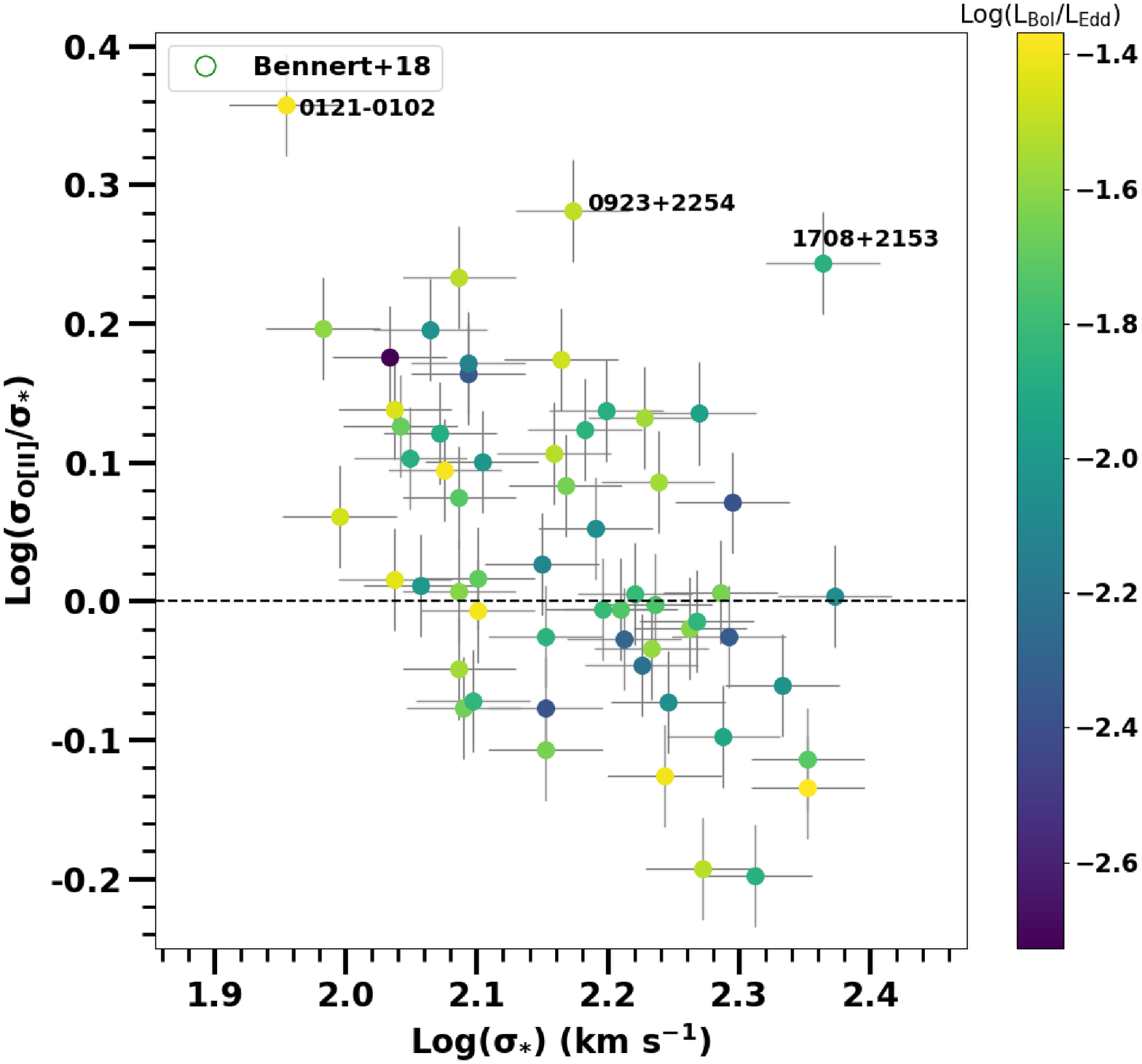}	
	\centering
	\caption{Difference between $\rm \sigma_{[OII]}$ and $\rm \sigma_{*}$ as a function of AGN properties, i.e., $\rm M_{BH}$, $\rm L_{bol}/L_{Edd}$, $\rm L_{bol}$, and $\rm \sigma_{*}$ for the sample of \citet{Bennert+18}. The color-scale is indicated in each color map panel. The dash-line shows the one-to-one relation. 	
\label{fig:comp_bennert_oii}}
\end{figure*}

Figure \ref{fig:comp_bennert_oiii} shows the comparisons between $\rm \sigma_{[OIII,core]}$$/$$\sigma_{*}$ and $\rm M_{BH}$, $\rm L_{bol}/L_{Edd}$, $\rm L_{bol}$ and $\sigma_{*}$ for the long-slit spatially resolved spectral sample of \citet{Bennert+18}. The logrithmic scale ratio of $\rm \sigma_{[OIII,core]}$$/$$\sigma_{*}$ is broadened, $\sim$-0.3$-$0.3 within a range of $\log \sigma_{*}$ = 1.8$-$2.6 (i.e., 60$-$400 \kms). We found that the good correlation with small scatter is shown between $\rm \sigma_{[OIII,core]}$ and $\sigma_{*}$ for those targets which have small \mbh\ and low $\rm L_{bol}$ i.e., $\rm \log(L_{bol}$) $<$ $\sim$43.5. While, the scatter is significantly broader when \mbh\ and $\rm L_{bol}$ of targets increase. In addition, we also found that for those targets which have higher $\rm L_{bol}/L_{Edd}$, $\rm \sigma_{[OIII,core]}$ tends to be larger than $\sigma_{*}$. Similar to the case of \OIII\ emission line, we found the good correlation between $\rm \sigma_{[OII]}$ and $\sigma_{*}$ for those targets which have small \mbh\ and low $\rm L_{bol}$. As $\rm L_{bol}/L_{Edd}$ increases, $\rm \sigma_{[OII]}$$/$$\sigma_{*}$ also grows (see Figure \ref{fig:comp_bennert_oii}). 

\begin{figure*}
	\includegraphics[width=0.455\textwidth, height=0.42\textwidth]{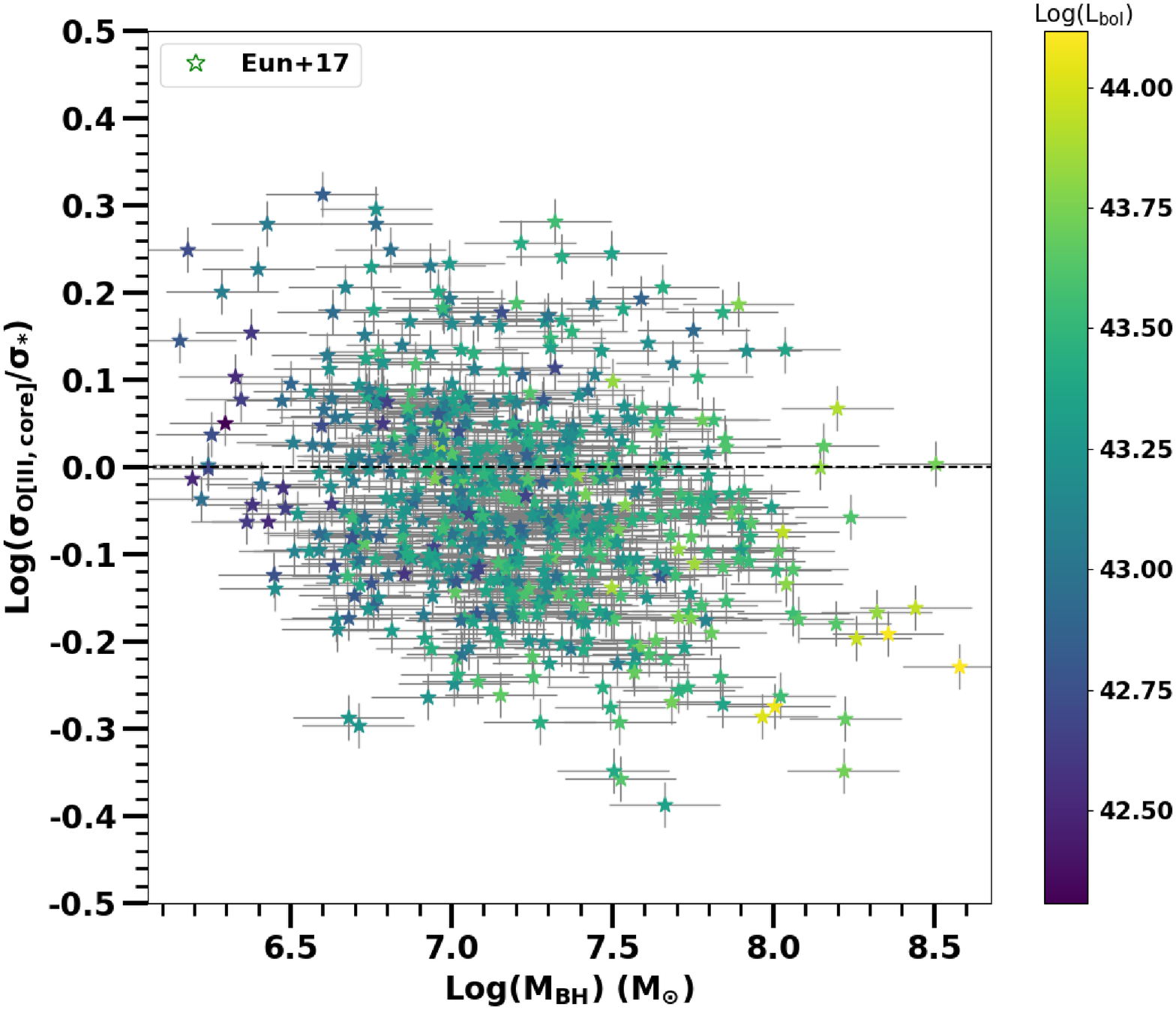}
	\includegraphics[width=0.45\textwidth, height=0.42\textwidth]{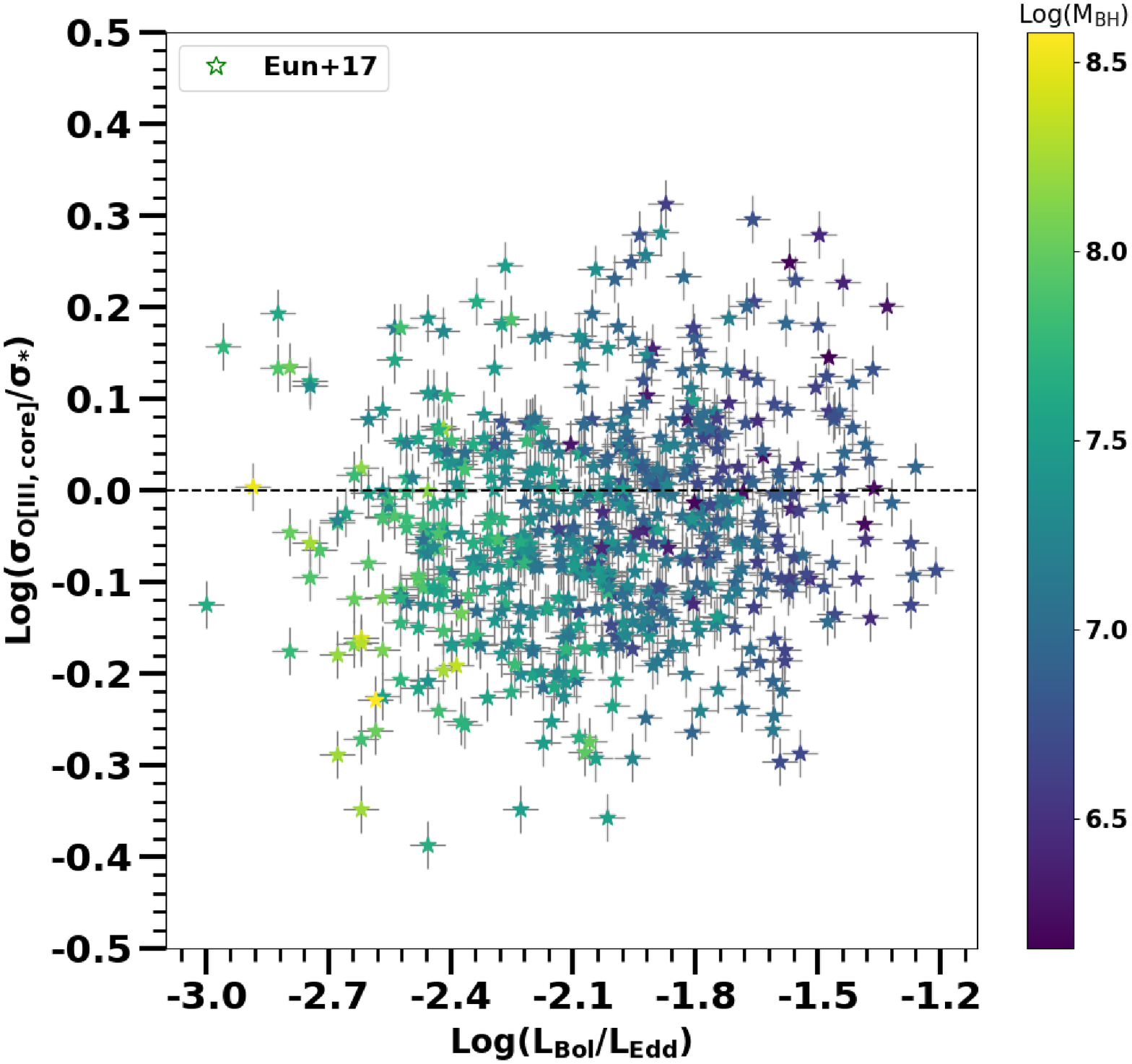}	
	\includegraphics[width=0.45\textwidth, height=0.42\textwidth]{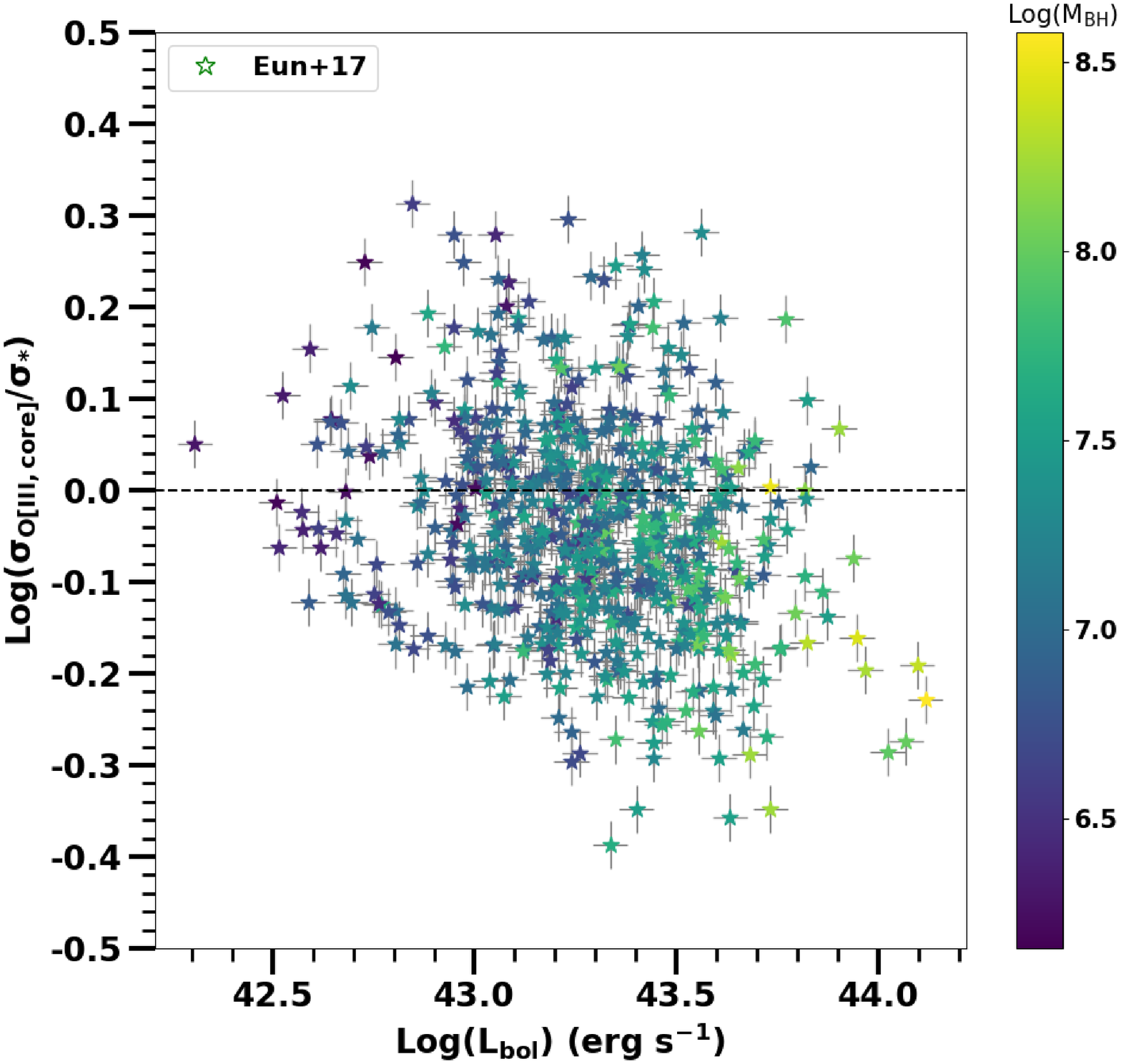}	
	\includegraphics[width=0.45\textwidth, height=0.42\textwidth]{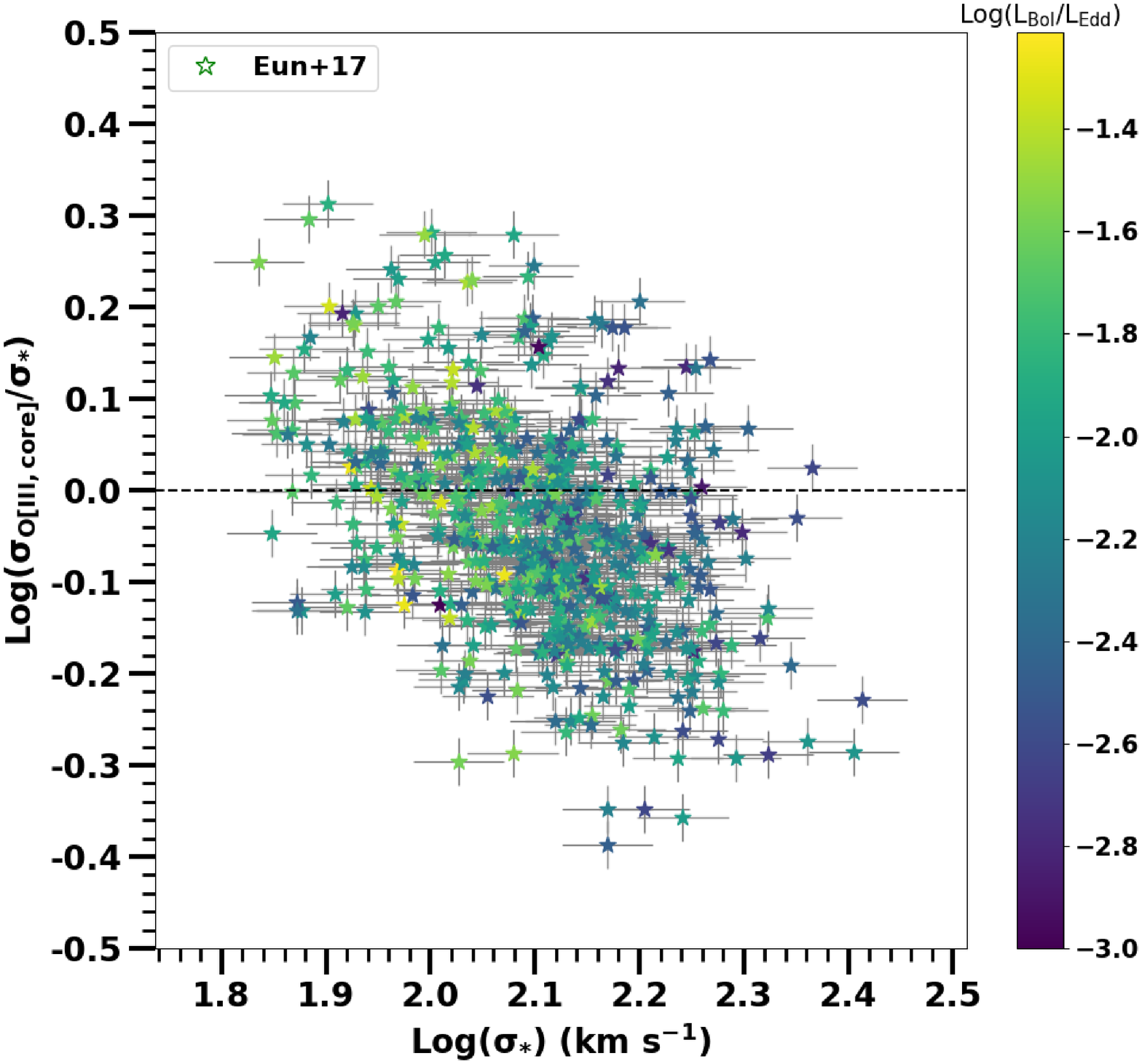}	
	\centering
	\caption{Same as Figure \ref{fig:comp_bennert_oiii} but for the local SDSS sample of \citet{Eun+17}. 
\label{fig:comp_eun}}
\end{figure*}

Figure \ref{fig:comp_eun} presents the comparisons of $\rm \sigma_{[OIII,core]}$ and $\sigma_{*}$ for the local SDSS sample. We found that those sources with low $\rm L_{bol}$ show the good correlation, while high bolometric luminosity sources show large discrepancy ($\sim$-0.4$-$0.4). Similar to the case of the long-slit spatially resolved sample of \citet{Bennert+18}, $\rm \sigma_{[OIII,core]}$$/$$\sigma_{*}$ tends to increase as $\rm L_{bol}/L_{Edd}$ increases. Also, for those targets which have larger \mbh\ and higher $\rm L_{bol}$, $\rm \sigma_{[OIII,core]}$$/$$\sigma_{*}$ tends to be smaller when $\rm L_{bol}/L_{Edd}$ are low in those targets. 

\begin{figure*}
	\includegraphics[width=0.455\textwidth, height=0.42\textwidth]{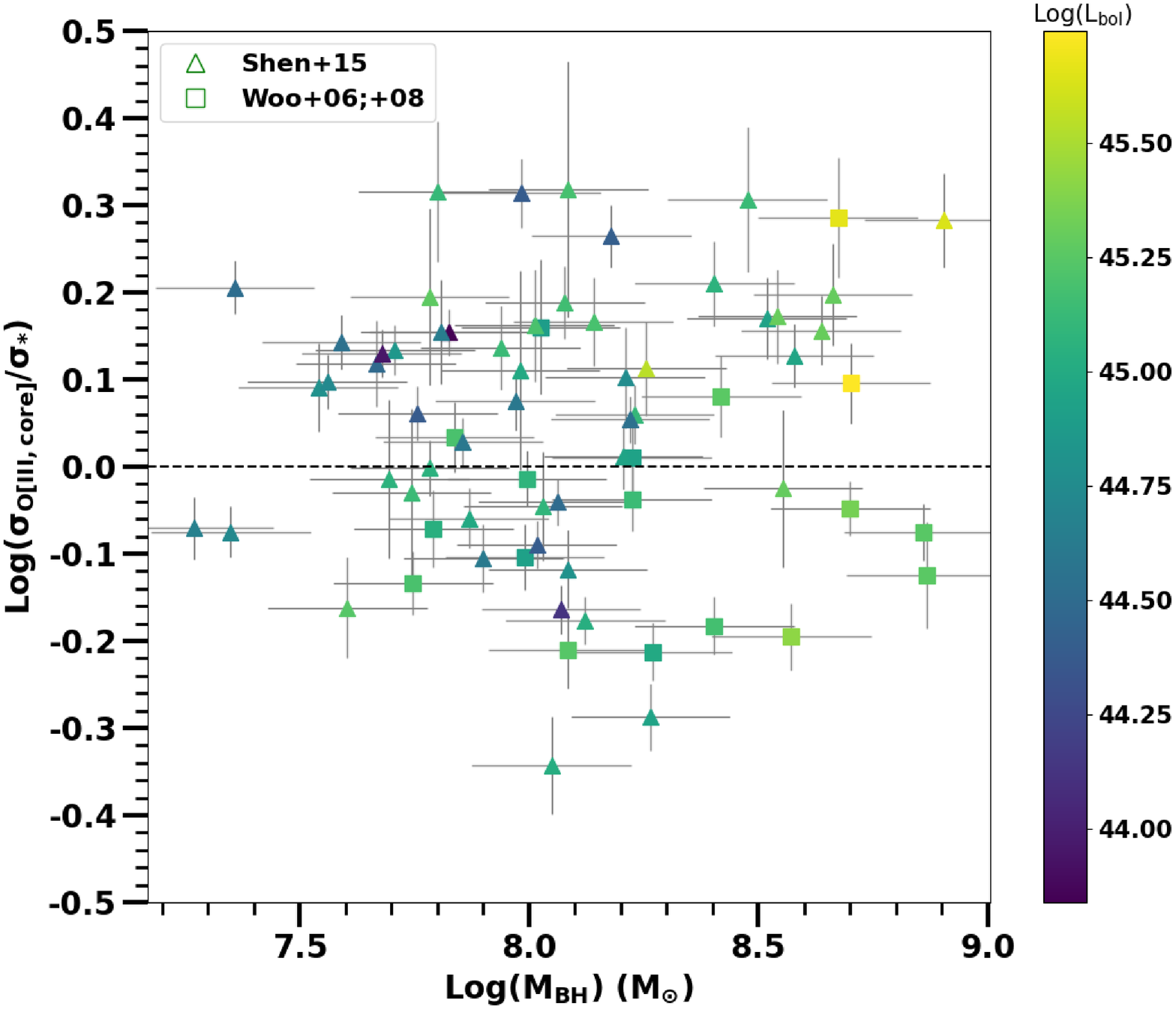}
	\includegraphics[width=0.45\textwidth, height=0.42\textwidth]{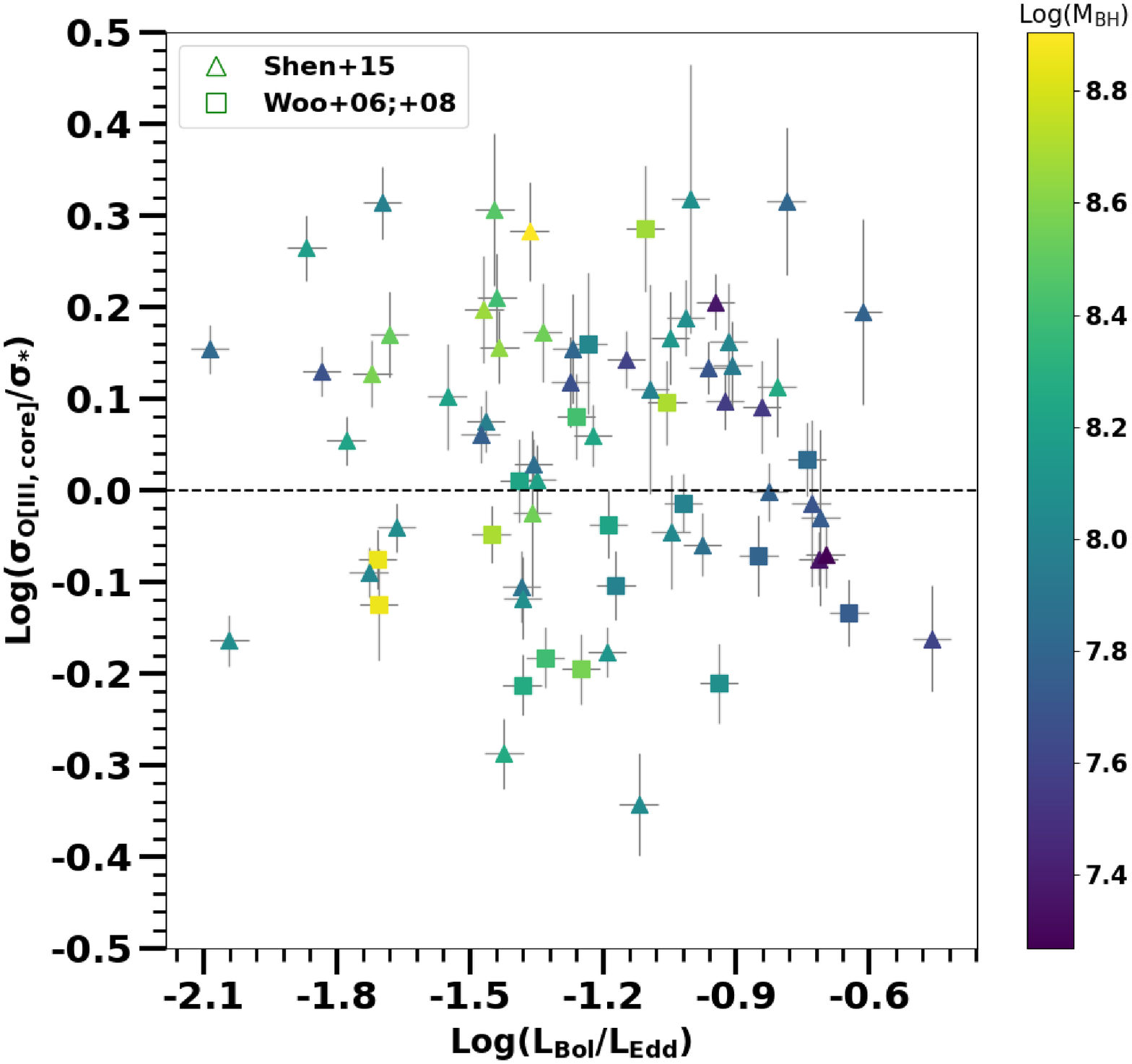}	
	\includegraphics[width=0.45\textwidth, height=0.42\textwidth]{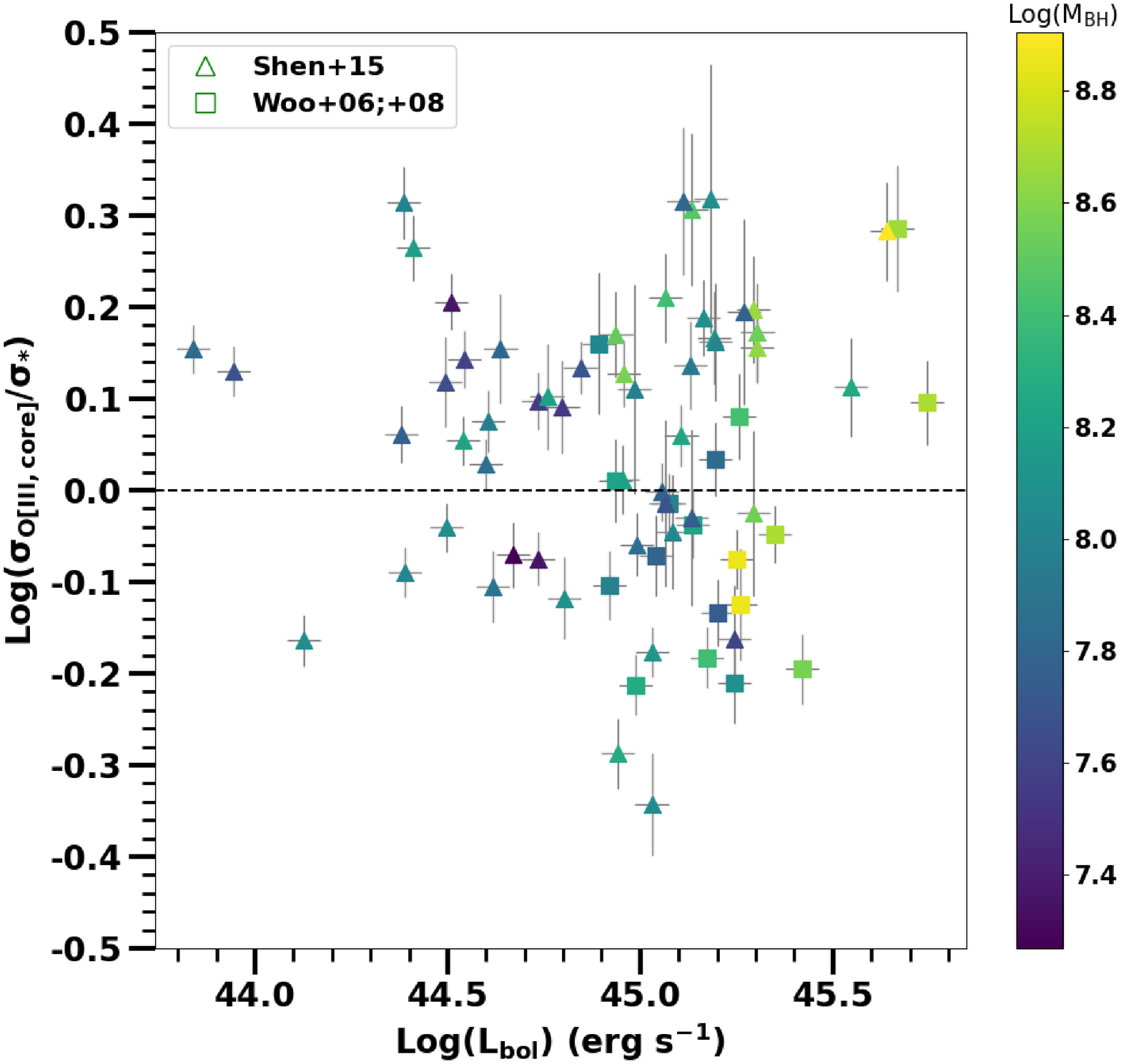}	
	\includegraphics[width=0.45\textwidth, height=0.42\textwidth]{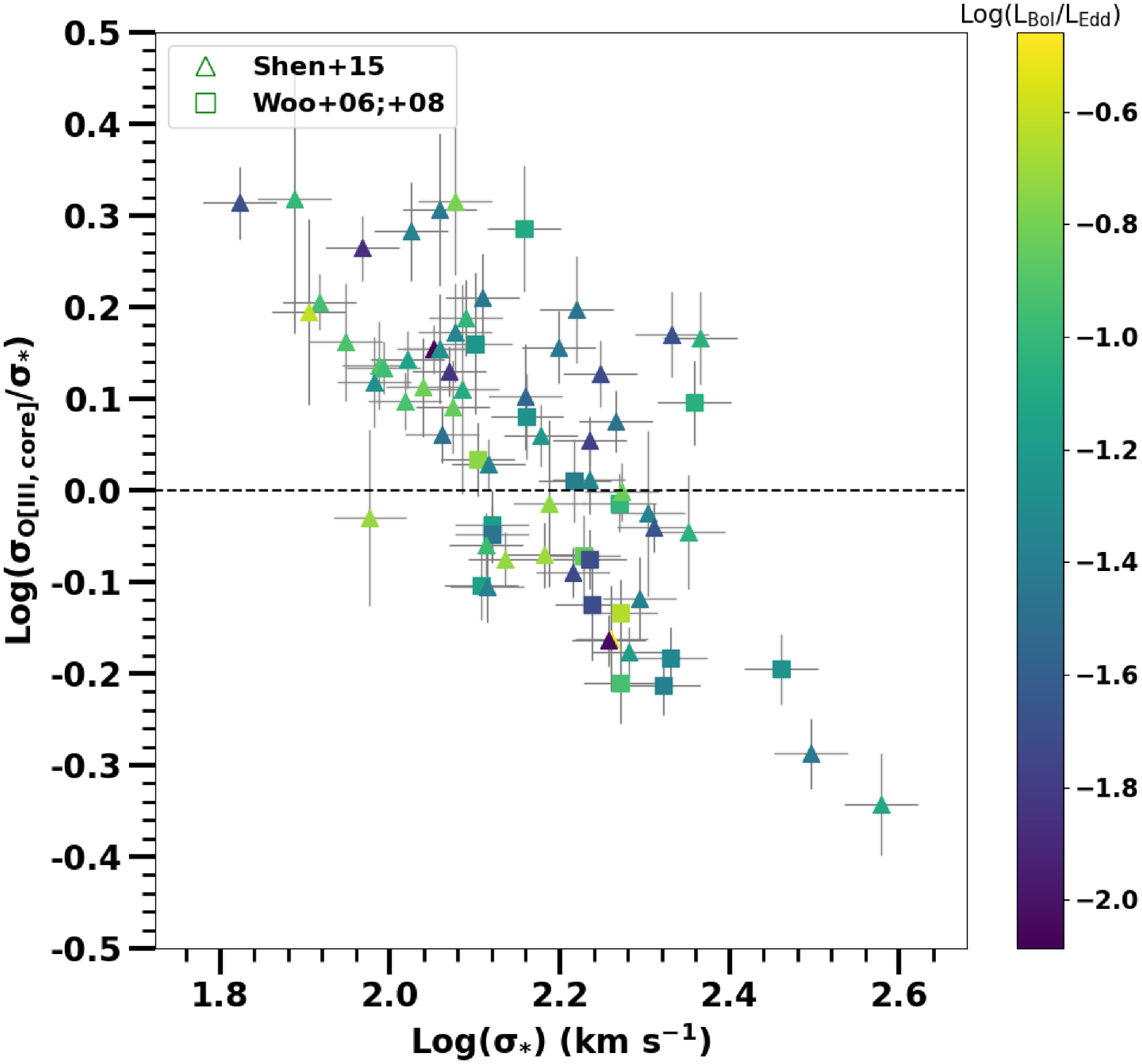}	
	\centering
	\caption{Same as Figure \ref{fig:comp_bennert_oiii} but for the high redshift sample of \citet[]{Woo06,Woo08} and \citet{Shen+15}.
\label{fig:comp_wooshen}}
\end{figure*}

In the case of the high redshift sample, the scatter is large between $\rm \sigma_{[OIII,core]}$ and $\sigma_{*}$ ($\sim$-0.4$-$0.4). Similar to the local objects, the low bolometric luminosity sources show small scatters between $\rm \sigma_{[OIII,core]}$ and $\sigma_{*}$ compared to those high luminosity sources. And, for sources which have higher $\rm L_{bol}/L_{Edd}$, $\rm \sigma_{[OIII,core]}$$/$$\sigma_{*}$ tends to be larger (see Figure \ref{fig:comp_wooshen}).  

The large scatters for those high bolometric luminosity objects may indicate the difficulty in the measurement of $\rm \sigma_{*}$ for sources with large contribution of AGN emission in the center or may also illustrate an additional significant effect which broadens the \OIII\ emission profile (e.g., outflows), leading to large uncertainty in measuring the gravitational component such as $\rm \sigma_{[OIII,core]}$ in the \OIII\ profile. Particularly, $\rm L_{bol}/L_{Edd}$ seems to play an important role in the relations between NLR gas and stellar velocity dispersions.

\subsection{Gas and Stellar Kinematics vs. Velocity Shifts and Outflow Strengths}\label{section:outflow}

To study the driving factor of the discrepancy between $\rm \sigma_{[OIII,core]}$ and $\rm \sigma_{*}$, we compared the ratio $\rm \sigma_{[OIII,core]}$$/$$\sigma_{*}$ with \OIII\ velocity shifts ($\rm V_{[OIII,core]}$ and $\rm V_{[OIII,out]}$) and outflow strengths such as $\rm V_{max}$ and $\rm \sigma_{[OIII,out]}$.

Figure \ref{fig:voutflow_properties} presents the comparison between $\rm V_{[OIII,core]}$ and $\rm V_{[OIII,out]}$ with respect to the systemic velocity as a function of $\rm M_{BH}$ and $\rm L_{bol}$. For the sources which have low $\rm M_{BH}$ and $\rm L_{bol}$, $\rm V_{[OIII,core]}$ and $\rm V_{[OIII,out]}$ are small. And, for the sources which have higher $\rm M_{BH}$ and $\rm L_{bol}$, $\rm V_{[OIII,core]}$ and $\rm V_{[OIII,out]}$ tend to be larger with large scatter. Interestingly, the line shifts of the broad component of \OIII\ show blueshifted with an average velocity of $-$135$\pm$150 \kms; while, the line shifts of the core component display a smaller offset with an average velocity of $-$14$\pm$47 \kms. This result may indicate that the kinematics of the core component of \OIII\ and stellar absorption lines are quite close to each other, but the offsets become larger for higher $\rm M_{BH}$ and $\rm L_{bol}$ sources which have stronger outflow signatures (with larger $\rm V_{[OIII,out]}$).

In Figure \ref{fig:voutflow_properties}, we also show the comparison between $\rm V_{max}$, $\rm M_{BH}$ and $\rm L_{bol}$. $\rm V_{max}$ indicates the maximum strength of the outflow component in the \OIII\ profile. It shows how large offset of the broad component compared to that of the core component of the \OIII\ emission line.  We found that when $\rm L_{bol}$ increases, $\rm V_{max}$ also grows. For those targets which have small $\rm M_{BH}$ and low $\rm L_{bol}$, $\rm V_{max}$ is low. In contrast, $\rm V_{max}$ tends to be larger for objects which have higher $\rm M_{BH}$ and higher $\rm L_{bol}$. This result may indicate that the AGN accretion plays a major role in outflow physical properties. 

Figure \ref{fig:comp_voutflow_oiiistar} shows the comparison of $\rm \sigma_{[OIII,core]}$$ $-$ $$\sigma_{*}$ with $\rm V_{max}$, $\rm \sigma_{[OIII,out]}$, $\rm V_{[OIII,core]}$ and $\rm V_{[OIII,out]}$, respectively. We found that when $\rm V_{max}$ increases, the discrepancy between $\rm \sigma_{[OIII,core]}$ and $\rm \sigma_{*}$ tends to be larger. Similarly, as $\rm \sigma_{[OIII,out]}$ is larger, the difference between $\rm \sigma_{[OIII,core]}$ and $\rm \sigma_{*}$ is also larger; when $\rm V_{[OIII,core]}$ and $\rm V_{[OIII,out]}$ increase, the difference between $\rm \sigma_{[OIII,core]}$ and $\rm \sigma_{*}$ also grows.

The difference between $\rm \sigma_{[OIII,core]}$ and $\rm \sigma_{*}$ is well connected with $\rm V_{max}$, $\rm \sigma_{[OIII,out]}$, $\rm V_{[OIII,core]}$ and $\rm V_{[OIII,out]}$. This result may indicate that outflows play a significant role in the discrepancy between $\rm \sigma_{[OIII,core]}$ and $\rm \sigma_{*}$.

\section{Discussions}\label{section:discuss}

\subsection{Outflow Strength Effect}

The main question in our work is: what are the physically driven factors for the scatters between ionized gas kinematics in the NLR and stellar velocity dispersions? If we assume that the gas kinematics in the NLR is governed by the gravitational potential of the bulge of the host galaxy, then we expect that the velocity field of NLR emission lines should trace the same velocity field as that of the stellar velocity dispersions. However, from previous studies, many authors found that $\rm \sigma_{[OIII]}$ and $\sigma_{*}$ show correlations but with large scatters. So, what are the factors for these substantial scatters between $\rm \sigma_{[OIII]}$ and $\sigma_{*}$? As discussed in \citet{Greene05}, the authors found that there is a strong correlation between the $\rm \sigma_{[OIII]}$$/$$\sigma_{*}$ ratio as a function of Eddington ratio, indicating that the gas kinematics of the NLR is not also governed by the primary role of the gravitational potential of the bulge but also followed by a secondary role of AGN activity such as outflows. Similar to \citet{Greene05}, \cite{Ho09} found the same conclusion by analyzing the line width of the \NII\ emission line. Later on, in the study of narrow-line Seyfert 1 sample, \citet{Komossa07} found that after correcting for the wing component in the \OIII\ profile, there is consistency between $\rm \sigma_{[OIII]}$ and $\sigma_{*}$. However, is there any caution in the validity of using $\rm \sigma_{[OIII,core]}$ as a surrogate for $\sigma_{*}$? To test this question, \citet{Bennert+18} used high-quality long-slit spatially resolved spectra from a sample of 52 Seyfert 1 galaxies to have a direct and simultaneous comparison between $\rm \sigma_{[OIII,core]}$ and $\sigma_{*}$. They found the large discrepancy in the comparison between object and object. So, even we use $\rm \sigma_{[OIII,core]}$ to replace $\sigma_{*}$, there is a large uncertainty to be considered.

In our study, we found the broad correlations with large scatters between $\rm \sigma_{[OIII,core]}$ and $\sigma_{*}$ for the local SDSS sample of \citet{Eun+17}.  For the high redshift sample from \citet[]{Woo06,Woo08} and \citet{Shen+15}, the correlation shows somewhat larger scatter compared to that of the local SDSS sample. By comparing the differences between $\rm \sigma_{[OIII,core]}$ and $\rm \sigma_{*}$ as a function of AGN properties, we found that $\rm \sigma_{[OIII,core]}$ and $\rm \sigma_{*}$ show small scatter in those targets which have low $\rm L_{bol}$. However, the scatter increases when $\rm L_{bol}$ grows. Particularly, we found that when $\rm L_{bol}/L_{Edd}$ increases, $\rm \sigma_{[OIII,core]}$ tends to be larger than $\rm \sigma_{*}$.

\begin{figure}
	\includegraphics[width=0.41\textwidth]{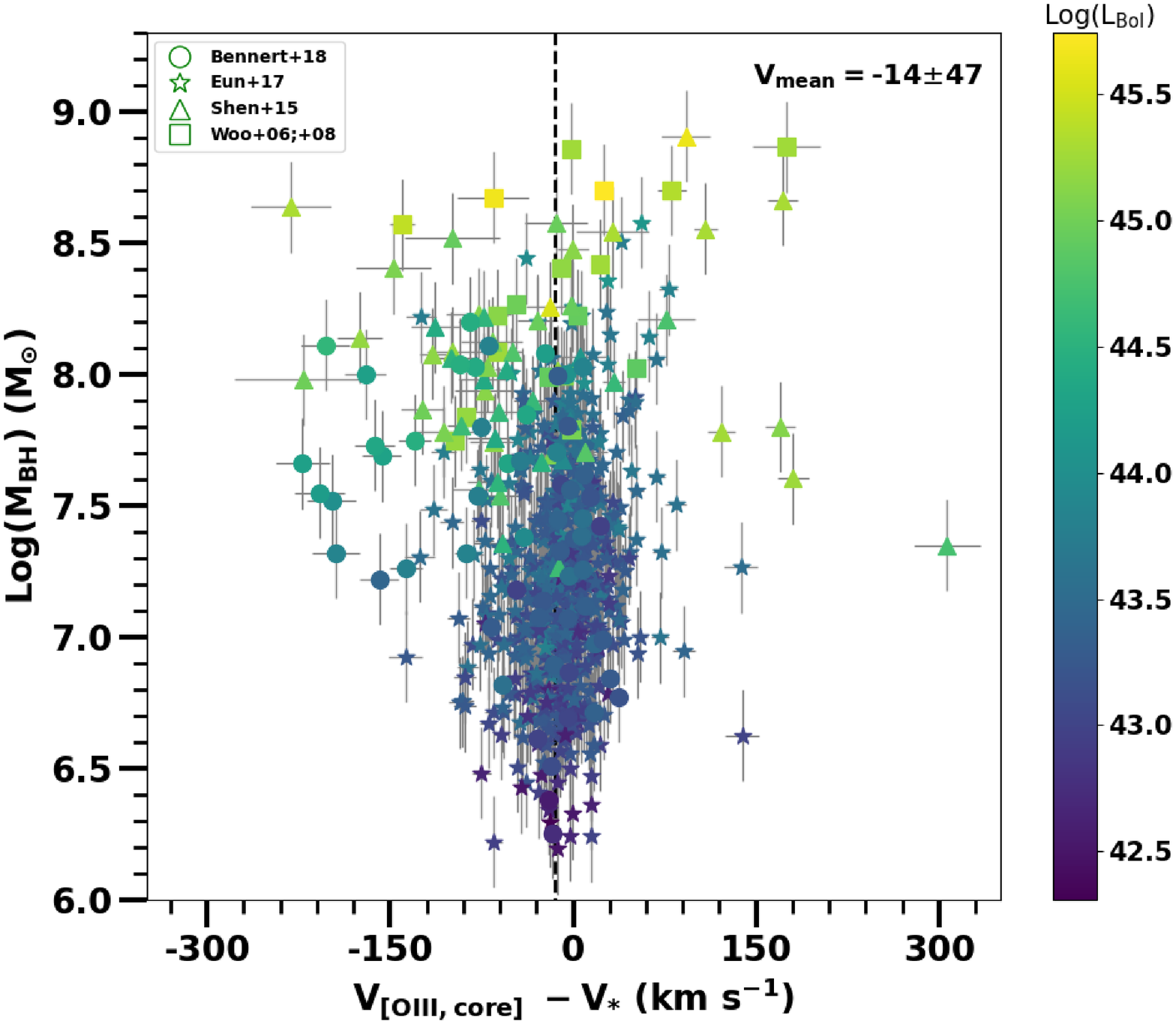}
	\includegraphics[width=0.41\textwidth]{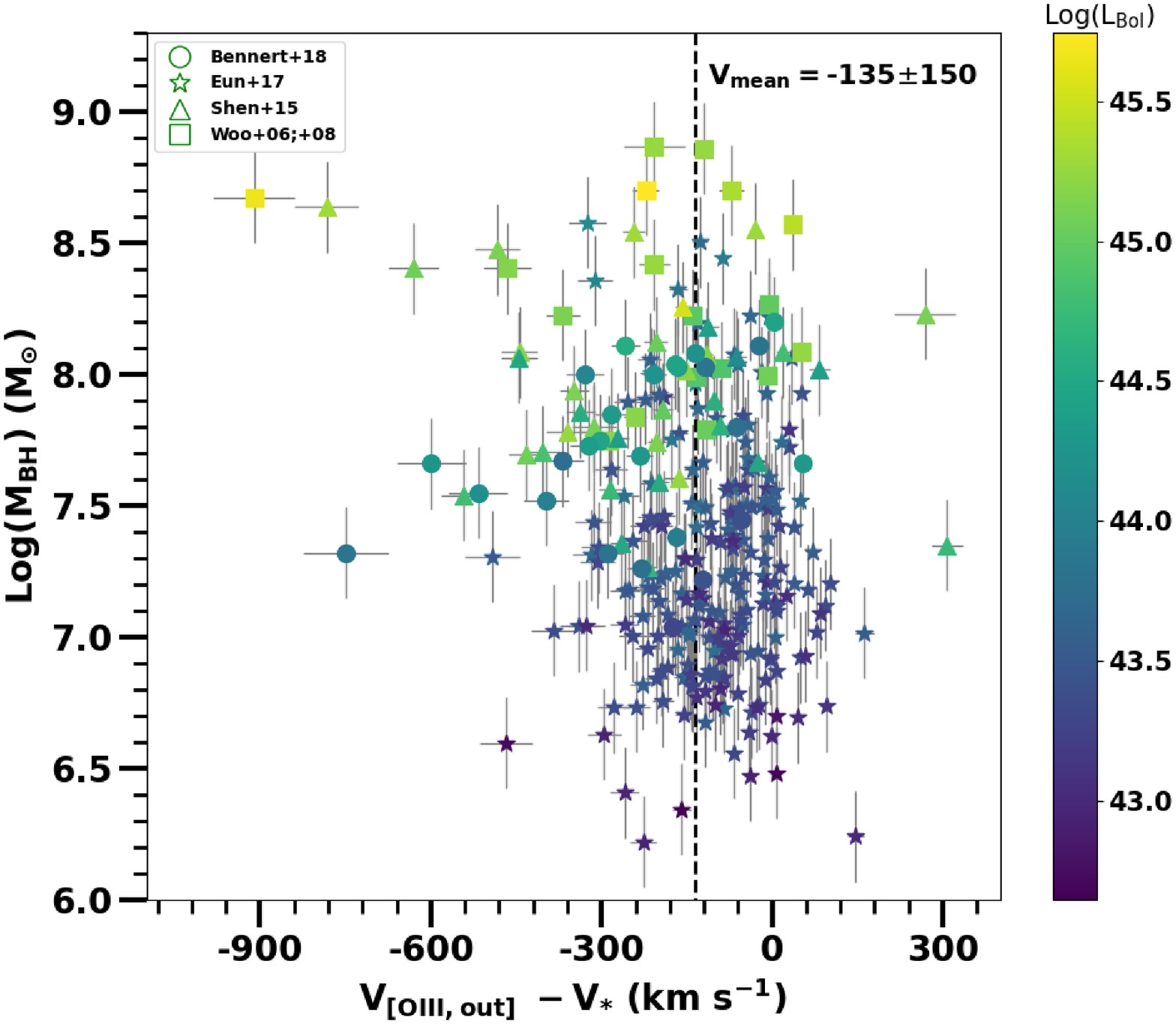}
	\includegraphics[width=0.41\textwidth]{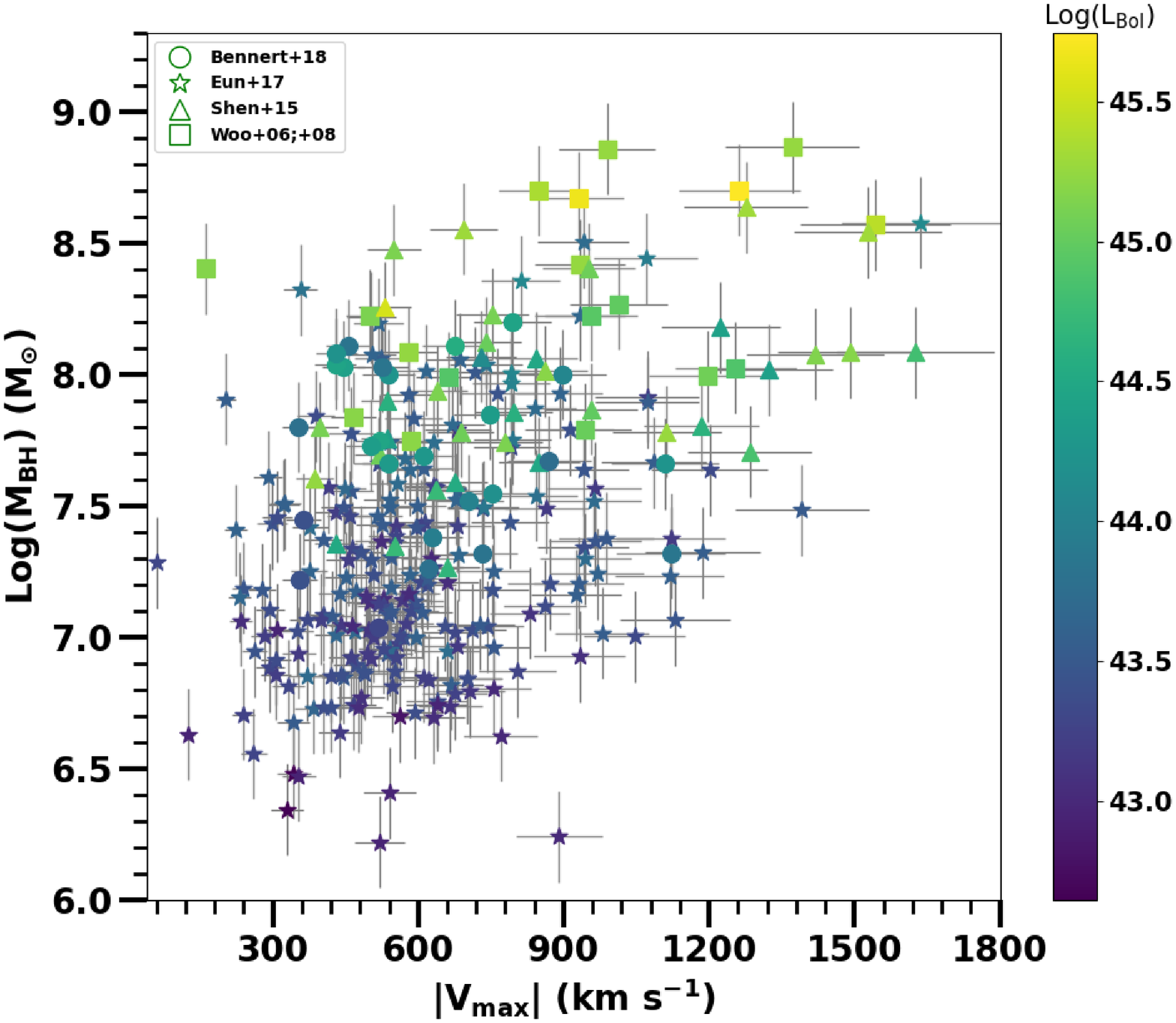}
	\centering
	\caption{Comparison between $\rm M_{BH}$ and $\rm V_{[OIII,core]}$ $-$ $\rm V_{*}$ (top panel), $\rm M_{BH}$ and $\rm V_{[OIII,out]}$ $-$ $\rm V_{*}$ (middle panel), and $\rm M_{BH}$ and $\rm V_{max}$ (bottom panel). The color scale is displayed for $\rm L_{bol}$. The samples of \citet{Bennert+18}, \citet{Eun+17}, \citet{Woo06,Woo08}, and \citet{Shen+15} are plotted as circle, star, square, and triangle symbols, respectively. The vertical dashed lines indicate the mean velocity shifts. 
\label{fig:voutflow_properties}}
\end{figure}

\begin{figure*}
	\includegraphics[width=0.49\textwidth]{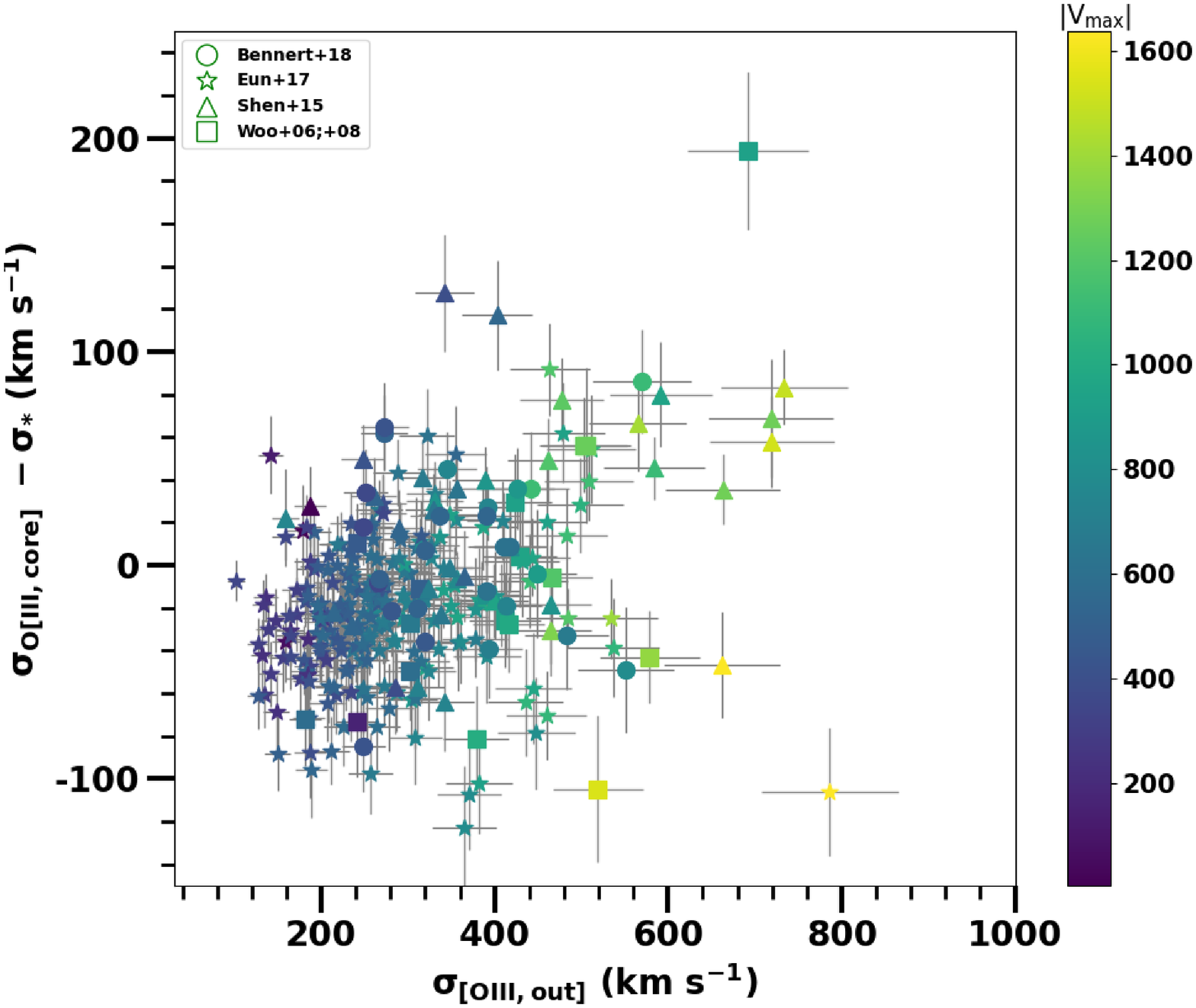}
	\includegraphics[width=0.49\textwidth]{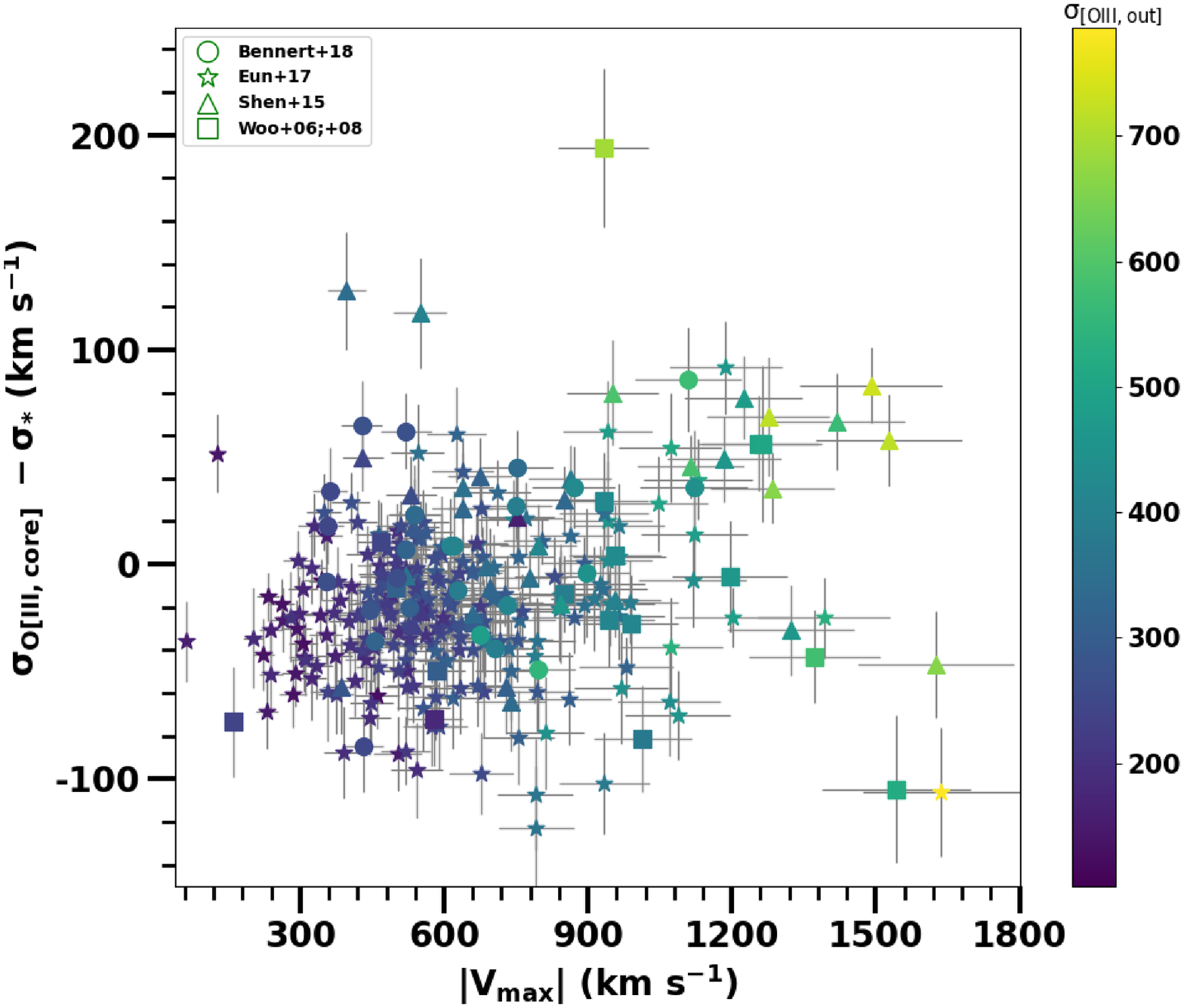}
	\includegraphics[width=0.49\textwidth]{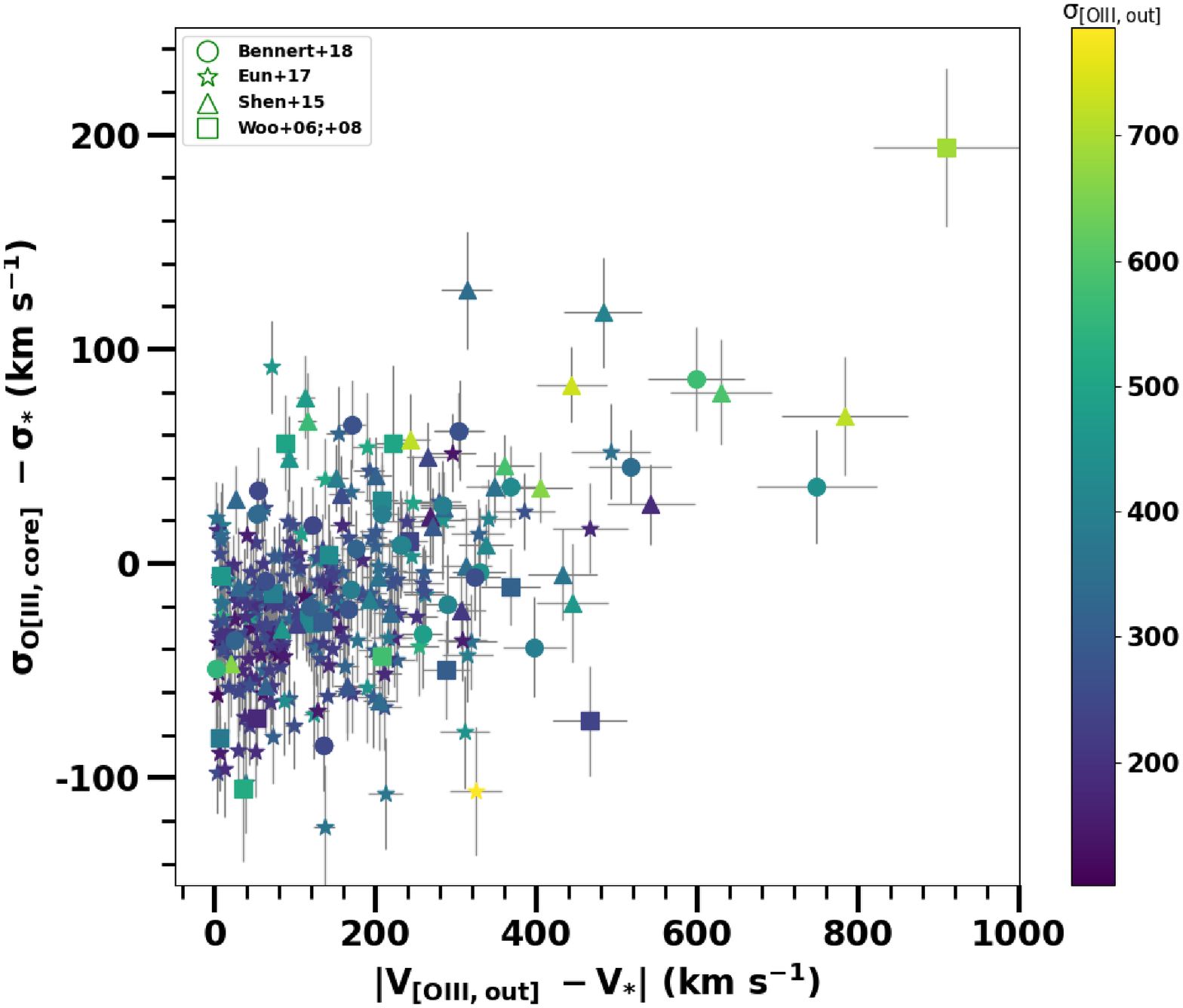}
	\includegraphics[width=0.49\textwidth]{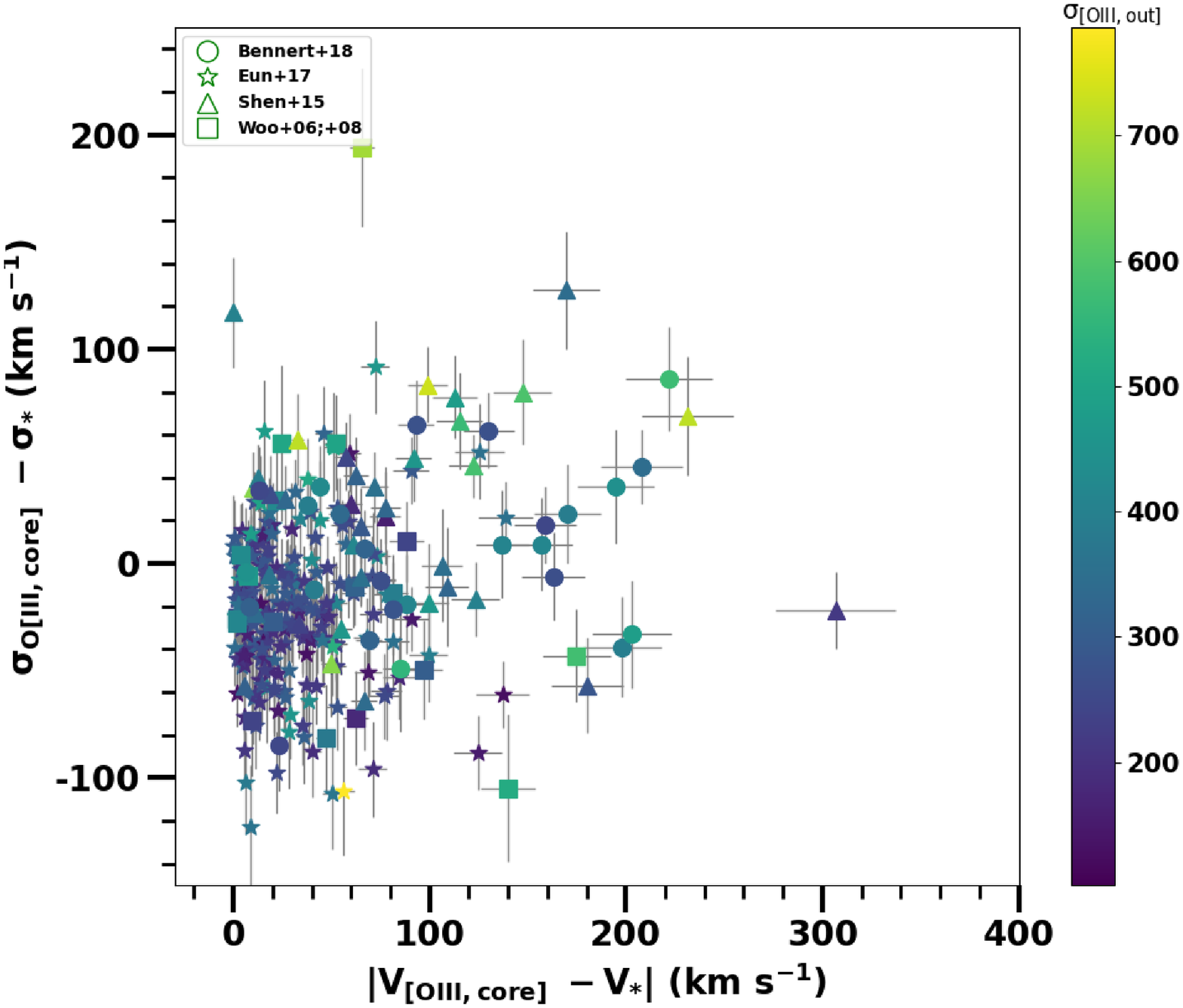}
	\centering
	\caption{Difference between $\rm \sigma_{[OIII,core]}$ and $\rm \sigma_{*}$ as a function of $\rm \sigma_{[OIII,out]}$ (top left), $\rm V_{max}$ (top right), $\rm V_{[OIII,out]}$ - $\rm V_{*}$ (bottom left), and $\rm V_{[OIII,core]}$ - $\rm V_{*}$ (bottom right). The samples of \citet{Bennert+18}, \citet{Eun+17}, \citet{Woo06,Woo08}, and \citet{Shen+15} are plotted as circle, star, square, and triangle symbols, respectively. 
\label{fig:comp_voutflow_oiiistar}}
\end{figure*}

In addition, we also tested the effects of velocity shifts ($\rm V_{[OIII,core]}$ and $\rm V_{[OIII,out]}$) and outflow strengths ($\rm \sigma_{[OIII,out]}$ and $\rm V_{max}$) on the scatter between $\rm \sigma_{[OIII,core]}$ and $\sigma_{*}$. In Figures \ref{fig:voutflow_properties} and \ref{fig:comp_voutflow_oiiistar}, we found that $\rm V_{[OIII,core]}$, $\rm V_{[OIII,out]}$, and $\rm V_{max}$ are connected with \mbh\ and $\rm L_{bol}$. As expected, we found that the discrepancy between $\rm \sigma_{[OIII,core]}$ and $\rm \sigma_{*}$ is also connected with $\rm V_{[OIII,core]}$ and $\rm V_{[OIII,out]}$, as well as $\rm V_{max}$ and $\rm \sigma_{[OIII,out]}$.

We also perform a Spearman's rank-order correlation to test the relationships between \OIII\ vs. stellar kinematics and AGN properties. Table \ref{table:2} shows the Spearman's rank-order coefficients (r) and the p-values of the null hypothesis that the two quantities are not correlated. We found that the discrepancy between $\rm \sigma_{[OIII,core]}$ and $\rm \sigma_{*}$ has mild correlations with Eddington ratio (r = 0.16) and $\rm V_{max}$ (r = 0.17). The correlations are stronger when compared with the broad component of \OIII\ such as $\rm \sigma_{[OIII,out]}$ (r = 0.31) and $\rm V_{[OIII,out]}$ (r = $-$0.33). And, we also found that in comparison with $\rm \sigma_{[OIII,core]}$ and $\rm \sigma_{*}$, the discrepancy between $\rm \sigma_{[OIII,out]}$ and $\rm \sigma_{*}$ has a stronger correlation with outflow strengths such as Eddington ratio (r = 0.25) and $\rm V_{max}$ (r = 0.80). In addition, there are mild correlations when comparing the discrepancy between $\rm V_{[OIII,core]}$ and $\rm V_{*}$ with Eddington ratio (r = $-$0.24), $\rm \sigma_{[OIII,out]}$ (r = $-$0.20), and $\rm V_{max}$ (r = 0.06). In the case of $\rm V_{[OIII,out]}$ and $\rm V_{*}$, the correlations tend to be stronger with Eddington ratio (r = $-$0.29), $\rm \sigma_{[OIII,out]}$ (r = $-$0.34), and $\rm V_{max}$ (r = 0.46). Our results for the correlations between velocity shifts and Eddington ratio are in agreement with that of \citet{Zhang11}. By using homogenous samples of 383 radio-quiet Seyfert 1 galaxies, they also found that there are weak correlations between $\rm V_{[OIII,core]}$ and $\rm V_{[OIII,out]}$ with AGN properties such as $\rm L_{5100}$, \mbh, and Eddington ratio. They interpreted results that although AGN activities are responsible for launching initial outflows, the surrounding interstellar medium (ISM) environment (e.g., density profile) may also have an impact on the later speed of outflow gas after their launch, and this may lead to the diversity of \OIII\ properties in AGNs.

Our results are consistent with those of \citet{Greene05} and \citet{Ho09} that although gas kinematics of the NLR mainly follows the gravitational potential of the bulge of the host galaxy, AGN activities such as outflows may also have impact on the gas kinematics of the NLR. For those targets with high $\rm L_{bol}$, we expect that the AGN activity is high, and there may be strong outflows acting on the gas kinematics of the NLR, leading to the large scatter between $\rm \sigma_{[OIII,core]}$ and $\rm \sigma_{*}$. As discussed in \citet{Ho09}, the kinematics of the NLR emission gas and stars are expected to be similar if both quantities are governed by the bulge gravitational potential. In the cases that the gas and stars suffer collisional hydrodynamic drags by the surrounding ISM, we may expect to see the results that the gas velocity dispersions are lower than that of the stars. However, if there is additional action from the AGN activity (i.e., outflows) on the relations between the gas and stars, the additional energy may accelerate the gas, then lead to the results that the gas velocity dispersions are comparable or larger than that of the stars.

\subsection{The Limitation of Double Gaussian Approach}

Regarding the large scatters between $\rm \sigma_{[OIII,core]}$ and $\sigma_{*}$ for high luminosity objects, we may need to consider the details of measurement of the \OIII\ line width. As suggested by previous studies (e.g., \citealp{Greene05}, \citealp{Bennert+18}), a double Gaussian is the best choice for fitting the \OIII\ profile and $\rm \sigma_{[OIII,core]}$ could be used as a surrogate for $\sigma_{*}$. But what is the {\it{true}} value of the gas kinematics following the gravitational potential of the stars and not disturbed by outflow effects? We have simply assumed the correction by removing the wing component in the \OIII\ profile. But for those high luminosity AGNs in which the fraction of AGN accretion is large, the measurement of the {\it{true}} $\rm \sigma_{[OIII,core]}$ as a surrogate for $\sigma_{*}$ may have high uncertainty even when we have corrected for non-gravitational effects by removing the broad wing component in the \OIII\ profile. 

In general, we suggest that it should be cautionary to use $\rm \sigma_{[OIII,core]}$ as a surrogate for $\sigma_{*}$ for individual sources (particularly for high luminosity AGNs). The replacement of $\rm \sigma_{[OIII,core]}$ for $\sigma_{*}$ could be applied for statistical studies albeit with large scatters.

\subsection{Aperture and Rotational Broadening Effects}\label{section:inclination}

Both gas and stellar velocity dispersions may vary with different extraction radius measurements. As we mentioned in Section \ref{section:result_sigma}, within the large aperture size of SDSS (3$\arcsec$), the extracted spectra of SDSS cover a size of $\sim$5.4 kpc in the center of the galaxy at redshift z $\sim$ 0.1. For the higher redshift sample at z $\sim$ 0.3-1.0, the radius size covers $\sim$3.3-7.8 kpc in the center of the galaxy. Within the large aperture size, it may be possible that integrated spectra of SDSS may not only extract the region within the bulge radius of the galaxy but also contain the disk component of the galaxy, then overestimate the measured velocity dispersion. Therefore, it is important to check whether the measurements of $\rm \sigma_{*}$ are well extracted within the bulge radius of the galaxy. Similarly, for the gas emission line, we should check whether the extraction radius for the \OIII\ emission is well defined within the NLR or not. 

\citet{Greene05} examined the extraction aperture size of SDSS for their study based on the local type 2 AGN sample. By using the relation between the NLR radius and the \OIII\ luminosity, they estimated that the typical size of the NLR for their objects at z $=$ 0.1 is $\sim$0.25-1.2 kpc within the angular size of $\sim$0.14$\arcsec$-0.67$\arcsec$. This angular size is well defined within the SDSS aperture size. Following their method, we estimated the NLR size for our local SDSS and high redshift samples by using the size-luminosity relation (e.g., \citealp{Bae+17}, \citealp{Le+17}). Based on the bolometric luminosity of our sample, we estimated the \OIII\ luminosity of our local SDSS and high redshift samples to be in the range of $\mathrm{L_{OIII} \sim10^{41.5} - 10^{43}}$ \ergs ($\rm L_{bol} = 700 \times L_{OIII}$, \citealp{LaMassa+09}). This luminosity range corresponds to the physical size of $\sim$0.25-3.5 kpc, well defined within the SDSS aperture size at redshift of z $\sim$ 0.1-1.0. Therefore, the measurements of the $\rm \sigma_{[OIII]}$ of our local and high redshift SDSS samples are well defined within the NLR. 

As for the extraction size of $\sigma_{*}$, as discussed by \citet{Greene05}, using the fundamental plane relation from \citet{Bernardi03}, a median $\rm \sigma_{*}$ of 150 \kms\ corresponds to an effective radius of $\sim$4.5 kpc. For the local SDSS sample, the integrated spectra of SDSS cover the size of $\sim$5.4 kpc at redshift of z $\sim$ 0.1. Therefore, for the local SDSS sample, the extraction size of $\rm \sigma_{*}$ is well defined within the size of the bulge of the galaxy. For the high redshift sample, the SDSS radius size includes $\sim$3.3-7.8 kpc at redshift of z $\sim$ 0.2-1.0. For the sample at high redshift z $>$ 0.5, the measurements of $\rm \sigma_{*}$ contain the light from both the bulge and disk components (see Section 4 in \citealp{Shen+15}), we expect that this may lead to the large scatters between $\rm \sigma_{[OIII,core]}$ and $\rm \sigma_{*}$ for the high redshift AGNs. As mentioned in \citet{Shen+15}, rotational velocity could bias their measured $\rm \sigma_{*}$ ($<$16$\%$).  

\citet{Greene05} checked the effect of rotational broadening in their sample. Based on their analysis, rotational broadening effect could overestimate the observed gas and stellar velocity dispersions in their sample by $\sim$12-15$\%$ at z $\sim$ 0.1. Accordingly, we expect that the rotational broadening effect could bias our measured gas and stellar velocity dispersions but with small fractions compared to the large scatters between $\rm \sigma_{[OIII,core]}$ and $\rm \sigma_{*}$. 

\begin{figure}
	\includegraphics[width=0.40\textwidth]{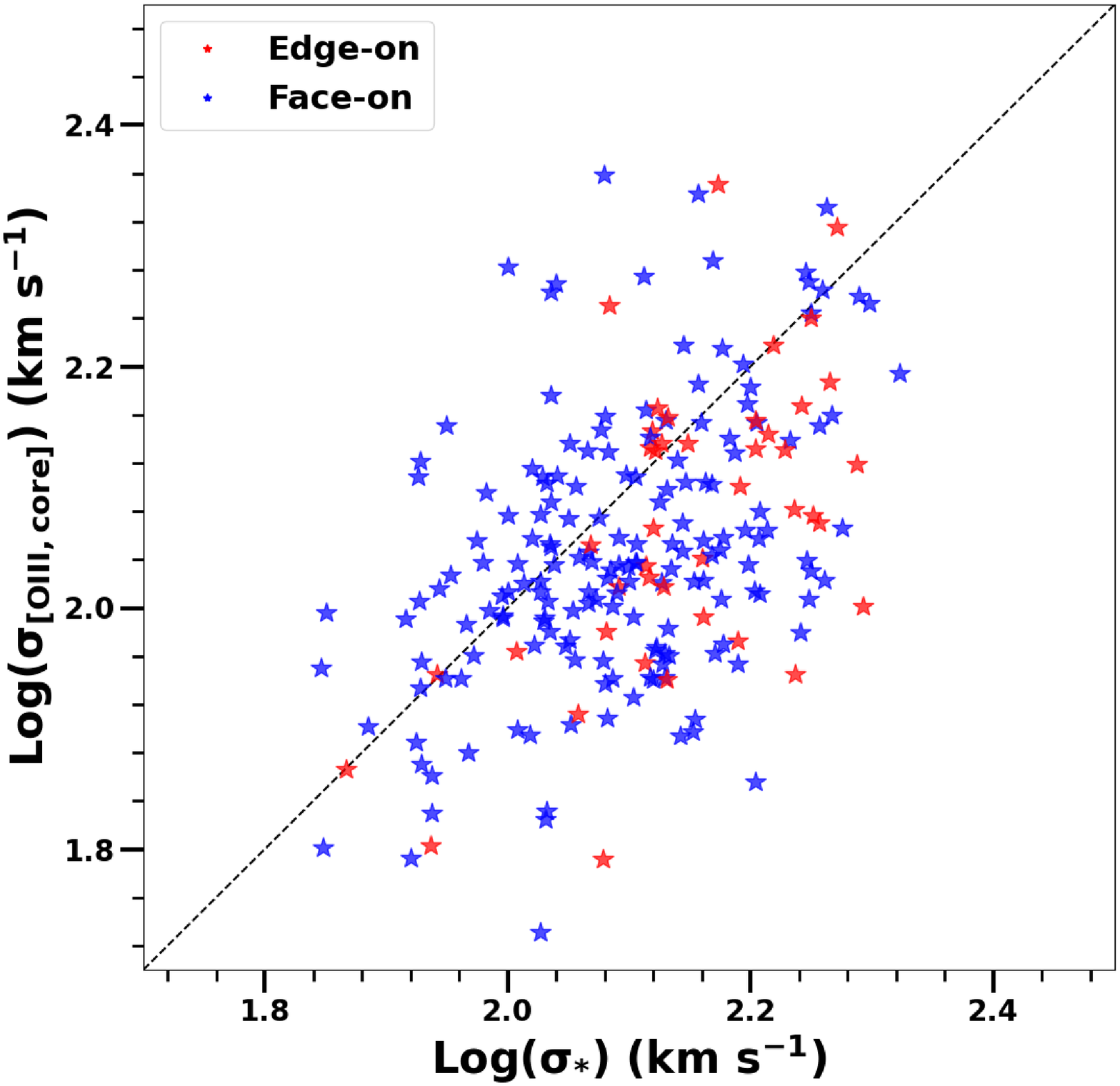}		
	\centering
	\caption{Comparison between $\rm \sigma_{*}$ and $\rm \sigma_{[OIII,core]}$ for the local SDSS sample (z $<$ 0.1) of \citet{Eun+17}. The red symbols indicate edge-on targets (b/a < 0.5), while, face-on targets are shown in blue symbols (b/a > 0.5).   
\label{fig:inclination}}
\end{figure}

For the local SDSS sample, \citet{Eun+17} classified 221 spiral galaxies among 611 objects in their sample. Based on the minor-to-major axis ratio (b/a) obtained from the SDSS-DR7, those sources are classified into face-on (b/a $>$ 0.5) and edge-on (b/a $<$ 0.5) spiral galaxies. Figure \ref{fig:inclination} shows the comparison of $\rm \sigma_{[OIII,core]}$ and $\rm \sigma_{*}$ for the 221 spiral galaxies from the local SDSS sample. Interestingly, for those edge-on targets, we found that $\rm \sigma_{*}$ is on average ($\sim$50 \kms) larger than $\rm \sigma_{[OIII,core]}$. While for those face-on sources, $\rm \sigma_{[OIII,core]}$ and $\sigma_{*}$ are comparable with a scatter of $\sim$0.12~dex. \citet{Eun+17} found that $\sigma_{*}$ for edge-on targets may be overestimated due to the contribution of the rotational velocity of the disk galaxy. In addition, for those face-on targets, the rotational velocity is relatively small compared to that of the edge-on objects. This may explain why $\sigma_{*}$ is larger than $\rm \sigma_{[OIII,core]}$ for those edge-on spiral galaxies in the local SDSS sample. Also, the rotational broadening effect may contribute less to the velocity dispersions of the NLR gas than to $\sigma_{*}$. 

\section{Summary}\label{section:sum}

We constructed our sample based on four different samples at redshift z $<$ 1.0. For local AGNs (z $<$ 0.1), we selected 59 high-quality long-slit Keck spectra \citep{Bennert+18} and 611 local hidden type 1 SDSS spectra  of \citet{Eun+17}. For the higher redshift sample (0.3 $<$ z $<$ 1.0), we chose the sample which is based on 18 high S/N Keck spectra \citep[]{Woo06, Woo08} and 52 co-added high S/N spectra from the SDSS-RM project \citep{Shen+15}. Consequently, our sample has the broad dynamic range of $\mathrm{\lambda L_{5100} \sim10^{41.5} - 10^{46.0}}$ \ergs and 6.5 $<$ $\log($\mbh) $<$ 9.5. In addition, the high S/N spectra of our sample allow us to achieve reliable measurements for $\sigma_{*}$ and $\rm \sigma_{[OIII]}$. We took advantage of this unique sample to test the validity of using $\rm \sigma_{[OIII,core]}$ as a surrogate for $\sigma_{*}$ not only for local AGNs but also for high redshift AGNs. We summarize our main results as follows: 

\smallskip
(1) From the comparisons of $\sigma_{*}$ and $\rm \sigma_{[OIII,core]}$, we found the broad correlation between $\rm \sigma_{[OIII,core]}$ and $\sigma_{*}$ with a scatter of 0.11~dex for the local objects, while, the scatter is somewhat larger for the high redshift targets, being $\sim$0.16~dex.

\smallskip
(2) We found that $\rm \sigma_{[OIII,core]}$ and $\sigma_{*}$ are well correlated for the low luminosity objects, while, for the targets which have higher luminosity ranges, the scatters of correlation between $\rm \sigma_{[OIII,core]}$ and $\sigma_{*}$ become larger.

\smallskip
(3) We found that the Eddington ratio plays an important role in the differences between $\rm \sigma_{[OIII,core]}$ and $\sigma_{*}$. For the targets which have high Eddington ratios, $\rm \sigma_{[OIII,core]}$ tends to be larger than $\sigma_{*}$. Also, we found that outflow strengths (i.e., $\rm V_{max}$, $\rm V_{[OIII,out]}$, and $\rm \sigma_{[OIII,out]}$) have significant effects on the differences between $\rm \sigma_{[OIII,core]}$ and $\sigma_{*}$. The discrepancies between $\rm \sigma_{[OIII,core]}$ and $\sigma_{*}$ are larger when $\rm V_{max}$, $\rm V_{[OIII,out]}$, and $\rm \sigma_{[OIII,out]}$ increase.

\smallskip
(4) For the local SDSS sample, $\sigma_{*}$ is on average larger than $\rm \sigma_{[OIII,core]}$ for the edge-on spiral galaxies. The result may indicate that the rotational broadening effect has large contribution to $\sigma_{*}$ compared to $\rm \sigma_{[OIII]}$. \\

\smallskip
Our results show that the Eddington ratios and outflow strengths may have significant effects on the discrepancies between the gas and stellar kinematics in AGNs. Therefore, using $\rm \sigma_{[OIII,core]}$ as a surrogate for $\sigma_{*}$ should be adopted with caution, and the validity of using $\rm \sigma_{[OIII,core]}$ to replace $\sigma_{*}$ is warranted only for the low luminosity AGNs.

\acknowledgements
We thank the anonymous referee for valuable suggestions and comments, which improved the paper. We thank Prof. Jong-Hak Woo and Ashraf Ayubinia for thoughtful discussions. This work has been supported by the National Natural Science Foundation of China (NSFC-12003031, NSFC-11890693, NSFC-12025303, NSFC-12203047), National Key R\&D Program of China No. 2022YFF0503401, the K.C. Wong Education Foundation, and the science research grants from the China Manned Space Project with NO. CMS-CSST-2021-A06. H. A. N. Le acknowledges the support from the "Fundamental Research Funds for the Central Universities".

\begin{deluxetable*}{cccccccccccc}
\tablecolumns{12}
\tablewidth{\textwidth}
\tabletypesize{\scriptsize}
\tablecaption{Targets and Measured Properties}
\tablehead{
\colhead{Target}&
\colhead{z}&
\colhead{$\rm \lambda L_{5100}$}&
\colhead{$\rm \sigma_{H\beta}$}&
\colhead{$\rm \log(M_{BH})$}&
\colhead{$\rm \sigma_{*}$}&
\colhead{$\rm \sigma_{[OIII,core]}$}&
\colhead{$\rm \sigma_{[OIII,out]}$} &
\colhead{$\rm V_{max}$} &
\colhead{$\rm V_{[OIII,core]}$}&
\colhead{$\rm V_{[OIII,out]}$} &
\colhead{S/N} 
\\
\colhead{}&
\colhead{}&
\colhead{($\rm 10^{44}\ erg~s^{-1}$)}&
\colhead{($\rm km~s^{-1}$)}&
\colhead{($\rm M_{\odot}$)}&
\colhead{($\rm km~s^{-1}$)}&
\colhead{($\rm km~s^{-1}$)}&
\colhead{($\rm km~s^{-1}$)} &
\colhead{($\rm km~s^{-1}$)} &
\colhead{($\rm km~s^{-1}$)} &
\colhead{($\rm km~s^{-1}$)} &
\colhead{}
\\
\colhead{(1)}&
\colhead{(2)}&
\colhead{(3)}&
\colhead{(4)}&
\colhead{(5)}&
\colhead{(6)}&
\colhead{(7)}&
\colhead{(8)} &
\colhead{(9)} &
\colhead{(10)} &
\colhead{(11)} &
\colhead{(12)}
}

\startdata
S01...........................		&	0.3590	&	1.37	&	2194	&	8.22	&	132	&	121	&	314	&	-500	&	-62	&	-368	&	72	\\
S04...........................		&	0.3579	&	1.19	&	1749	&	7.99	&	186	&	180	&	466	&	1196	&	-8	&	-8	&	46	\\
S05...........................		&	0.3530	&	2.23	&	3333	&	8.70	&	132	&	118	&	391	&	-850	&	81	&	-72	&	119	\\
S06...........................		&	0.3684	&	1.10	&	1413	&	7.79	&	169	&	143	&	412	&	-944	&	-2	&	-117	&	31	\\
S07...........................		&	0.3517	&	1.81	&	2547	&	8.42	&	145	&	175	&	423	&	-934	&	22	&	-208	&	105	\\
S08...........................		&	0.3585	&	1.59	&	1217	&	7.75	&	187	&	138	&	303	&	-586	&	-97	&	-288	&	54	\\
S09...........................		&	0.3542	&	1.76	&	1748	&	8.08	&	187	&	115	&	181	&	581	&	-62	&	52	&	39	\\
S11...........................		&	0.3558	&	1.57	&	1354	&	7.84	&	127	&	137	&	242	&	-467	&	-88	&	-241	&	115	\\
S12...........................		&	0.3574	&	1.82	&	4256	&	8.87	&	173	&	130	&	579	&	-1327	&	175	&	-207	&	40	\\
S23...........................		&	0.3511	&	1.78	&	4251	&	8.86	&	172	&	145	&	416	&	-990	&	-2	&	-120	&	106	\\
S24...........................		&	0.3616	&	1.49	&	2635	&	8.40	&	214	&	141	&	241	&	-161	&	-10	&	-466	&	103	\\
S26...........................		&	0.3691	&	0.83	&	1914	&	7.99	&	128	&	101	&	302	&	-661	&	-20	&	-134	&	50	\\
S28...........................		&	0.3678	&	0.97	&	2532	&	8.27	&	210	&	129	&	379	&	1014	&	-47	&	-6	&	74	\\
W09...........................		&	0.5654	&	2.64	&	2747	&	8.57	&	289	&	184	&	519	&	1543	&	-140	&	35	&	63	\\
W11...........................		&	0.5650	&	0.78	&	2026	&	8.02	&	126	&	182	&	503	&	-1255	&	52	&	-89	&	18	\\
W14...........................		&	0.5617	&	5.56	&	2616	&	8.70	&	228	&	284	&	506	&	-1263	&	25	&	-222	&	77	\\
W17...........................		&	0.5617	&	0.86	&	2483	&	8.22	&	165	&	169	&	429	&	-956	&	4	&	-141	&	25	\\
W22...........................		&	0.5622	&	4.65	&	2654	&	8.67	&	144	&	338	&	692	&	-932	&	-65	&	-910	&	82	\\
141359.51+531049.3	&	0.8982	&	0.97	&	1817	&	7.98	&	121	&	157	&	0	&	0	&	-222	&	0	&	15	\\
141324.28+530527.0	&	0.4559	&	1.07	&	2080	&	8.12	&	191	&	127	&	342	&	-739	&	-67	&	-205	&	381	\\
141323.27+531034.3	&	0.8492	&	1.97	&	3299	&	8.66	&	166	&	261	&	0	&	0	&	172	&	0	&	14	\\
141532.36+524905.9	&	0.7147	&	1.76	&	1005	&	7.60	&	182	&	126	&	285	&	-387	&	180	&	-165	&	33	\\
141522.54+524421.5	&	0.5263	&	0.63	&	1229	&	7.54	&	119	&	146	&	187	&	4	&	-60	&	-543	&	48	\\
141417.69+532810.8	&	0.8077	&	1.55	&	1927	&	8.14	&	232	&	340	&	0	&	0	&	-175	&	0	&	32	\\
141151.78+525344.1	&	0.5165	&	1.16	&	2817	&	8.40	&	128	&	208	&	591	&	-952	&	-147	&	-630	&	62	\\
141408.76+533938.3	&	0.1919	&	0.32	&	1188	&	7.36	&	82	&	132	&	248	&	-430	&	-58	&	-265	&	291	\\
141324.66+522938.2	&	0.8122	&	1.36	&	2941	&	8.48	&	114	&	231	&	403	&	-550	&	0	&	-483	&	75	\\
141724.59+523024.9	&	0.4818	&	0.90	&	2398	&	8.20	&	172	&	176	&	0	&	0	&	-29	&	0	&	129	\\
141721.80+534102.6	&	0.1934	&	0.54	&	1025	&	7.35	&	137	&	115	&	215	&	551	&	307	&	307	&	163	\\
141645.58+534446.8	&	0.4418	&	0.47	&	971	&	7.27	&	152	&	129	&	338	&	-660	&	-12	&	-219	&	72	\\
141018.04+532937.5	&	0.4696	&	0.41	&	2077	&	7.90	&	130	&	102	&	236	&	-536	&	-33	&	-102	&	97	\\
141751.14+522311.1	&	0.2806	&	0.13	&	3407	&	8.07	&	181	&	124	&	311	&	-730	&	6	&	-63	&	124	\\
141112.72+534507.1	&	0.5872	&	1.35	&	1585	&	7.94	&	97	&	133	&	356	&	-638	&	-72	&	-348	&	39	\\
140943.01+524153.1	&	0.4211	&	0.40	&	2268	&	7.97	&	184	&	219	&	0	&	0	&	33	&	0	&	108	\\
141409.44+535648.2	&	0.8251	&	1.36	&	1264	&	7.74	&	95	&	88	&	357	&	-778	&	-65	&	-203	&	195	\\
141941.11+533649.6	&	0.6457	&	3.52	&	1767	&	8.25	&	109	&	142	&	261	&	-531	&	-19	&	-157	&	93	\\
142010.25+524029.6	&	0.5477	&	0.91	&	3681	&	8.58	&	177	&	237	&	0	&	0	&	-14	&	0	&	68	\\
141004.27+523141.0	&	0.5266	&	1.28	&	2248	&	8.23	&	150	&	173	&	159	&	754	&	-77	&	268	&	147	\\
142038.52+532416.5	&	0.2647	&	0.24	&	2633	&	7.98	&	66	&	137	&	0	&	0	&	-73	&	0	&	183	\\
141658.28+521205.1	&	0.6018	&	2.01	&	3184	&	8.64	&	158	&	227	&	719	&	-1276	&	-231	&	-783	&	63	\\
141514.15+540222.9	&	0.8488	&	1.97	&	2910	&	8.55	&	201	&	190	&	324	&	-694	&	109	&	-29	&	60	\\
142043.53+523611.4	&	0.3368	&	0.24	&	2034	&	7.76	&	115	&	133	&	289	&	-535	&	-65	&	-272	&	167	\\
141320.05+520527.9	&	0.5998	&	0.87	&	2590	&	8.26	&	313	&	182	&	0	&	0	&	-1	&	0	&	378	\\
142124.36+532312.5	&	0.8255	&	1.21	&	1813	&	8.03	&	225	&	203	&	0	&	0	&	-71	&	0	&	40	\\
142112.29+524147.3	&	0.8425	&	1.52	&	1818	&	8.09	&	77	&	161	&	734	&	-1491	&	-99	&	-444	&	91	\\
141031.33+521533.8	&	0.6076	&	1.14	&	1388	&	7.78	&	187	&	186	&	348	&	-687	&	-106	&	-313	&	149	\\
141049.76+540040.6	&	0.8343	&	1.16	&	1247	&	7.69	&	154	&	149	&	364	&	-521	&	-18	&	-432	&	23	\\
142209.14+530559.8	&	0.7542	&	4.35	&	3532	&	8.91	&	106	&	203	&	0	&	0	&	93	&	0	&	11	\\
141058.78+520712.2	&	0.3910	&	0.26	&	3249	&	8.18	&	93	&	170	&	477	&	1224	&	-113	&	-113	&	224	\\
141852.64+520142.8	&	0.4398	&	0.31	&	1713	&	7.67	&	96	&	126	&	331	&	850	&	-26	&	-26	&	52	\\
141553.09+541816.5	&	0.7305	&	1.86	&	1218	&	7.78	&	80	&	126	&	584	&	-1113	&	122	&	-361	&	32	\\
140904.43+540344.2	&	0.6585	&	0.87	&	3478	&	8.52	&	215	&	318	&	0	&	0	&	-99	&	0	&	31	\\
141115.19+515209.0	&	0.5716	&	1.45	&	1822	&	8.08	&	123	&	189	&	565	&	1420	&	-115	&	-115	&	37	\\
141135.89+515004.5	&	0.6500	&	1.30	&	1365	&	7.80	&	119	&	247	&	343	&	-397	&	170	&	-313	&	19	\\
140715.49+535610.2	&	0.6827	&	2.02	&	2851	&	8.54	&	119	&	177	&	719	&	-1528	&	33	&	-244	&	78	\\
140551.99+533852.1	&	0.4548	&	0.64	&	2291	&	8.08	&	197	&	150	&	662	&	1625	&	-50	&	19	&	139	\\
140923.42+515120.1	&	0.8532	&	1.07	&	1915	&	8.05	&	379	&	172	&	337	&	-520	&	184	&	-161	&	8	\\
141419.84+533815.3	&	0.1645	&	0.70	&	1446	&	7.71	&	98	&	134	&	663	&	-1284	&	10	&	-404	&	91	\\
140915.70+532721.8	&	0.2584	&	0.35	&	3147	&	8.22	&	172	&	195	&	0	&	0	&	-73	&	0	&	86	\\
141253.92+540014.4	&	0.1871	&	0.09	&	2431	&	7.68	&	117	&	158	&	0	&	0	&	-9	&	0	&	191	\\
140759.07+534759.8	&	0.1725	&	0.40	&	1995	&	7.86	&	130	&	139	&	418	&	796	&	-61	&	-337	&	146	\\
140812.09+535303.3	&	0.1161	&	0.07	&	3076	&	7.83	&	112	&	160	&	0	&	0	&	-3	&	0	&	274	\\
142103.53+515819.5	&	0.2634	&	0.35	&	1521	&	7.59	&	105	&	146	&	317	&	-675	&	-62	&	-200	&	107	\\
141318.96+543202.4	&	0.3623	&	0.98	&	1592	&	7.87	&	130	&	113	&	399	&	-956	&	-123	&	-192	&	82	\\
141644.17+532556.1	&	0.4253	&	0.43	&	1843	&	7.81	&	114	&	163	&	462	&	1185	&	-92	&	-92	&	127	\\
141729.27+531826.5	&	0.2374	&	0.31	&	2695	&	8.06	&	204	&	186	&	465	&	-844	&	-100	&	-445	&	86	\\
142100.04+532139.6	&	0.6766	&	1.57	&	1657	&	8.01	&	89	&	129	&	390	&	-862	&	-13	&	-151	&	27	\\
141308.10+515210.4	&	0.2882	&	0.54	&	1307	&	7.56	&	104	&	130	&	329	&	-638	&	-78	&	-285	&	68	\\
141645.15+542540.8	&	0.2439	&	0.25	&	2733	&	8.02	&	164	&	134	&	465	&	1323	&	-55	&	83	&	118	\\
142225.62+533426.3	&	0.7569	&	0.57	&	2717	&	8.21	&	144	&	183	&	0	&	0	&	77	&	0	&	10	\\
\enddata
\label{table:sample}
\tablecomments{Col. (1): Name of target. Col. (2): Redshift.  Col. (3): Continuum luminosity at 5100\AA. Col. (4): Line dispersion of $\rm {H\beta}$ emission line. Col. (5): Black-hole mass. Col (6): Stellar velocity dispersion. Col. (7): Line dispersion of the core component of \OIII\ emission line. Col. (8): Line dispersion of the broader component of \OIII\ emission line. Col. (9): Outflow velocity. Col. (10): Velocity shift of the core component of \OIII\ emission line with respect to the stellar systemic velocity. Col. (11): Velocity shift of the broader component of \OIII\ emission line with respect to the stellar systemic velocity. Col. (12): Signal-to-noise ratio of spectra at 5100\AA\ continuum. Error values are 0.04 dex for all measured properties (except that black-hole mass uncertainty is 0.4 dex).}
\end{deluxetable*}


\begin{turnpage}
\begin{deluxetable*}{cccccccccc}
\tablecolumns{10}
\tabletypesize{\scriptsize}
\tablecaption{Spearman's rank-order Correlation Results between \OIII\ vs. Stellar Kinematics and AGN Properties}
\tablehead{
\colhead{  }&
\colhead{$\rm \log(M_{BH})$}&
\colhead{$\rm \log(L_{Bol}/L_{Edd})$}&
\colhead{$\rm \log(\lambda L_{5100}$)}&
\colhead{$\rm V_{max}$} &
\colhead{$\rm \sigma_{[OIII,out]}$} &
\colhead{$\rm V_{[OIII,out]}$ - $\rm V_{*}$} &
\colhead{$\rm V_{[OIII,core]}$} - $\rm V_{*}$&
\colhead{$\rm \sigma_{[OIII,core]}$ - $\rm \sigma_{*}$}&
\colhead{$\rm \sigma_{[OIII,out]}$ - $\rm \sigma_{*}$} 
\\
\colhead{}&
\colhead{($\rm M_{\odot}$)}&
\colhead{}&
\colhead{($\rm 10^{44}\ erg~s^{-1}$)}&
\colhead{($\rm km~s^{-1}$)}&
\colhead{($\rm km~s^{-1}$)}&
\colhead{($\rm km~s^{-1}$)} &
\colhead{($\rm km~s^{-1}$)} &
\colhead{($\rm km~s^{-1}$)} &
\colhead{($\rm km~s^{-1}$)} 
\\
\colhead{ } &
\colhead{(1)}&
\colhead{(2)}&
\colhead{(3)}&
\colhead{(4)}&
\colhead{(5)}&
\colhead{(6)}&
\colhead{(7)}&
\colhead{(8)} &
\colhead{(9)} 
}
\startdata

$\rm \sigma_{[OIII,core]}$ $-$ $\rm \sigma_{*}$	&	$-$0.11 (5e$-$03)	&	0.16 (9e$-$06)	&	$-$0.08 (3e$-$02)	&	0.17 (3e$-$03)	&	0.31 (4e$-$08)	&	$-$0.33 (6e$-$09)	&	$-$0.09 (1e$-$02)	&	1.00 (0e+00)	&	0.53 (8e$-$23)	\\
$\rm \sigma_{[OIII,out]}$ $-$ $\rm \sigma_{*}$	&	0.29 (3e$-$07)	&	0.25 (1e$-$05)	&	0.35 (5e$-$10)	&	0.80 (3e$-$67)	&	0.94 (2e$-$140)	&	$-$0.34 (2e$-$09)	&	$-$0.20 (6e$-$04)	&	0.53 (8e$-$23)	&	1.00 (0e+00)	\\
$\rm V_{[OIII,core]}$ $-$ $\rm V_{*}$	&	0.03 (4e$-$01)	&	$-$0.24 (1e$-$10)	&	$-$0.11 (3e$-$03)	&	0.06 (3e$-$01)	&	$-$0.20 (6e$-$04)	&	0.30 (1e$-$07)	&	1.00 (0e+00)	&	$-$0.09 (1e$-$02)	&	$-$0.20 (6e$-$04)	\\
$\rm V_{[OIII,out]}$ $-$ $\rm V_{*}$	&	$-$0.15 (9e$-$03)	&	$-$0.29 (4e$-$07)	&	$-$0.31 (5e$-$08)	&	0.46 (2e$-$17)	&	$-$0.34 (2e$-$09)	&	1.00 (0e+00)	&	0.30 (1e$-$07)	&	$-$0.33 (6e$-$09)	&	$-$0.34 (2e$-$09)	\\

\enddata
\label{table:2}
\tablecomments{Col. (1): Black-hole mass. Col. (2): Eddington ratio. Col. (3): Continuum luminosity at 5100\AA. Col. (4): Outflow velocity. Col. (5): Line dispersion of the broader component of \OIII\ emission line. Col (6): Velocity shift of the broader component of \OIII\ emission line with respect to the stellar systemic velocity. Col. (7): Velocity shift of the core component of \OIII\ emission line with respect to the stellar systemic velocity. Col. (8): Line dispersion of the core component of \OIII\ emission line with respect to the stellar dispersion. Col. (9): Line dispersion of the broad component of \OIII\ emission line with respect to the stellar dispersion. The p-values of the null hypothesis that the two quantities are not correlated are shown in parentheses, following the Spearman's rank-order coefficients. The outflow velocities are used in absolute values for comparing with the line dispersions. The numbers of the sources in the correlation calculations for the core and broad components of \OIII\ emission line are 740, and 300, respectively.} 

\end{deluxetable*}
\end{turnpage}




\end{CJK*}


\begin{thebibliography}{}

\bibitem[Abazajian et al.(2009)]{Abazajian09} Abazajian, K. N., Adelman-McCarthy, J. K., Ag{\"u}eros, M. A., et al. 2009,
ApJS, 182, 543

\bibitem[Ayubinia et al.(2022)]{Ayubinia+22} Ayubinia, A., Xue, Y. Q., Woo, J-.H. et al. 2022, Universe, in press

\bibitem[Bae \& Woo(2014)]{Bae+14} Bae, H.-J., \& Woo, J.-H. 2014, ApJ, 795, 30

\bibitem[Bae et al.(2017)]{Bae+17} Bae, H-J., Woo, J.-H., Karouzos, M. et al. 2017, ApJ, 837, 91

\bibitem[Bernardi(2003)]{Bernardi03} Bernardi, M. 2003, AJ, 125, 1866

\bibitem[Bennert et al.(2018)]{Bennert+18} Bennert V. N. et al., 2018, MNRAS, 481, 138

\bibitem[Bennert et al.(2015)]{Bennert+15} Bennert V. N. et al., 2015, ApJ, 809, 20

\bibitem[Bennert et al.(2010)]{Bennert+10} Bennert, V.~N., Treu, T., Woo, J.-H., et al.\ 2010, \apj, 708, 1507 

\bibitem[Bentz et al.(2006)]{Bentz+06} Bentz, M.~C., Peterson, 
B.~M., Pogge, R.~W., Vestergaard, M., \& Onken, C.~A.\ 2006, \apj, 644, 133 

\bibitem[Bentz et al.(2009a)]{Bentz+09} Bentz, M.~C., Peterson, 
B.~M., Netzer, H., Pogge, R.~W., \& Vestergaard, M.\ 2009, \apj, 697, 160 


\bibitem[Bentz et al.(2013)]{Bentz+13} Bentz, M. C., Denney, 
K. D., Grier, C. J., et al.\ 2013, \apj, 767, 149


\bibitem[Boroson(2003)]{Boroson03} Boroson, T. A., 2003, ApJ, 585, 647

\bibitem[Boroson \& Green(1992)]{BG92} Boroson, T.~A., \& Green, R.~F.\ 1992, \apjs, 80, 109 

\bibitem[Boroson(2005)]{Boroson05} Boroson, T. 2005, AJ, 130, 381

\bibitem[Bonning et al.(2005)]{Bonning05} Bonning, E.W., Shields, G. A., Salviander, S., McLure, R. J., 2005, ApJ, 626,
89

\bibitem[Cappellari \& Emsellem(2004)]{Cappellari04} Cappellari, M., \& Emsellem, E. 2004, PASP, 116, 138

\bibitem[Crenshaw \& Kraemer(2000)]{Crenshaw00} Crenshaw, D. M., \& Kraemer, S. B. 2000, ApJL, 532, L101

\bibitem[Eun, Woo \& Bae(2017)]{Eun+17} Eun, D., Woo, J.-H. \& Bae, H.-J. 2017, ApJ, 842, 5

\bibitem[Eracleous \& Halpern(2004)]{Eracleous04} Eracleous, M. \& Halpern, J. P. 2004, ApJS, 150, 181

\bibitem[Ferrarese \& Merrit(2000)] {Ferrarese00} Ferrarese, L., \& Merritt, D. 2000, ApJ, 539, L9


\bibitem[Gebhardt et al.(2000)]{Gebhardt00} Gebhardt, K., Bender, R., Bower, G., et al. 2000, ApJ, 539, L13



\bibitem[Greene 
\& Ho(2005)]{Greene05} Greene, J. E., \& Ho, L. C. 2005a, ApJ, 627, 721

\bibitem[Grupe \& Mathur(2004)]{Grupe04} Grupe, D. \& Mathur, S., 2004, ApJ, 606, L41

\bibitem[Harris et al.(2012)]{Harris+12} Harris, C. E., Bennert V. N., Auger M. W., Treu T., Woo J.-H., Malkan M.
A., 2012, ApJS, 201, 29

\bibitem[Harrison et al.(2014)]{Harrison+14} Harrison, C. M., Alexander, D. M., Mullaney, J. R., Swinbank, A. M., 2014,
MNRAS, 441, 3306

\bibitem[Heckman et al.(1984)]{Heckman84} Heckman, T. M., Miley, G. K., \& Green, R. F. 1984, ApJ, 281, 525

\bibitem[Ho(2009)]{Ho09} Ho, L.~C. 2009, ApJ, 699, 638

\bibitem[Hu et al.(2008)]{Hu08} Hu, C., Wang, J. M., Ho, L. C. et al. 2008, ApJ, 687, 78

\bibitem[Komossa \& Xu(2007)]{Komossa07} Komossa, S. \& Xu, D., 2007, ApJ, 667, L33

\bibitem[Komossa et al.(2008)]{Komossa08} Komossa, S., Xu, D., Zhou, H., Storchi-Bergmann, T., \& Binette, L. 2008, ApJ,
680, 926

\bibitem[Kormendy \& Ho(2013)]{Kormendy&Ho13} Kormendy, J., \& Ho, L.~C.\ 2013, ARA\&A, 51, 511

\bibitem[LaMassa et al.(2009)]{LaMassa+09} LaMassa, S. M., Heckman, T. M., Ptak, A., et al. 2009, ApJ, 705, 568

\bibitem[Le et al.(2014)]{Le+14} Le, H. A. N., Pak, S., Im, M., et al. 2014, JASR, 54, 6 

\bibitem[Le et al.(2017)]{Le+17} Le, H. A. N., Woo, J-.H., Karouzos, M. et al. 2017, ApJ, 851, 8

\bibitem[Le \& Woo(2019)]{Le+19} Le, H. A. N. \& Woo, J-.H.\ 2019, \apj, 887, 246

\bibitem[Le, Woo \& Xue(2020)]{Le+20} Le, H. A. N., Woo, J-.H., \& Xue, Y. Q. 2020, ApJ, 901, 35

\bibitem[Lin et al.(2022)]{Lin+22} Lin, X., Xue, Y. Q., Fang, G. et al. 2022, RAA, 22, 015010

\bibitem[McLure 
\& Dunlop(2004)]{MD04} McLure, R.~J., \& Dunlop, J.~S.\ 2004, \mnras, 352, 1390 

\bibitem[McLure 
\& Jarvis(2002)]{MJ02} McLure, R.~J., \& Jarvis, M.~J.\ 2002, \mnras, 337, 109 

\bibitem[Nelson \& Whittle(1996)]{Nelson+96} Nelson, C. H. \& Whittle, M., 1996, ApJ, 465, 96

\bibitem[Nelson(2000)]{Nelson20} Nelson, C. H., 2000, ApJ, 544, 91

\bibitem[Netzer \& Trakhtenbrot(2007)]{Netzer+07} Netzer, H., Trakhtenbrot, B., 2007, ApJ, 654, 754

\bibitem[Oke et al.(1995)]{Oke+95} Oke, J.~B., Cohen, J.~G., Carr, M., et al.\ 1995, \pasp, 107, 375 

\bibitem[Park et al.(2015)]{Park+15} Park, D., Woo, J.-H., Bennert, V. et al.\ 2015, \apj, 799, 164

\bibitem[Peng et al.(2006)]{Peng+06} Peng, C. Y., Impey, C. D., Rix, H.-W., et al. 2006, ApJ, 649, 616

\bibitem[Phillips et al.(1986)]{Phillips86} Phillips, M. M., Jenkins, C. R., Dopita, M. A., Sadler, E. M., \& Binette, L.
1986, AJ, 91, 1062

\bibitem[Rice et al.(2006)]{Rice+06} Rice, M., Martini, P., Greene J. E., Pogge, R. W., Shields J. C., Mulchaey, J.
S., Regan, M. W., 2006, ApJ, 636, 654

\bibitem[Salviander et al.(2007)]{Salviander07} Salviander, S., 
Shields, G.~A., Gebhardt, K., \& Bonning, E.~W.\ 2007, \apj, 662, 131 

\bibitem[Salviander \& Shields(2013)]{Salviander13} Salviander, S. \& Shields, G. A, 2013, ApJ, 662, 13

\bibitem[Salviander, Shields \& Bonning(2015)]{Salviander15} Salviander, S., Shields, G. A., Bonning, E. W., 2015, ApJ, 799, 173

\bibitem[Shankar et al.(2016)] {Shankar+16} Shankar, F., Bernardi, M., Sheth, R. K., et al.
2016, MNRAS, 460, 3119

\bibitem[Shankar et al.(2019)] {Shankar+19} Shankar, F., Weinberg, D. H., Marsden, C. et al.\ 2019, MNRAS, 493, 1500

\bibitem[Shen et al.(2015)]{Shen+15} Shen, Y. et al.\ 2015, \apj, 805, 96 

\bibitem[Shen et al.(2016)]{Shen+16} Shen, Y. et al.\ 2016, \apj, 831, 7 

\bibitem[Shields et al.(2003)]{Shields+03} Shields, G. A., Gebhardt, K., Salviander, S., Wills, B. J., Xie, B., Brotherton, M. S., Yuan, J., \& Dietrich, M. 2003, ApJ, 583, 124

\bibitem[Silk \& Rees(1998)]{Silk98} Silk, J., \& Rees, M. J. 1998, A\&A, 331, L1

\bibitem[Sexton et al.(2020)]{Sexton+20} Sexton, R. O., Matzko, W., Darden, N., Canalizo, G., \&
Gorjian, V. 2020, MNRAS, 500, 2871

\bibitem[Sun et al.(2015)]{Sun+15} Sun, M., Trump, J. R., Brandt, W. N., et al. 2015, \apj, 802, 14

\bibitem[Terlevich, Diaz \& Terlevich(1990)] {Terlevich90} Terlevich, E., Diaz, A. I., Terlevich, R., 1990, MNRAS, 242, 271

\bibitem[Trakhtenbrot \& Netzer(2012)]{Trakhtenbrot+12} Trakhtenbrot, B., \& Netzer, H.\ 2012, \mnras, 427, 3081

\bibitem[Tremaine et al.(2002)]{Tremaine+02} Tremaine, S., Gebhardt, K., Bender, R., et al.\ 2002, \apj, 547, 2


\bibitem[Valdes et al.(2004)]{Valdes04} Valdes, F., Gupta, R., Rose, J. A., Singh, H. P., \& Bell, D. J. 2004, ApJS,
152, 251

\bibitem[Wang \& Lu(2001)]{Wang01} Wang, T. \& Lu, Y. 2001, A\&A, 377, 52

\bibitem[Wilson \& Heckman(1985)]{Wilson85} Wilson, A. S., \& Heckman, T. M. 1985, in Astrophysics of Active Galaxies and Quasi-Stellar Objects, ed. J. S. Miller (Mill Valley: University Science Books), 39

\bibitem [Whittle(1992)] {Whittle92} Whittle, M., 1992, ApJ, 387, 109

\bibitem[Woo et al.(2006)]{Woo06} Woo, J.-H., Treu, T., 
Malkan, M.~A., \& Blandford, R.~D.\ 2006, \apj, 645, 900 

\bibitem[Woo et al.(2008)]{Woo08} Woo, J.-H., Treu, T., 
Malkan, M.~A., \& Blandford, R.~D.\ 2008, \apj, 681, 925 

\bibitem[Woo et al.(2010)]{Woo+10} Woo, J.-H., Treu, T., Barth, A.~J., et al.\ 2010, \apj, 716, 269 

\bibitem[Woo et al.(2015)]{Woo15} Woo, J.-H., Yoon, Y., Park, S. et al.\ 2015, \apj, 801, 38

\bibitem[Woo et al.(2016)]{Woo+16} Woo, J.-H., Bae, H.-J., Son, D., \& Karouzos, M.\ 2016, \apj, 817, 108 

\bibitem[Woo et al.(2018)]{Woo+18} Woo, J.-H., Le, H. A. N., Karouzos, M., et al.\ 2018, \apj, 859, 138 

\bibitem[Woo et al.(2019)]{Woo+19b} Woo, J.-H., Cho, H., Gallo, E., et al. 2019, Nature
Astronomy, 3, 755

\bibitem[Xue et al. (2010)]{Xue+10} Xue, Y. Q., Brandt, W. N., Luo, B., et al. 2010, \apj, 720, 368

\bibitem[Xue(2017)]{Xue+17} Xue, Y. Q. 2017, NewAR, 79, 59 

\bibitem[York et al.(2000)]{York+20} York, D. G., Adelman, J., Anderson, Jr., J. E., et al. 2000, AJ, 120, 1579

\bibitem[Zamanov et al.(2002)]{Zamanov02} Zamanov, R., Marziani, P., Sulentic, J. W., et al. 2002, ApJL, 576, L9

\bibitem[Zhang et al.(2011)]{Zhang11} Zhang, K., Dong, X-. B., Wang, T-. G. \& Gaskell, C. M. 2011, ApJ, 737, 71

\end{thebibliography}
\end{document}